\documentclass[leqno]{article}
\usepackage{color,mathtools, amsmath, amssymb, enumitem, bbm, amsthm,
  hyperref, graphicx, setspace, tikz-cd,placeins, changepage,xspace,
  shuffle, blkarray,booktabs}
\usepackage[margin=1in]{geometry}
\usepackage[font=small,margin=0.5in]{caption}
\usepackage[textsize=small]{todonotes}
\usepackage{comment}
\usepackage{svg}
\usetikzlibrary{matrix}
\usetikzlibrary{shapes.misc}
\tikzset{cdlabel/.style={execute at begin node=$\scriptstyle,execute at end node=$}}
\usepackage{chngcntr}
\usepackage{thm-restate}

\hypersetup{bookmarks=true, bookmarksopen=true,%
    bookmarksdepth=3,%
    colorlinks=true,%
    linkcolor=blue,%
    citecolor=blue,%
    filecolor=blue,%
    menucolor=blue,%
    urlcolor=blue}

\numberwithin{equation}{section}
 
\newtheorem{theorem}{Theorem}[section]
\newtheorem{corollary}[theorem]{Corollary}
\newtheorem{lemma}[theorem]{Lemma}

\newtheorem{proposition}[theorem]{Proposition}

\newtheorem{conj}[theorem]{Conjecture}

\theoremstyle{definition}
\newtheorem{definition}[theorem]{Definition}
\newtheorem{remark}[theorem]{Remark}
\newtheorem{example}[theorem]{Example}

\newcommand{\R}{\mathbb{R}}

\newcommand{\C}{\mathbb{C}}

\newcommand{\Sh}{\mathrm{Sh}} %shiftahedron

\newcommand{\st}{ST} % spanning trees
\newcommand{\dd}{\Delta} % distributions
\renewcommand{\SS}{\mathsf{s}} % vector s=(s_1,...,s_n)
\newcommand{\MSTZ}{{\rm MST}_0} % MST_0
\newcommand{\prob}{\mathbb{P}} % probability
\renewcommand{\P}{\mathbb{P}}
\newcommand{\M}{\mathcal{M}}
\newcommand{\D}{\mathcal{D}}

\DeclareMathOperator{\Edges}{E}
\newcommand{\MST}{{\mathrm{MST}}}
\newcommand{\UST}{{\mathrm{UST}}}
\newcommand{\Kru}{{\mathrm{Kru}}}
\newcommand{\RD}{{\mathrm{RD}}}
\newcommand{\wt}[1]{\widetilde{#1}}

\newcommand{\cH}{\mathcal{H}}
\newcommand{\cW}{\mathcal{W}}
\newcommand{\cM}{\mathcal{M}}

\newcommand{\cF}{\mathcal{F}}

\newcommand{\bP}{\mathbb{P}}

\newcommand{\al}{\alpha }

\newcommand{\si}{\sigma }

\newcommand{\Om}{\Omega }

\newcommand{\ones}{\mathbbm{1}}

\newcommand{\rank}{\operatorname{rank}}

\newcommand{\bi}{\begin{itemize}}
\newcommand{\ei}{\end{itemize}}

\usepackage[linesnumbered,algoruled,boxed,lined]{algorithm2e}
\definecolor{myOrange}{HTML}{FF8000}
\definecolor{myGreen}{HTML}{009900}
\definecolor{myBlue}{HTML}{3333FF}
\definecolor{myGold}{HTML}{CCCC00}
\definecolor{myRed}{HTML}{FF0000}
\definecolor{myTurquoise}{HTML}{00CCCC}
\definecolor{myPurple}{HTML}{9933FF}
\definecolor{myBrown}{HTML}{D9B857}
\definecolor{myMaroon}{HTML}{800000}
\definecolor{green}{HTML}{009900}

\newcommand{\rr}{\mathbb R}

\newcommand{\abs}[1]{\left|{#1}\right|}

\newcommand{\suchthat}{\ | \ }

\newcommand{\tth}{^\text{th}}
\newcommand{\genseq}[3]{{#1}_1 {#3} {#1}_2 {#3} \dots {#3} {#1}_{#2}}
\newcommand{\seq}[2]{\genseq{#1}{#2}{,}}

\newcommand{\txt}[1]{\text{#1}}

\makeatletter 
\g@addto@macro{\@algocf@init}{\SetKwInOut{Parameter}{Parameters}} 
\makeatother
\makeatletter
\@addtoreset{algocf}{section}

\makeatother

\newcommand{\bC}{\mathbb{C}}
\newcommand{\bR}{\mathbb{R}}
\newcommand{\bQ}{\mathbb{Q}}
\newcommand{\oeis}[1]{\href{http://oeis.org/#1}{{#1}}}

\newcommand{\abcd}[4]{{\begin{bsmallmatrix}
#1\\#2\\ #3 \\ #4 \end{bsmallmatrix}}}

\definecolor{alizarin}{rgb}{0.82, 0.1, 0.26}
\definecolor{tealgreen}{rgb}{0.0, 0.51, 0.5}

\newcommand{\Both}{{\tt Both}\xspace}
\newcommand{\Right}{{\tt Right}\xspace}
\newcommand{\Left}{{\tt Left}\xspace}
\newcommand{\Neither}{{\tt Neither}\xspace}

\newcommand{\bari}[1]{\overline{e_{#1}}}

\newcommand{\ordab}[2]{\scalebox{.85}{$(#1 < #2)$}}
\newcommand{\ordabc}[3]{\scalebox{.85}{$(#1 < #2 < #3)$}}
\newcommand{\ordabcd}[4]{\scalebox{.85}{$(#1 < #2 < #3 < #4)$}}

\newcommand{\E}{\mathcal{E}}

\newcommand{\supp}{\txt{supp}}

\graphicspath{{./images/}}

\begin{document}
\title{Models of random spanning trees}
% Quantitative properties of random spanning trees
% Product measures on (spanning) trees
% Generalizing MST
% Minimum spanning trees with general random weights
% Random minimal spanning tree distributions 
% Random spanning trees 
% Quantitative properties of generalized Kruskal 
%
\author{
Eric Babson\thanks{Department of Mathematics, University of California, Davis, Davis, CA, USA}%
\and
Moon Duchin\thanks{Department of Computer Science and Data Science, University of Chicago, Chicago, IL, USA}%
\and
Annina Iseli\thanks{Institut de mathématiques, École polytechnique fédérale de Lausanne (EPFL), Lausanne, Switzerland}%
\and
Pietro Poggi-Corradini\thanks{Department of Mathematics, Kansas State University, Manhattan, KS, USA}%
\and
Dylan Thurston\thanks{Department of Mathematics, Boston College, Chestnut Hill, MA, USA}%
\and
Jamie Tucker-Foltz\thanks{Yale School of Management and Computer Science Department, Yale University, New Haven, CT, USA}%
}

\maketitle
\begin{abstract}
  There are numerous randomized algorithms to generate spanning trees in a given ambient graph; several target the uniform distribution on trees (UST), while in practice the fastest and most frequently used draw random weights on the edges and then employ a greedy algorithm to choose the minimum-weight spanning tree (MST). Though MST is a workhorse in applications, the mathematical properties of random MST are far less explored than those of UST. In this paper we develop tools for the quantitative study of random MST. We consider the standard case that the weights are drawn i.i.d.\ from a single distribution on the real numbers, as well as successive generalizations that lead to \emph{product measures}, where the weights are independently drawn from arbitrary distributions.
\end{abstract}

%\tableofcontents

\section{Introduction}\label{secIntro}

The theoretical computer science literature includes a wide variety of randomized algorithms for choosing a spanning tree of a graph $G$.  First, it is natural to try to sample uniformly: We know how to count the number of spanning trees of any graph from Kirchhoff's Matrix--Tree formula \cite{MatrixTreeTheorem}, and from this it is possible to build an efficient uniform sampler. In the late 1980s and 1990s, Aldous \cite{Aldous}, Broder \cite{Broder}, and most notably Wilson \cite{wilson} devised simple, faster algorithms that build uniformly random trees via random walks on $G$. In a series of papers starting in the late 2010s, Anari, Liu, Oveis Gharan, Vinzant and various collaborators have had a series of results leveraging matroid theory to define a Markov chain process that they showed to be rapidly mixing, converging to the uniform distribution \cite{Matroid1,Matroid2}.

However, a different algorithm is ubiquitous in practical applications, and it is so fundamental that it is often at the heart of a first course in discrete mathematics: assign every edge a random weight and then find a minimum spanning tree with respect to those weights, which can be computed greedily using  Kruskal's algorithm \cite{kruskal:1956}.  Unlike the previous methods, this does not target the uniform distribution.  Our motivation in this paper is to advance the systematic study of the differences between uniform spanning trees (UST) and different versions of minimum spanning trees (MST).  
It is already of interest simply to learn more properties of ordinary MST (with weights drawn i.i.d.\ from $[0,1]$), since it is widely used but not (quantitatively) well studied.

The simplest example of a graph in which UST is different from MST is a square with a diagonal (four vertices, five edges).  Here, one easily verifies that there are eight possible spanning trees; four involve the diagonal and four do not.  Under UST, they are all equally likely, but under ordinary MST, the ones with the diagonal collectively have probability $8/15$.  
It turns out that a small modification to the edge weights will restore the greedy process to uniformity:  if the outer edges have weights drawn from the same interval $[0,1]$ (say), but the diagonal is weighted from a slightly perturbed interval $[\varepsilon,1+\varepsilon]$, then the minimum-weight tree is uniformly distributed.\footnote{The exact value of $\varepsilon \approx 0.03$ needed to recover UST has no nice closed form; it is the real root of the quintic polynomial $6x^5 - 20x^3 + 30x - 1$.}  This is our first example of a non-i.i.d.\ {\em product measure} on the edges---that is, a choice of real-valued distribution on each edge.   This example is a special form of generalized MST, that of drawing uniformly from shifted intervals.

To motivate pushing to the next level of complexity, we next consider $K_4$, the complete graph on four vertices.  In this case, we can once again construct a product measure on the edges so that the minimum-weight tree recovers UST.  However, shifted intervals are provably not enough and we must resort to more general measures.  Fix a small $\varepsilon$ and draw the edge weight on one edge from $\frac 12 \delta_\varepsilon+\frac 12 \delta_1$ (where $\delta_x$ denotes the atomic probability distribution supported on $\{x\}$) and on the opposite edge (the one that shares no endpoint with the first) from 
$\frac 12 \delta_0+\frac 12 \delta_{1-\varepsilon}$.  The other four are drawn uniformly from $\left[ \frac 14,\frac 34\right]$.  This is a more general example of a product measure, and the reader can verify that this weighting scheme makes all spanning trees equally likely.

Stepping back, we can seek to understand the space of probability distributions on minimum spanning trees by first understanding the achievable distributions of weights. The question of what distributions can be achieved on $\R^m$ from $m$ independent random variables is fundamental, and one aspect of the question has received significant study for many decades:  the projection onto pairwise comparisons.  In 1959, Steinhaus and Trybu\l a began to study what they called ``the paradox of three random variables," namely the fact that triples exist for which $\P(X>Y)$, $\P(Y>Z)$, $\P(Z>X)$ are all more than $\frac 12$ \cite{ST:ParadoxApplProb}.  Later this would be called the problem of ``intransitive dice" (see \S\ref{subLitReview} for further discussion). 
For our application, we require more information than the projection onto pairwise comparisons.  
Since Kruskal-style selection of the minimum-weight tree is greedy, it depends only on the order of the edge weights. That is, a draw from a product measure induces a permutation via the order of the weights, and the permutation suffices to specify a choice of tree.  
So if we write $\M=\M_m$ for the space of non-colliding product measures on $m$ real-valued random variables (i.e., such that the probability is zero that any two draws are equal) and 
$\Delta(S_m)$ for the probability distributions on permutations of $m$ symbols, then 
we have defined a map $\psi\colon\M \to \Delta(S_m)$ and 
we set out to describe the image 
$P_m=\psi(\M)$ of product measures.   Surprisingly, though this appears to be a very  natural and elementary question, we have not found this much discussed in the probability literature.

\subsection{Summary of contributions}

We study successive generalizations of the ordinary MST theory. Given a finite graph $G=(V,E)$ with vertex set $V$ and edge set $E$, write $(X_i)_{i\in E}$ for a collection of real-valued random variables indexed by $E$.  We will focus much of our study on random spanning trees which arise as minimum-weight trees given edge weights drawn independently from the $X_i$.

\begin{itemize}
    \item \emph{Ordinary MST} (also denoted $\MSTZ$) is a
      specific product measure in which there is a single measure $X$ and all $X_i\sim X$.
      Without loss of generality, the weights are drawn uniformly from $[0, 1]$.  (\S\ref{secOrdinaryMST})
\end{itemize}

Theorems~\ref{thm:rd-global} and \ref{thmInternalFormula} give exact
formulas for the $\MSTZ$ probability of any particular spanning tree
in a graph, but
they involve sums over permutations of edges outside
and inside the tree, respectively.
Though this can be improved from factorial time to exponential time,
it is still
more useful for insight than for practical calculation in large graphs. We show that the probability of
sampling any (labeled) star in a $K_n$ is exactly $1/(2n-3)!!$. We
then develop a pair of ``rotation tricks" that let us show in
Theorem~\ref{thm:stars-paths} that in fact stars have the highest
probability weight of any labeled trees and paths have the lowest. In
the process we also apply rotations to show in
Theorem~\ref{thm:most-graphs-mst-ust} that random graphs have
$\MSTZ\neq \UST$.

\begin{itemize}      
    \item  \emph{Shifted-interval MST} is the class of product measures where each  $X_i$ is uniform on
      a unit interval $[s_i,s_i+1]$. We briefly consider more general
      \emph{connected-support} measures, where the support of
      each edge distribution is a connected subset of $\R$. (\S\ref{secConnectedSupport})
\end{itemize}

We build a parameter space for shifted-interval measures $M_\SS\subset \M$ and show in Theorem~\ref{MST_not_UST_Kn} that shifts do not suffice to recover the uniform distribution on spanning trees for $K_n$, $n\ge 4$.  In the process, we motivate the study of arbitrary product measures by showing some limitations of measures with connected support.

\begin{itemize}   
    \item \emph{Arbitrary product measures} have no restrictions other than the non-colliding property (the probability of drawing two identical values is zero).
    (\S\ref{secArbitraryProductMeasures})
\end{itemize}

To analyze arbitrary product measures, we develop a discrete abstraction called  {\em weighted words} that helps isolate the combinatorial aspects of the problem.  We obtain
Theorem~\ref{thm:ProductMeasuresDiscrete}, which states that every non-colliding product measure on $m$ variables can be represented by a weighted word of finite length (bounded in terms of $m$). 
We construct universal words that suffice to hit all of $P_m$ as their weights vary, and we employ classical quadrature schemes (from the theory of integration) to give efficient words that induce the uniform measure on permutations.  
Finally, we study the dimension of $P_m$ and prove an upper bound for
it in Theorem~\ref{thm:dim-Pn-upper}, also verifying computationally
that the upper bound is an equality for $m\le 7$.

%%%%
\subsection{An application of generalized MST}\label{subsec:ReCom-motivation}

Some of the original motivation for this study came from shifted-interval MST because it is already in practical use in a popular graph algorithm.  Random spanning trees are a key ingredient in so-called {\em recombination} algorithms that step from one random graph partition to another \cite{recom}.  
Recombination runs by iterating a merge-split procedure:
two parts of a graph partition are fused, and then a random spanning tree is used to divide them in a new way.  A principal application is for generating random political districting plans.  
When users desire to make it more likely that two nodes are placed in {\em different} parts of a partition, they can add a positive ``surcharge" to the weight on the edge between those nodes.
When a collection of contiguous nodes makes up a region that users prefer to keep in the {\em same} piece, the same idea can be used to surcharge the boundary edges of the region.  This has the effect that minimum spanning tree is more likely to restrict to a tree on the designated region; in the bipartition step, that means the region will be kept whole or split at most once.  

\begin{figure}[bht!]
    \centering
    \includegraphics[width=6in]{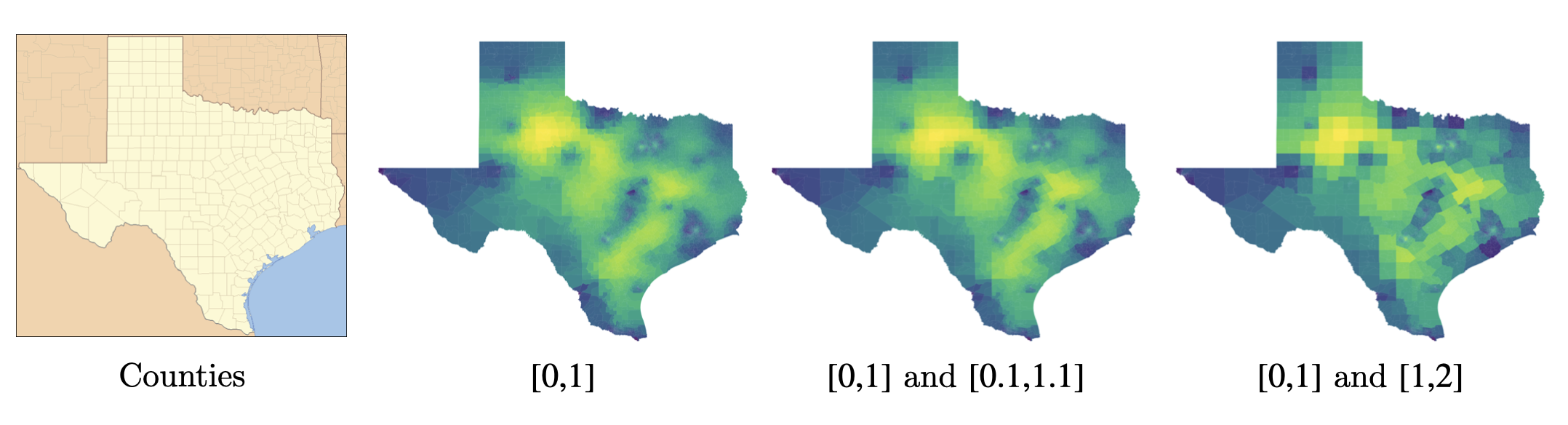}
    \caption{In \cite{MRL}, a ``recombination" Markov chain is run to draw random partitions of Texas into 150 legislative districts.  The Markov chain combines and re-splits two districts at a time by drawing and bisecting a random spanning tree. 
    These figures show heatmaps from three different runs, with the $\mathnormal{\sim}\relax 9000$ precincts of Texas colored on a scale from dark blue (reassigned rarely) to yellow (reassigned frequently).  
    When the spanning tree step uses MST weights drawn from $[0,1]$, there is no particular relationship to county boundaries.  As between-county edges are surcharged by $s=0.1$ (middle) and then $s=1.0$ (right), county boundaries become visible, since the steps tend to keep counties intact within the partition and so reassign whole counties at a time.}
    \label{fig:Texas}
\end{figure}

A primary example is keeping counties together, which is a traditional preference in forming districting plans.  If within-county edges are drawn from $[0,1]$ and between-county edges from a shifted interval $[s,s+1]$ for $0< s\le 1$, then for higher values of $s$ far fewer counties will be split over the long term, a phenomenon illustrated in Figure~\ref{fig:Texas}.  (There is no further effect for $s$ larger than 1, because at that point the surcharged weights are necessarily larger than the $[0,1]$ weights, which is all that is seen by Kruskal's algorithm.)

Empirically, this works very well to keep designated regions intact in a random partition process.  But theoretically, it is difficult to characterize the distribution induced by the shift.

%%%
\subsection{Related work}\label{subLitReview}

The literature giving quantitative contrasts between UST and ordinary MST is limited.
The Lyons--Peres book {\em Probability on Trees and Networks} \cite{lyons-peres} is a valuable and thorough reference in which Chapter 11 is devoted to minimal spanning forests in the context of percolation theory.  In the setting of ``wired" and ``free" minimal spanning forests on infinite graphs, the authors give results on degree distribution, component sizes, and the number of ends.
A paper by Garban, Pete, and Schramm~\cite{garban-pete-schramm:2018} explores the scaling limit of MST on triangular lattices in the plane. The authors show invariance under rotations, scalings and translations, but do not expect full conformal invariance (as in the case of SLE) to hold.
Simultaneous recent works by Makowiec-Salvi-Sun~\cite{RSTinRandomEnv} and K\'usz~\cite{Kusz} compare MST to UST by creating Boltzmann-like distributions on spanning trees that interpolate between them.

In settings closer to ours, one of the striking quantitative comparison results known to us is in a paper by  Addario-Berry, Broutin, and Reed~\cite{addarioberry-broutin-reed:2009}, where
the authors show that the expected diameter of a spanning tree of the complete graph $K_n$ is on the order of $n^{1/3}$ under $\MSTZ$---in contrast to the classical result that the expected UST diameter is $n^{1/2}$.
%Earlier, Kruskal's algorithm  was used by Chopra \cite{chopra:1989} to describe the spanning tree polyhedron obtained by taking the convex hull of the indicator functions for spanning trees in~$\R^E$.
More recently, Tapp~\cite{TappGridMST} considers $\MSTZ$ on square grid graphs, giving upper and lower bounds for the probability of sampling a given tree. He also provides conjectures about the most and least likely tree structures on grid graphs---we  settle the corresponding conjecture for random spanning trees in complete graphs (Theorem~\ref{thm:stars-paths}).

\begin{figure}[htb!]
  \centering
  \includegraphics[height=2in]{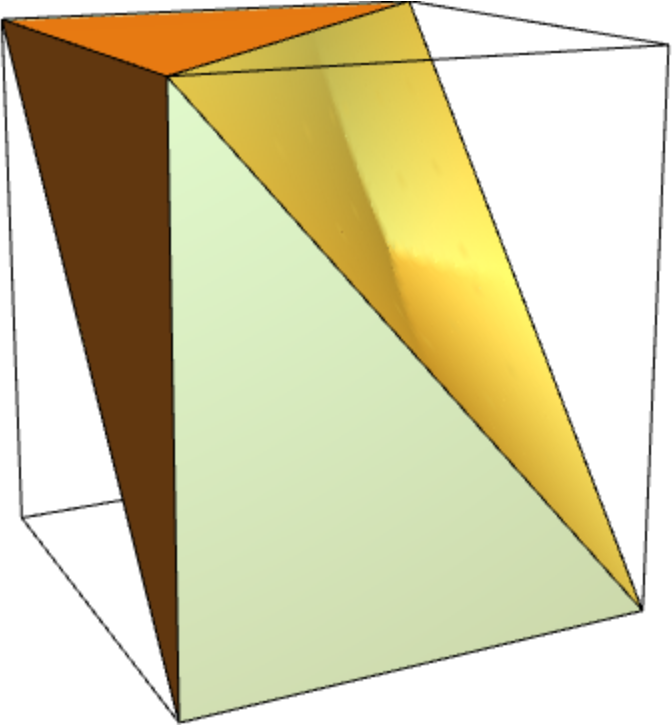}
  \caption{The Trybu\l a region $T_3$.
    Given independent random variables $X_1,X_2,X_3$ that are the components of a non-colliding product measure $\D$, the three coordinate axes are $x = \prob_\D(X_1>X_2)$,
$y= \prob_\D(X_2>X_3)$, and $z = \prob_\D(X_3>X_1)$.  The vertices of the cube  that are hit by $T_3$ correspond to the pure permutations (for example, $(1,1,0)$ comes from $X_1>X_2>X_3$), whereas $(0,0,0)$ and $(1,1,1)$ are not hit.}
  \label{fig:Trybula}
\end{figure}

Above, we described a question about $P_m \subset \Delta(S_m)$  that is well-studied:
{\em Can you build
$n$ weighted dice so that each one beats the next (cyclically) with probability
greater than $50\%$?}  
After Steinhaus-Trybu\l a gave a positive answer in 1959 \cite{ST:ParadoxApplProb},
Trybu{\l}a launched a quantitative investigation in
\cite{Trybula:Three,Trybula:N} whose development was later surveyed by Savage in \cite{Savage:Dice}. 
To state some results, consider the set $T_m$ which is the projection of
$P_m \subset \Delta(S_m) \subset \R^{m!-1}$ to the much
smaller-dimensional set $[0,1]^{\binom{m}{2}}$, only keeping track of
the pairwise probabilities. 
For example, the point $\left( \frac 12, \frac 12, \frac 12\right) \in T_3$ is realized by standard dice, and the paradox amounts to the observation that all three coordinates can exceed $\frac 12$.
The Trybu\l a region $T_3$ is shown in  Figure~\ref{fig:Trybula}. 

On general principles, the feasible locus $T_3$ is
contained in the convex hull of the six vertices defined by pure permutations, but $T_3$ is not quite
convex: as formulated by Suck \cite{Suck:Utility}, it is the region defined by
% \begin{align*}
%   \min \bigl(p_{12}+p_{23}p_{31}, p_{23}+p_{31}p_{12},
%     p_{31}+p_{12}p_{23}\bigr) &\le 1 \\
%   \min \bigl(p_{21}+p_{32}p_{13}, p_{32}+p_{13}p_{21},
%     p_{13}+p_{21}p_{32}\bigr) &\le 1.
% \end{align*}
$$
  \min \bigl(x+yz, y+xz,
    z+xy\bigr) \le 1; \qquad  
  \min \bigl(\bar x +\bar y \bar z, \bar y + \bar x \bar z,
    \bar z + \bar x \bar y\bigr) \le 1,
$$
where $x,y,z$ are the probabilities that $X_1>X_2$, $X_2>X_3$, $X_3>X_1$, respectively, and $\bar x=1-x$, $\bar y = 1-y$, $\bar z= 1-z$ are the complementary probabilities.  
There are still open questions about $T_m$; for instance, 
Komisarski only recently (2021) found the maximum extent of intransitivity 
 \cite{komisarski}. Concretely, for $m \ge 3$, Komisarski maximizes $p_m$ for which there exist
dice $(X_i)_{i=1}^m$ where $X_i$ beats $X_{i+1}$
with probability $p_m$. 

Much less is known about the full permutation locus $P_m$, even for $m=3$.  We will discuss it in \S\ref{subsec:Dimension}.

%%%
\subsection*{Acknowledgements}
We gratefully acknowledge  the Simons-Laufer Mathematical Sciences Institute (SLMath) for hosting subsets of the authors in two recent semesters, during the thematic programs on the Analysis and Geometry of Random Spaces (AGRS) and Algorithms, Fairness, and Equity (AFE).
The authors thank the National Science Foundation for partial support through  grants DMS-2005512 (MD), 
DMS-2154032 (PPC), DMS-2110143 (DT), 
and  DGE-1745303 (JTF), and the University of Fribourg for postdoc travel funding (AI).
We also thank Peter Rock for his work on the replication repository
and Nantel Bergeron, Adi Gl\"ucksam, and Russell Lyons for many extremely helpful conversations.

%%%%%%%%%
%%%%%%%%%

\section{Preliminaries}\label{secPreliminaries}

We first set some notation.  Recall that $\Delta(A)$ is the set of distributions on a set $A$; if $A$ is a finite set, then a distribution is a non-negative function on $A$ with total mass one. For a distribution $\mathcal D \in \dd(A)$ we denote a random variable $X$ sampled from $\mathcal D$ by writing $X \sim \mathcal{D}$.
All graphs we consider are simple and undirected, with  $V(G)$ and $E(G)$ for the vertex and edge sets, and write $n=|V|$ and $m=|E|$. We write $\st(G)$ for the set of all spanning trees of $G$ and let $\UST \in \dd(\st(G))$ denote the uniform distribution on spanning trees. In other words, 
$$\forall T,T'\in \st(G)\colon\quad\prob_\UST(T)=\prob_\UST(T').$$
We will also adopt the mild abuse of notation allowing us to write
$e\in G$ rather than the fuller $e\in E(G)$, and $H\setminus e$ or
$H\cup e$ for the
subgraph of $G$ where an edge $e$ is removed from or
added to a subgraph $H$.  That is, we will minimize the use of set
notation except where the sets themselves are being emphasized, so
that $H$ can flexibly refer to a subgraph or its edge set.

For any finite nonempty set $A$, we let $S(A)$ denote the set of orderings (permutations) of $A$, in particular writing $S_n$ for the symmetric group on $n$ symbols. For a graph $G$ and a choice of distinct real-valued weights on edges, there is a unique 
{\em minimum spanning tree}, defined by the smallest possible sum of weights of included edges.  Kruskal's algorithm (1956) is based on the observation that a greedy process suffices to find it:
begin with the lowest-weighted edge, and successively add the other edges in order of weight, rejecting an edge only if it forms a cycle with others already selected \cite{kruskal:1956}.  Since this is proven to globally minimize total weight, it follows that the order of the $m$ edge weights sufficed:  the map from weights to trees factors through the permutations $S_m$ that record the order of the weights.

A \emph{(non-colliding) product measure} on a set $S$ is defined using a collection of real-valued random variables $\{X_i \suchthat i \in S\}$ such that, for all distinct $i, j \in S$, $\prob[X_i = X_j] = 0$; a set of independent draws gives the product measure. A product measure on $m$ variables induces a distribution in $\dd(S_m)$ via the order of the weights.
Therefore, as described above, a product measure on the edge set $E$ of a graph induces a distribution in $\dd(\st(G))$.

A key tool to understand variants of MST will be the use of what we call {\em broken cycles}.

\begin{definition}[Broken cycles]\label{def:broken_cycle}
Let $T$ be a spanning tree of a graph $G$ and $e\in G\setminus T$ a non-edge of the tree.  Then there is a unique cycle
$C_e=C_{T,e}$ in $G$ consisting of $e$ and the unique non-backtracking path $P_e=P_{T,e}$ in $T$ between the endpoints of $e$; we call $C_e$ 
the {\em broken cycle} for $e$ in the graph and call $P_e$ the {\em broken cycle} for $e$ in the tree.
\end{definition}

The relevance of broken cycles to MST is that Kruskal's algorithm rejects a proposed edge $e$ for inclusion in a minimum spanning tree $T$ if and only if its weight is higher than those in $P_{T,e}$.

% A second key tool that relates this work to existing literature is that of {\em edge utilization}.

% \begin{definition}[Edge utilization]\label{rmk:equal_edge_inclusion}
% For any distribution $D\in\dd(\st(G))$ on spanning trees of $G$, the {\em utilization} of an edge $e\in G$ is defined to be $\prob_D(e\in T)$.  We say that $D$ has 
% {\em equal edge utilization} if 
% $\prob_D(e\in T)$ is constant over $e\in G$.
% \end{definition}

% For any distribution on any graph $G$, the sum over $e$ of the $\prob_D(e\in T)$ is the expected number of edges in $T$, which is identically $n-1$, so if $D$ has equal edge utilization, then $\prob_D(e\in T)=(n-1)/m$. For complete graphs, clearly $\MSTZ$ and UST both have this property.

Next, we set up notation for a key criterion about when $\MST$ induces a tree $T$.

\begin{definition}[Cycle relation]
  Given a connected graph $G$ and a spanning tree $T \subset G$,
  define the \emph{cycle relation} $R_T$ between the edges and non-edges of $T$ 
(i.e., between $E(T)$ and $E(G\setminus T)$) by
  \[
    e_i \mathbin{R_T} e_j \qquad\text{iff}\qquad e_i \in P_{T,e_j},
  \]
  for $e_i \in T$, $e_j \notin T$.
Sometimes it is useful to think of $R_T$ as a subset of $E(T)\times E(G\setminus T)$ and write $(e_i,e_j)\in R_T$.
\end{definition}
In terms of this relation, it is easy to check the following.

\begin{proposition}[Cycles versus weights]\label{prop:MST-cycle-char}
  A set of distinct
  weights $\{w_i\}$ on the edges of $G$ induces  $T$ as a minimum spanning tree if and only if for all  $e \in T$,  $e' \notin T$ and corresponding weights $w,w'$,
$$e \mathbin{R_T} e' \implies w < w'.$$
\end{proposition}
\begin{proof}
Assume first that $T$ is the MST, $e \in T$, $e' \notin T$, and $e,e'$
are in the same broken cycle. If $w>w'$, then we could swap $e$ with
$e'$ and obtain a tree $T'$ with lower total weight than $T$,
contradicting the minimality of $T$.
Conversely, suppose that $T$ has the property in the statement.
If any edge is rejected in the process of running Kruskal's algorithm, it must be because it completes a cycle whose other edges have lower weight.  No edge of $T$ can have this property by assumption.  So $T$ must be the tree  constructed in a run of Kruskal's algorithm.
\end{proof}

%%%%%%%%%
%%%%%%%%%

\section{Ordinary MST}\label{secOrdinaryMST}

\subsection{Inductive formulas}\label{subClosedForm}

Let $G$ be a connected graph with $n$ vertices and $m$ edges. Given a spanning tree $T \subset G$,
%Given a connected graph~$G$ and a spanning tree $T \subset G$, 
we wish
to find $\prob_{\MST}(T;G)$, the chance that $T$ is chosen as the
minimum spanning tree when $\MST$ is applied to~$G$ with i.i.d. random edge
weights. We will give two
different inductive formulas for this, based on two algorithms from
the same paper of
Kruskal \cite{kruskal:1956}:
the first, now known as \textit{Kruskal's algorithm},
  starts with an empty forest $F$ and successively adds the  minimal-weight remaining edge  that does not create a  cycle until $F$ becomes a spanning tree.
The second, now known as the \textit{reverse-delete
    algorithm}, starts with the entire graph $H = G$ and successively
  deletes the maximal-weight edge of~$H$ that does not disconnect $H$ until $H$ becomes a tree. 

A {\em spanning subgraph} $H \subset G$ is a not-necessarily-connected subgraph with the full vertex set.  For $H \subset G$ a spanning subgraph, set
\begin{align*}
  \partial H &\coloneqq \{e \in G\setminus H \mid
             \text{$e$ connects different components of $H$}\}\\
  H^\circ &\coloneqq \{e \in H \mid
            \text{$e$ is not a separating edge of $H$}\}.
\end{align*}
We can regard $\partial H$ as the cut-set of~$H$ as a partition of $G$, and $H^\circ$ 
as the set consisting of edges lying totally within some biconnected (2-connected) component of~$H$.
Suppose that $\si\in\R^E\setminus \Delta$ is a set of distinct edge weights, i.e.,  avoiding the ``fat diagonal" $\Delta$, so that all
$\sigma(e)$ are distinct.  Let  $\cF_k$ be the family of all forests
$F\subset G$ with $|F|=k$ edges, for $k=0,\dots,n-1$. 
Let $\cH_j$ be the family of all connected spanning subgraphs $H\subset G$ with
$|H|=n-1+j$ edges, for $j=0,\dots,m-n+1$.

Kruskal's algorithm can now be phrased as the greedy process that, for
a given~$\sigma$, outputs the  (nested) tuple of spanning forests $(F_k)_{k=0}^{n-1}$ with
$F_k\in \cF_k$ given by starting with the empty forest $F_0$
(containing all the nodes but no edges) and successively taking
$F_i\coloneqq F_{i-1}\cup e$ for $e\in\partial F_k$ with minimum~$\sigma(e)$.
Likewise, the reverse-delete algorithm outputs nested connected
subgraphs $(H_j)_{j=0}^{m-n+1}$ (generated in the reverse order),
where $H_{m-n+1} = G$ and
$H_{j+1}\setminus H_j$ is the edge maximizing
$\si(e)$ for $e\in H_{j+1}^\circ$.
Kruskal proves that if all edge weights are distinct,
then $F_{n-1}=H_0$, and this is the unique minimum-weight spanning tree. 
We can introduce graph-valued random variables, $X_k$ for the $k$-edge
forests seen in the first process and $Y_k$ for the $(n-1+k)$-edge spanning
subgraphs seen in the second.
Along the way to computing $\prob_{\MST}(T)$, we will compute
intermediate probabilities
$$
 \prob_\Kru(F; G) \coloneqq \prob(X_{|E(F)|} = F)
 \quad {\rm and} \quad 
  \prob_\RD(H; G) \coloneqq \prob(Y_{|E(H)| - (n-1)} = H),
$$
suppressing $G$ from the notation when it is clear.

% For the computation based on Kruskal's algorithm, for $F \in \cF_k$ not a tree ($k < n-1$), set
% \[
%   \wt\prob_\Kru(F) \coloneqq \frac{\prob(F)}{\#\partial F}.
% \]
% It turns out that, for any $e \in \partial F$, this is $\prob(X_k = F
% \mathbin{\&} X_{k+1} = F \cup \{e\})$. That is, all feasible edges are equally likely to be added next in a randomized Kruskal's algorithm. 
% %See Lemma~\ref{lem:kruskal-next-edge}.

\begin{theorem}[Kruskal induction]\label{thm:kruskal-induction}
  For $F \subset G$ a forest consisting of $k\geq 1$ edges in a connected ambient graph $G$, the probability of encountering it in a run of Kruskal's algorithm can be computed by adding each of its edges ``last," as follows:
$$\prob_\Kru(F) =\sum_{e\in F}\prob(X_k = F \mid X_{k-1} = F \backslash e)\cdot\prob( X_{k-1} = F \backslash e) = \sum_{e \in \Edges(F)} 
    \frac{\prob_\Kru(F\setminus e)}{|\partial(F\setminus e )|}.$$
\end{theorem}
The idea is to condition on the last edge $e$ added to get $F$, and
then notice that each such edge $e$ is equally likely in $\partial(F\setminus e).$

For a tree~$T$, we can leverage this to give an inductive algorithm for computing $\prob_{\MST}(T)$ with a runtime bound on the order of $n\cdot 2^n$.
Namely, we compute $\prob_\Kru(F)$ for all of the $2^{n-1}$ subforests $F \subsetneq T$ in order from the
fewest edges to the most, using
Theorem~\ref{thm:kruskal-induction} initialized with probability $1$ of encountering the empty forest.
This needs less than $2^n$ space; 
in fact, we only need to remember one generation back, so this can be refined.   At the stage of choosing $k$ edges in the forest,  there are $k$ numbers to add, 
so the time complexity is 
$\sum k \binom{n-1}{k}=(n-1)2^{n-2}$. 
For the reverse-delete algorithm, we have an entirely analogous statement.
\begin{theorem}[Reverse-delete induction]\label{thm:rd-induction}
  For $H \subsetneq G$ a connected proper spanning subgraph of $G$,
  the probability of encountering it in a run of reverse-delete can be computed by considering the removal of each of its missing edges in turn, as follows:
$$\prob_\RD(H) =\sum_{e\in G\setminus H}\prob(Y_k=H \mid  Y_{k+1}=H\cup e)\cdot\prob(Y_{k+1}=H\cup e) = \sum_{e \in \Edges(G\setminus H)} \frac{\prob_\RD(H\cup e)}{|(H\cup e)^\circ|}.$$
\end{theorem}
For the corresponding algorithm, it suffices to store all the connected supergraphs of the tree, and there are at most $2^{m-n+1}$ of these.  For runtime, we have to add $j$ terms in each of the $\binom{m-n+1}{j}$ terms, and the sum simplifies to $(m-n+1)2^{m-n}$.

Comparing the runtimes, we find that the Kruskal induction is more efficient for computing $\prob_{\MST}(T)$ unless the graph is very sparse. In fact, in the reverse-delete case, we can often speed up this computation: frequently, the probabilities are independent of~$H$.

\begin{proposition}
  If $H \in \cH_j$ is biconnected, then all of its supergraphs in $G$ are as well, so we could have deleted down to it in any order.  That is, 
$\prob_\RD(H) = 1/\binom{m}{m-(n-1+j)}$.
\end{proposition}

There is a similar ``dual'' proposition for the Kruskal induction; instead of considering graphs for which no single edge can disconnect if removed (i.e., biconnected graphs), we would consider graphs for which no single edge could form a cycle if added.  

%%%
\subsection{Global formulas}
The formulas above are inductive; we now write global formulas directly, first  an ``external'' formula for $\prob_\MST(T)$ that corresponds to the reverse-delete algorithm and then an ``internal" formula that follows the steps of Kruskal.

Let $E=E(G\setminus T)$ be the external edges, i.e., the edges of $G$ that are not in $T$.  We first give the formula supposing we fix an order $\pi_E = (e_1,\dots,e_{m-n+1})$  of weights on $E$.
We will compute $\prob_\MST(T, \pi_E)$, by which we mean the probability in an i.i.d.\
choice of weights~$\si$ that the external
edges occur in the relative order $\pi_E$ and the minimum spanning tree is $T$.
For an edge $e\in E$, recall that $C_e \subset T\cup e$ is the
corresponding broken cycle in $G$. Define $D_k(\pi_E)= D_k \coloneqq
\bigcup_{j=1}^k C_{e_j}$ to be the union of broken cycles up to stage~$k$.

\begin{theorem}[External formula]\label{thm:rd-global}
    For any spanning tree $T \subset G$ and permutation $\pi_E$ of its non-edges,
$$\prob_\MST(T, \pi_E) = \prod_{j=1}^{m-n+1} \frac{1}{\abs{D_j}}, \qquad {\rm giving } \qquad 
    \prob_\MST(T) = \sum_{\pi_E \in S(E(G \setminus T))}
                  \prod_{j=1}^{m-n+1} \frac{1}{\abs{D_j(\pi_E)}}.$$
\end{theorem}
Note that, for given weights $\sigma$ inducing 
$\pi_E$ in the expression for $T$, the weight $\si(e_j)$ must be the largest among
any of the edges in the set~$D_j$. For a single~$j$, this happens with
probability $1/\abs{D_j}$, and the crux of the proof of
Theorem~\ref{thm:rd-global} will be to see that all of these events are
independent.

To connect this external formula to the inductive expression for reverse-delete,  set
$H_j = D_j \cup T = T \cup \{e_1,\dots,e_j\}$. These $H_j$ are the graphs that appear in the reverse-delete process, while on the other hand it is easy to see that $D_j = H_j^\circ$.

For the internal global formula, instead of fixing the order of the external edges we order the weights within the
tree: for a spanning tree and any permutation $\pi_I = (e_j)_{j=1}^{n-1}$ of the
edges in~$T$, let $\prob_\MST(T,\pi_I)$ be the probability for an i.i.d.\
choice of weights~$\si$  that the edges in $T$ are in 
weight order $\pi_I$ and  the minimum spanning tree is~$T$. Given such an
order, let $F_k = \bigcup_{j=1}^k \{e_j\} \subset T$ be the forest (or partial spanning tree) constructed from the first $k$ edges.

\begin{theorem}[Internal formula]\label{thmInternalFormula}
  For a spanning tree $T \subset G$ and a permutation $\pi_I$  of its edges,
$$\prob_\MST(T,\pi_I) = \prod_{j=0}^{n-2} \frac{1}{\abs{\partial F_j}}, \quad {\rm giving} \quad 
    \prob_\MST(T) = \sum_{\pi_I\in S(E(T))} \prod_{j=0}^{n-2} \frac{1}{\abs{\partial F_j(\pi_I)}}.$$
\end{theorem}
The proof is similar: for given weights $\sigma$ that are counted
in $\prob_\MST(T,\pi_I)$, the weight $\si(e_j)$ must be the smallest
among any of the edges in the set~$\partial F_{j-1}$. For a single~$j$,
this happens with probability $1/\abs{\partial F_{j-1}}$, and again we
must check that these are independent.
We note that the internal formula appears as Proposition 11.2 of the Lyons-Peres book \emph{Probability on Trees
and Networks} \cite{lyons-peres}, but without an argument for independence.

To illustrate the independence argument, we now run the proof of the internal formula on an example before remarking on how this generalizes to  arbitrary $T\subset G$.

%%% EXAMPLE

\begin{example}[An example in $K_5$] 
We compute the probability that the spanning tree $T$ shown in Figure~\ref{fig:Y-graph} is selected from a $K_5$ under ordinary $MST$,
with the edges appearing in the indicated order $(a,c),(b,c),(d,e),(c,d)$.
This happens iff
\begin{itemize}
\item $(c,d)$ has the smallest weight among the $6$ edges joining
  $\{a,b,c\}$ to $\{d,e\}$;
\item $(d,e)$ has the smallest weight among the $7$ edges joining any two
  of $\{a,b,c\}$, $\{d\}$, and $\{e\}$;
\item $(b,c)$ has the smallest weight among the $9$ edges joining any two
  of $\{a,c\}$, $\{b\}$, $\{d\}$, and~$\{e\}$; and
\item $(a,c)$ has the smallest weight among all $10$ edges.
\end{itemize}
That is, each edge highlighted in white on the right-hand side of Figure~\ref{fig:Y-graph}  must have the lowest weight
among the edges in its smallest enclosing region.

\begin{figure}[bht!]
    \centering
    \includegraphics[width=6in]{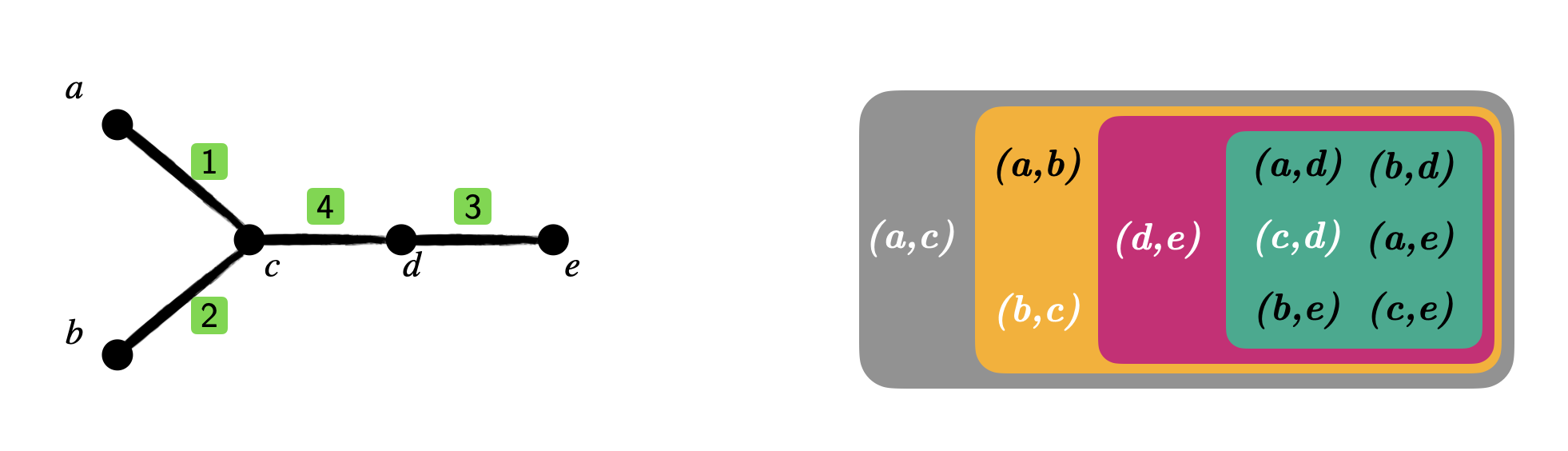}
    \caption{A  tree on five vertices, regarded as belonging to an ambient $K_5$, will be used to illustrate the independence argument for the internal formula. We suppose the edges are added in the indicated order $(a,c),(b,c),(d,e),(c,d)$ (white text).  On the right is a tiered diagram used in the argument, where the nested sets from the inside out  are $\partial F_3 \subset \partial F_2 \subset \partial F_1 \subset \partial F_0$ for the partial forests $F_j$ constructed in the course of Kruskal's algorithm.}
    \label{fig:Y-graph}
\end{figure}

These events are all independent of each other---for example, 
conditioning on $(d,e)$ having lower weight than the six edges grouped to its right 
says nothing about the relative weights among those six.
Thus the $\MSTZ$ probability of realizing $T$ with this edge order is
$\frac16 \cdot \frac17 \cdot \frac19 \cdot \frac1{10}$.  
To get the overall probability of this tree, we sum over all
permutations on these 4 edges, performing a similar calculation. 
\end{example}

\begin{proof}[Proof of Theorems~\ref{thm:rd-global} and~\ref{thmInternalFormula}]
The proof of the general internal formula follows the identical structure to the example above; given a permutation of the tree edges, it is necessary and sufficient that each edge has the lowest weight in its corresponding $\partial F_j$.  Each condition that  $e_j$ has lower weight than the $\partial F_{j-1}$ is independent of the relative ordering of the $\partial F_{j-1}$.
Theorem~\ref{thm:rd-global} has a similar structure, but using nested
sets $D_1 \subset D_2 \subset \dots \subset D_{m-n+1}$ instead of
$\partial F_{n-2} \subset \dots \subset \partial F_{0}$.
\end{proof}

%%%%%%%

\begin{corollary}[Probability of stars]
    For any (labeled) spanning tree $T_\star$ of $K_n$ that is a star, 
    $$\P_\MST(T_\star)=\frac{1}{(2n-3)!!} \quad 
    {\rm while}\quad \P_\UST(T_\star)=\frac{1}{n^{n-2}}.$$  
\end{corollary}

\begin{proof}
We apply the internal formula.
Because of the symmetry in star graphs and complete graphs, it suffices to do the computation for one ordering of the edges in the star graph and multiply by $(n-1)!$.  
Once $j-1$ edges have been added from within the star by Kruskal's
algorithm, the $j$th edge must be the lowest from $|\partial
F_{j-1}|=\binom n2 - \binom j2$ possibilities. (That is, $\partial
F_{j-1}$ is all edges of the $K_n$ minus those between the $j$ vertices touching
the $j-1$ edges added already---these can be ignored because they've
either been added already or are forbidden for completing a triangle.)
To complete the calculation, we will use two easily verified facts.  First, $\binom n2 - \binom j2 = \frac{(n+j-1)(n-j)}{2}$.
Second, using double-factorial notation for the parity-restricted factorial
$(2k-1)!!\coloneqq (2k-1)(2k-3)\cdots 3\cdot 1$, we have the identity 
$(2k-1)!! = \frac{(2k-1)!}{2^{k-1}(k-1)!}.$
With these, we get

\begin{align*}
\prob_\MST(T_\star)&=(n-1)!\prod_{j=1}^{n-1}\frac{1}{\binom n2 - \binom j2} = (n-1)!\prod_{j=1}^{n-1}\frac{2}{(n+j-1)(n-j)} \\
& = \frac{(n-1)!\cdot 2^{n-1}}{\left[ n(n+1)\cdots (2n-2)\right]
  \left[ (n-1)(n-2)\cdots 1 \right]} \\
& = \frac{2^{n-1}(n-1)!}{(2n-2)!} = 
\frac{2^{n-1}(n-1)(n-2)!}{(2n-2)(2n-3)!}=
\frac{2^{n-2}(n-2)!}{(2n-3)!}= \frac{1}{(2n-3)!!}.\qedhere
\end{align*}
\end{proof}

Unfortunately, we have no similar closed-form expression for the
probability of a given path. Nevertheless, we may compute these
probabilities for small $n$ using Theorem~\ref{thmInternalFormula}.
Each probability can be thought of as the number of permutations on
the $\binom{n}{2}$ edges that result in the given path being selected,
divided by $\binom{n}{2}!$.
Table~\ref{tabPathProbabilities} lists the first several probabilities.\footnote{The list of all numerators computed so far has been added to the Online Encyclopedia of Integer Sequences (OEIS) as entry \oeis{A374293} \cite{EIS}. We note an intriguing conjecture based on the match between this numerator sequence and a subsequence of the previously existing OEIS entry \oeis{A253950}:  for $n$ vertices and $m$ edges, each numerator is equal to the number of ordered monoids of order $m-n+3$ with a neutral element that is also the top element. The pattern holds as far as the available terms allow us to compare, which is $n = 5$.}

\begin{table}[bht!]
	\centering
	\begin{tabular}{r | r | r | l}
		$n$ & $m = \binom{n}{2}$ & Numerator: \oeis{A374293}$(n)$ & Denominator: $m!$\\\hline
        1 & 0 & 1 & 1\\
        2 & 1 & 1 & 1\\
        3 & 3 & 2 & 6\\
        4 & 6 & 44 & 720\\
        5 & 10 & 27120 & 3628800\\
        6 & 15 & 882241920 & 1307674368000\\
        7 & 21 & 2443792425984000 & 51090942171709440000\\
	\end{tabular}
	\caption{Probabilities (as numerators and denominators) of sampling a labeled path from $K_n$ under $\MSTZ$.}
	\label{tabPathProbabilities}
\end{table}

% \begin{conj}
%     The probability under $\MSTZ$ of sampling a path from a complete graph with $n$ vertices and $m$ edges is precisely $\frac{1}{m!}$ times the number of semi-groups of order $m - n + 3$ with a neutral element that is also the top element.
% \end{conj}

Replication code can be found in \cite{GitHub}.
By $m=14$, the $\UST$ probability of a path is more than three times its $\MSTZ$ probability.
In the following sections, we will introduce several  ``rotation moves" that will allow us to derive probability inequalities between related trees.  
In \S\ref{subsec:path-rotation} this will let us show rigorously what is suggested by the discussion in this section: a star is the most likely (labeled) spanning tree in a complete graph under ordinary MST, and a path is the least.

%%%%%%%%%%
%%%%%%%%%%
\subsection{Triangle-edge rotation}\label{subRotationTrick}

\begin{figure}[bht!]\centering 
    \includegraphics[width=6in]{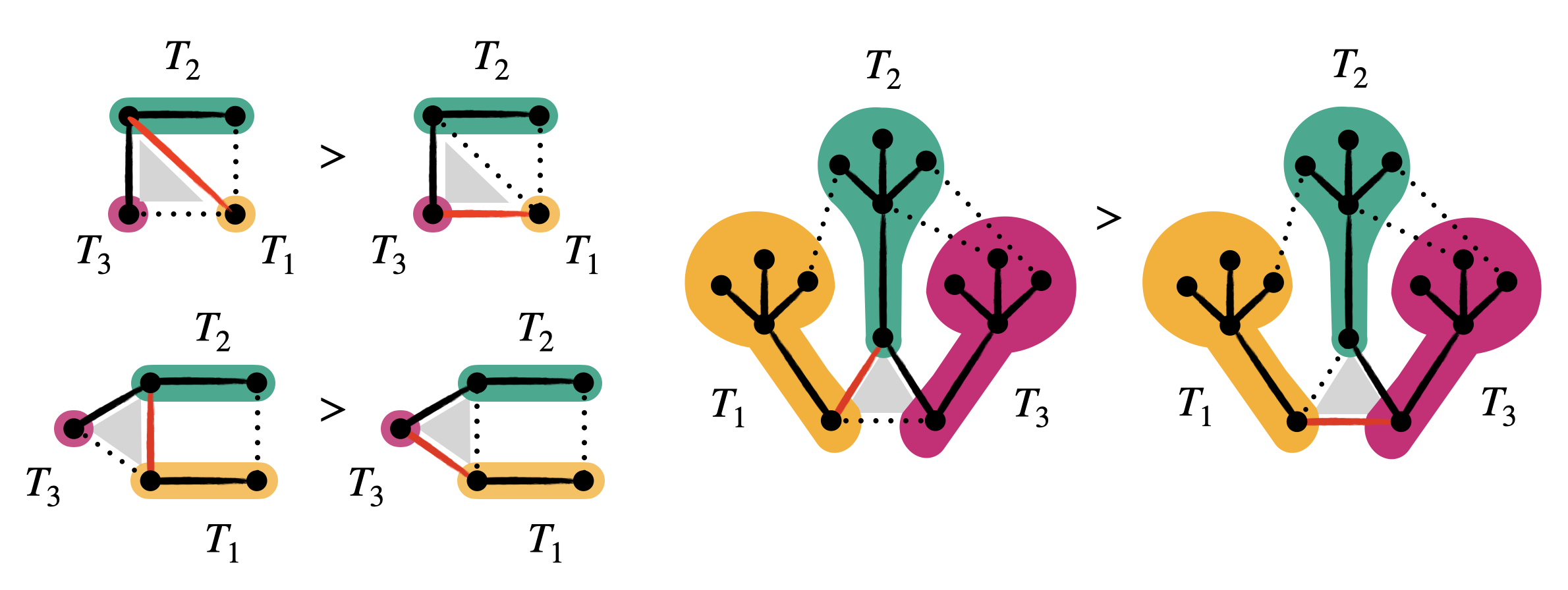}
\caption{Three examples of triangle-edge rotation, where the ambient graphs are a square with a diagonal, a house, and a 17-edge graph.  Edges in the graph but not included in the spanning trees are denoted with dotted lines. 
In each case, the three trees cited in the lemma are highlighted and the spanning trees $S,S'$ differ only by the ``rotation" of the red edge. In each case, the left-hand spanning tree $S$ is strictly more likely than the right-hand spanning tree $S'$ under ordinary MST.}
\label{fig:tri_rotation} 
\end{figure}

We start with an edge rotation lemma that is motivated by the simple square-with-a-diagonal graph.
It is not the most general rotation move we will employ, but it is useful and easy to explain. See Figure~\ref{fig:tri_rotation} for the accompanying illustration.

\begin{lemma}[Triangle-edge rotation in general graphs]
  \label{lem:tri-rotation}

Suppose a graph $G$ includes all three edges of a triangle on vertices $v_1,v_2,v_3$, denoted $e_{12},e_{23},e_{13}$.  Suppose there are disjoint trees $T_1,T_2,T_3$ containing $v_1,v_2,v_3$, respectively, so that $S=T_1\cup T_2 \cup T_3\cup e_{12}\cup e_{23}$ and $S'=T_1\cup T_2 \cup T_3\cup e_{13}\cup e_{23}$ are spanning trees of $G$, related by an ``edge rotation" based at $v_1$.
Suppose the ambient graph $G$ has at least one other edge between $T_1$ and $T_2$ and has no edges other than $e_{13}$ between $T_1$ and $T_3$ (with no conditions on edges between $T_2$ and $T_3$).  
Then there is a strict inequality for ordinary MST:
$$\prob_{\MSTZ}(S)>\prob_{\MSTZ}(S').$$
\end{lemma}
We postpone the proof of Lemma \ref{lem:tri-rotation} to the end of this section.

Note that this triangle-edge rotation has  limited usefulness for working in a complete graph, because $T_1$ and $T_3$ would have to be singletons in order to satisfy the hypothesis that there are no other edges between them than $e_{13}$.  It is not applicable to square grid-graphs, for instance, which have no triangles.  But it can be applied in many planar graphs and in random graphs, as we will see in the next section.

Recall from Proposition \ref{prop:MST-cycle-char} that a set of weights $w$ on $E(G)$ induces $T$ via Kruskal's algorithm if and only if, for every $e\in T$ and $e'\not\in T$, 
$e \mathbin{R_T} e' \implies w(e)<w(e')$. 
To motivate the next definition, note that any two spanning trees of
the same graph have the same number of edges ($n-1$), and therefore
the same number of non-edges ($m-n+1$), which are in one-to-one
correspondence with broken cycles.  In a star graph in $K_n$, all
broken cycles have length 3.  In a path graph, they come in all
lengths from 3 to $n$.  This means in the latter, there are many more
pairs $e,e'$ in the cycle relation, which means many more weight
inequalities to be satisfied.  We leverage this by defining bijections
of edges that increase broken-cycle lengths.
The number of inequalities is twice the number of non-edges, or $n^2-3n-2$, in a star, compared to $\frac 16(n^3-7n+6)$ in a path.  Of course there are various dependencies among these inequalities, so we will need controlled comparisons to prove probability bounds.

\begin{definition}[Cycle-expanding]
Let   $T_1$ and $T_2$ be two spanning trees of a graph $G$, with
  respective cycle relations $R_1$ and $R_2$. A bijection $\beta\colon E(G)\to E(G)$
  is \emph{cycle-expanding from $T_1$ to $T_2$} if
  $\beta(T_1) = T_2$ and $\beta(R_1) \subseteq R_2$, i.e., $e \mathbin{R_1} e' \implies \beta(e) \mathbin{R_2} \beta(e')$.  Equivalently, this says that  $\beta(P_{T_1,e'}) \subseteq P_{T_2,\beta(e')}$ for all $e'\notin T$---broken cycles can only get longer after applying $\beta$.
 The correspondence $\beta$ is \emph{strictly} cycle-expanding if the inclusion $\beta(R_1) \subsetneq R_2$ is strict, i.e., some cycles get strictly longer.
\end{definition}

It is easy to confirm that the examples in Figure~\ref{fig:tri_rotation} are all cycle-expanding from left to right:  the broken cycle lengths go from $(3,3)$ to $(3,4)$ for the square with a diagonal; from $(3,4)$ to $(3,5)$ for the house; and from $(3,5,6,6)$ to $(3,5,6,7)$ for the 18-edge graph.  (Here the bijection exchanges $e_{12}$ and $e_{13}$, fixing all other edges.)

\begin{proposition}[Cycle expansion  inequality]\label{prop:tri_rotation}
  If there is a cycle-expanding bijection~$\beta$ from $T$ to $T'$, then the tree with the shorter cycles is more likely:
$$\prob_{\MSTZ}(T) \ge \prob_{\MSTZ}(T').$$
Moreover, if $\beta$ is strictly cycle-expanding, then the inequality is strict.
\end{proposition}

\begin{proof}
To show that $T$ is at least as likely as $T'$, it suffices to show that it is the minimum tree for at least as many draws of edge weights.  So we will take an arbitrary set of distinct weights $w$ that is carried by $\beta$ to a permuted set of weights $\{w_i'\}_{i=1}^m$. Without loss of generality, we may order these weights so that $w_1'<\dots <w_m'$ and we will label the edges correspondingly as $e_1,\dots,e_m$.  By this convention, 
we have $w_i=w_{\beta^{-1}(i)}$ as a weight on $e_i$ before $\beta$ has been applied, noting that the $w_i$ need not be in increasing order.  We assume that the $w'$ induce $T'$ and we want to show that the $w$ induce $T$.

Since $\beta$ maps between $T$ and $T'$, any $e \mathbin{R_T} e'$ is sent  by $\beta$ to some $\beta(e)=e_i$ and $\beta(e')=e_j$ with $e_i \mathbin{R_{T'}} e_j$.  The weights satisfy $w_i'<w_j'$ because $T'$ is assumed  minimum (appealing to Proposition~\ref{prop:MST-cycle-char}).
But then we are done:  the weights on $e$ and $e'$ are precisely $w_i'<w_j'$ by construction, and we've verified that edges in $T$ have lower weight than the non-edges that complete their broken cycle.

To handle the  case of strict inequality, we will find a
set of weights $w_i$ that induces $T$, but so that $w_{\beta(i)}$ does not induce $T'$. Fix any edge $e_0\notin T$ so that
$\beta(P_{T,e_0}) \subsetneq P_{T',\beta(e_0)}$. Assign weights
$w_i$ so that all weights $w_i$ for $e_i \in P_{T,e_0}$ are clustered near $1$;
the weight $w_0$ of our chosen edge $e_0$ is $2$;
the weights $w_i$ for all other edges $e_i$ of $T$ are clustered near $3$; and the weights $w_i$ for non-edges of $T$ are clustered near $4$.
Then these weights $\{w_i\}$ induce $T$, because Kruskal's algorithm will reject $e_0$, but $\beta(e_0)$ has some edge on the  $T'$ path completing its broken cycle that did not come from $P$, and therefore has weight more than 2.  This means Kruskal will accept $\beta(e_0)$, so the weights 
$\{w_{\beta(i)}\}$ do not induce $T'$.
\end{proof}

Now the proof of the rotation trick is nearly immediate: we only need to show that the triangle-edge rotation is cycle-expanding.

\begin{proof}[Proof of Lemma~\ref{lem:tri-rotation}]
As before, $\beta$ exchanges $e_{13}$ and $e_{23}$ and preserves all other edges.
We must check that $\beta$ is strictly cycle-expanding from $S$ to $S'$.  This is an easy case analysis on the non-edges $e\notin S$:
\begin{itemize}
    \item if $e$ is an edge of $G$ with endpoints in a single tree $T_i$, then its broken cycle is within the tree before and after $\beta$ and its length does not change;
    \item if $e=e_{13}$, its broken cycle stays at length 3;
    \item if $e$ connects $T_2$ to $T_3$, its broken cycle contains $e_{23}$ and its length is  unchanged;
    \item if $e$ connects $T_1$ to $T_2$, then in $S'$ it must go the
      long way around the triangle rather than the short way, so its
      length is increased by one.
\end{itemize}
Note that there are no $G$-edges between $T_1$ and $T_3$ (a needed hypothesis, as these would have cycle lengths reduced by $\beta$).
This completes the proof, because all cycles stay the same length or are increased by one.
\end{proof}

%We can also prove a version of Lemma~\ref{lem:tri-rotation} with
%longer rotations. \thingtodo{And maybe this should become the main
%  version}
%\begin{lemma}\label{lem:poly-rotation}
%  Suppose we have a graph $G$ with a $(k+1)$-cycle ($k \ge 2$)
%  connecting the vertices
%  $v_0,v_1,\dots,v_k$, and a spanning forest $H$ with two components, one
%  component $T_a$ containing the vertex $v_0$ and the other component
%  $T_b$ containing the path connecting $v_1,v_2,\dots,v_k$. If we
%  remove the path, the tree $T_b$ decomposes into further components,
%  with $T_{b,i}$ containing $v_i$. With this setup, if there is at
%  least one other edge in the ambient graph between $T_a$ and $T_{b,1}$,
%  but no other edge between $T_a$ and any other $T_{b,i}$, then
%  \[
%    \prob_{\MSTZ}(H \cup e_{01}) > \prob_{\MSTZ}(H \cup e_{0k}).
%  \]
%\end{lemma}

%In particular, the condition in Lemma~\ref{lem:poly-rotation}
%requires that $G$ have no diagonals connecting $v_0$ to any other
%$v_i$ other than $e_{01}$ and $e_{0k}$.

% \begin{proof}
%   Apply Proposition~\ref{prop:MST-cycle-char} with
%   $\beta$ switching $e_{01}$ and $e_{0k}$.
% \end{proof}

% \moon{These are orphaned comments from earlier...

% \begin{itemize}
% \item Proof of weak inequality holds as written as long as random
%   variables $X_i$ are preserved by $\beta$
% \item Strict inequality more delicate if not all permutations are realized
% \end{itemize}}

\subsection{Application to random graphs}\label{subGnp}

We can apply the rotation trick to derive a simple sufficient condition for $\MSTZ$ to fail to capture the uniform distribution on spanning trees of a given graph. As a consequence, for large $n$, we can conclude that $\MSTZ$ is different from UST on almost all graphs of order $n$.  Recall the famous fact that $\log n/n$ is a sharp threshold for connectivity in the Erd\H{o}s-Renyi 
 random graph model $\mathcal G(n,p)$, where $\log$ is the natural logarithm.

\begin{theorem}[Random graphs have $\MSTZ\neq \UST$]\label{thm:most-graphs-mst-ust}
    If $p=c \log n /n$ for any constant $c>1$, the probability that $\MSTZ \neq \UST$ on a random $\mathcal G(n, p)$ graph tends to one as $n \to \infty$.
\end{theorem}

\begin{proof}
We will use two main facts about random graphs in this probability regime:  they almost surely have many triangles, and they are almost surely $k$-connected for every $k$.   In particular, the threshold for almost sure $k$-connectivity is roughly $p=(\log n + k\log\log n)/n$, with detail about lower-order terms found in  \cite[\S7.2]{Bollobas}.  The expected number of triangles is $\binom{n}{3}p^3$, which is on the order of $(\log n)^3$, and the variance is small because many triangle events are independent.  Consider a particular triangle $T$ and one of its vertices $v_1$; the probability that $v_1$ has no edges to $G\setminus T$ is 
$(1-p)^{n-3}$, so it has at least one such edge with high probability.  Putting these observations together, the probability that there exists a triangle $T$ with $v_1$ connected to $G\setminus T$ tends to 1, which means the probability also tends to 1 of finding such a triangle in a 2-connected graph.  
But then we can apply the rotation trick (Lemma~\ref{lem:tri-rotation}) with $T_1=\{v_1\}$, $T_3=\{v_3\}$, and $T_2$ any spanning tree of $G\setminus\{v_1,v_3\}$ (which is connected because $G$ is 2-connected).
\end{proof}

\subsection{MST on complete graphs via path rotation}\label{subsec:path-rotation}

To prove that labeled trees with the highest probability weight for
$\MSTZ$ on $K_n$ are
stars and those with the lowest probability weight are paths,
we will develop a more powerful rotation move.
On one hand, these path rotations will be far more general than single-edge rotations, but the ambient graph must be a $K_n$ (while Lemma~\ref{lem:tri-rotation} works in an arbitrary $G$).

\subsubsection{Statement and main application}

\begin{figure}[bht!]
    \centering
    
    \includegraphics[width=6in]{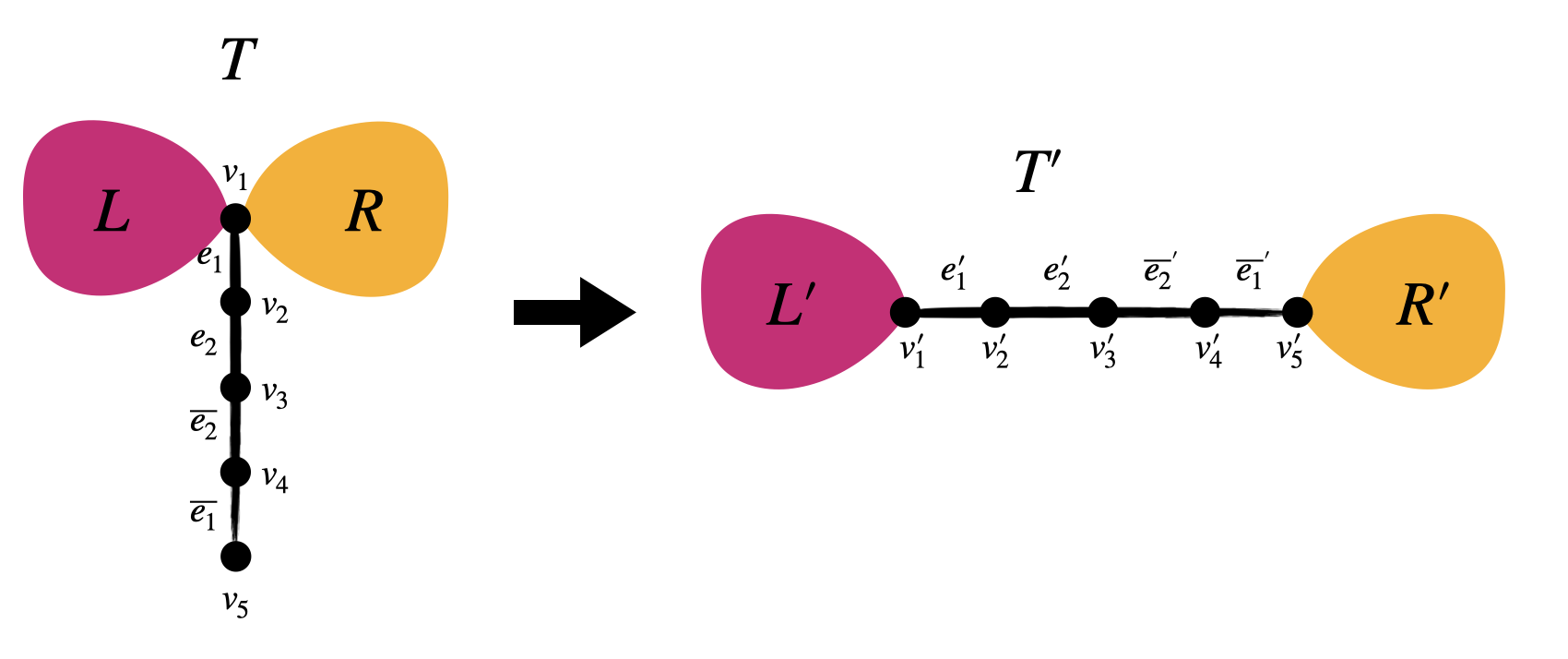}
    \caption{A path rotation operation from $T$ to $T'$ that rotates a path $P$ from $v_1$ to $v_5$.}
    \label{fig:PathRotation}
\end{figure}

\begin{theorem}[Path rotation in complete graphs]\label{thm:PathRotation}
  Let $T=L\cup P \cup R$ and $T'=L'\cup P' \cup R'$ be spanning trees of $K_n$ obtained by rotating a
  path~$P$ of $\ell \geq 2$ vertices as depicted in Figure
  \ref{fig:PathRotation}, where $L \cong L'$, $R \cong R'$, and both have at least one edge.  Then
  \[\prob_{\MSTZ}(T) > \prob_{\MSTZ}(T').\]
\end{theorem}

\FloatBarrier

The path $P\subset T$ is on vertices $v_1,\dots,v_\ell$.  
It will be useful for the later arguments in this section to label the edges of $P$ so as to keep track of each edge and its (central) reflection.
Namely, letting $r\coloneqq \lfloor \ell/2 \rfloor$, we label the edges $e_1,\dots,e_r,\bari r, \dots, \bari 1$, with $e_r=\bari r$ if $\ell$ is even.  
We will similarly use the notation $\overline{v_j}=v_{\ell-j+1}$ where it simplifies the exposition.
Even though this path rotation move only applies in a $K_n$, we give two
different names $G \cong G' \cong K_n$ to the ambient graph for clarity in the proof, and write $T,P \subset G$
and $T',P' \subset G'$, fixing this notation for the remainder of the section. Note that we do not consider $v_1$ to be part of $L$ or $R$ or $v_1',v_\ell'$ to be part of $L'$ or $R'$; we regard $v_1$ as part of $P$ and $v_1',v_\ell'$ as part of $P'$.

The proof of Theorem \ref{thm:PathRotation} requires new machinery and is postponed to the end of this section.  First, we derive the main application.

\begin{theorem}[Most and least likely trees]
\label{thm:stars-paths}
  In a complete graph $K_n$ with labeled vertices, let $T_\star$ and $T_P$ be any spanning trees isomorphic to a star and a path, respectively.  Then for all spanning trees $T$ of $K_n$, 
  $$\prob_{\MSTZ}(T_\star)\ge \prob_{\MSTZ}(T)
  \ge \prob_{\MSTZ}(T_P),$$
  with equality on either side if and only if $T$ is itself a star or path.
\end{theorem}

%let $T_*$ be the star spanning tree with $v_1$ connected to all others, and let $T_P$ be the path spanning tree with $v_i$ connected to $v_{i+1}$ for $i = 1,\dots,n-1$. Then for any spanning tree~$T_0$,
%\[
%    \prob_{\MSTZ} (T_*) \ge \prob_{\MSTZ}(T_0) \ge \prob_{\MSTZ}(T_P)
%  \]
%with equality in the two cases iff $T_0$ is itself a star tree (resp.~a path).

\begin{figure}[bht!]
    \centering
    
    \vspace{-1cm}
    
    \includegraphics[width=6in]{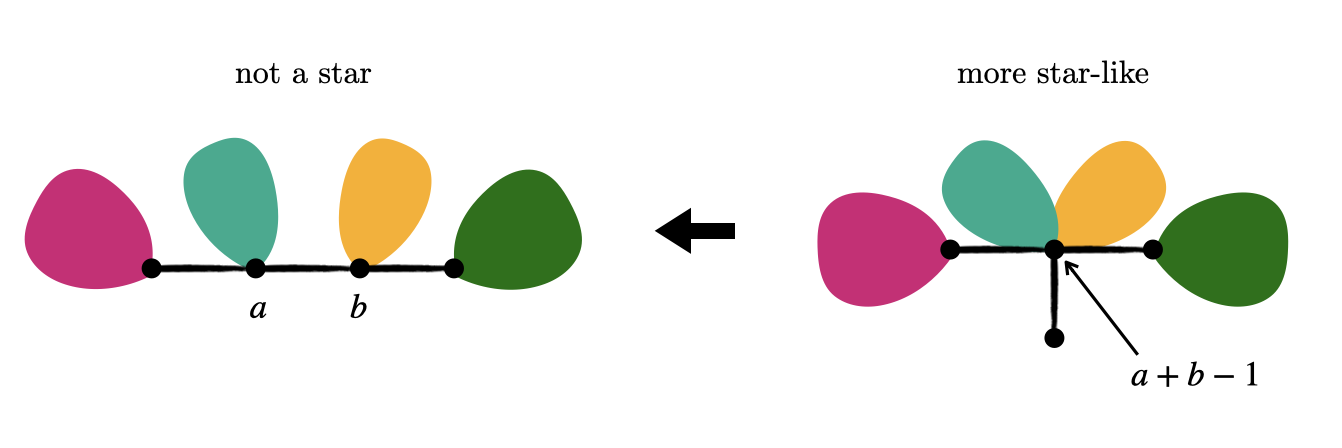}
    
    \vspace{-1cm}
    
    \includegraphics[width=4.5in]{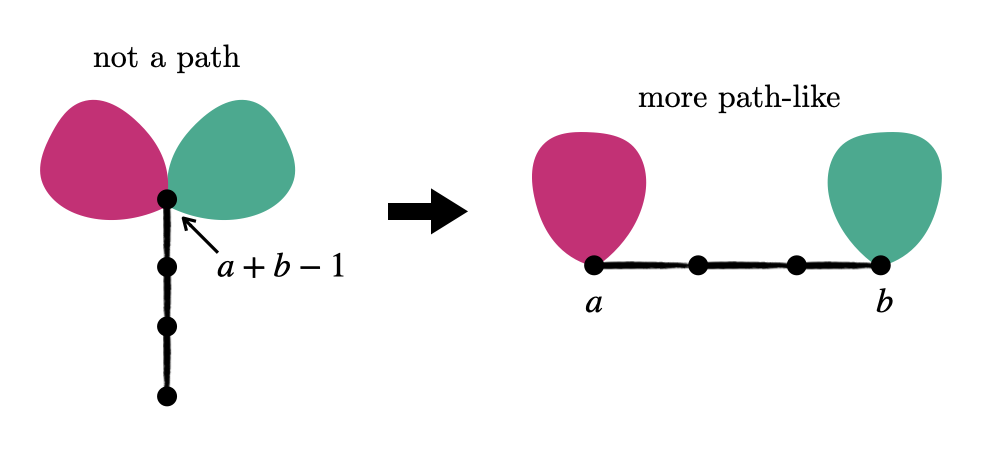}
    
    \vspace{-.6cm}
    
    \caption{Moves that interpolate from arbitrary trees to stars and
      paths, strictly monotonic with respect to both $\MSTZ$
      probability and the product~$D$ of vertex degrees.}
    \label{fig:star-to-path}
\end{figure}

\begin{proof}
For any graph $G$, let $D(G)$ be the product of all the vertex degrees.  Stars that span $K_n$ have $D=n-1$ while spanning paths have $D=2^{n-2}$.  We will show that these are the minimum and maximum among trees, and that there are sequences of path rotation operations interpolating from stars to arbitrary trees and from arbitrary trees to paths, with each step increasing $D$ and decreasing the probability of selection under $\MSTZ$.

If $T$ is not a star, then it contains some path of length 3, in which
each vertex has some (possibly empty) tree attached
(Figure~\ref{fig:star-to-path}).  This means it comes from a (reverse)
path rotation from another tree involving a single edge. All the vertex degrees are
the same in the two trees except at the central vertices, which
changed from $(a,b)$ to $(a+b-1,1)$ with $a,b\ge 2$.  Since $ab>a+b-1$
for $a,b\ge 2$, this reverse rotation operation decreases~$D$.

If $T$ is not a path, then it contains some vertex of degree $\ge 3$,
one of whose incident edges is part of a simple path ending in a leaf
(Figure~\ref{fig:star-to-path}).  Since its degree is at least three,
its other incident edges can be separated into two non-empty trees.
Rotating the path induces a graph for which $P'$ has endpoints with
vertex degrees $(a,b)$.  This means that on the original graph the
corresponding vertices had
degree $(a+b-1,1)$.  The same inequality holds, so the path rotation
operation increases~$D$.

Since any non-star can have $D$ strictly decreased while any non-path can have $D$ strictly increased, it follows that stars and paths are extremal for $D$.  The path rotation theorem tell us that moves that increase $D$ decrease the $\MSTZ$ probability and vice versa, and it follows that stars and paths are also extremal for the probability. \end{proof}

\FloatBarrier

From the proof we see that the passage from stars to arbitrary trees can be accomplished with only single-edge rotations. 
However, to get from an arbitrary tree to a path, longer path rotations may be required, which in turn requires a significant adaptation to the previous proof approach of constructing a cycle-expanding bijection.\footnote{To see what goes wrong with this approach, consider the edges in Figure~\ref{fig:PathRotation}. We could try to choose an edge to pair with  $\overline{e_1}$, but choosing either $e'_1$ or $\overline{e_1}'$ fundamentally breaks symmetry, which will cause trouble later on. For instance, suppose we choose to pair up $\overline{e_1}$ with $\overline{e_1}'$. Let $w \in R$ be a neighbor of $v_1$ in $T$, and consider the permutation where the first three edges are $\overline{e_1}$, $(v_1, w)$, and $(v_4, w)$. Note that this final edge is not in $T$, but it will still be added as it does not create a cycle. Consider the three corresponding edges in $G'$. We are already assuming that $\overline{e_1}$ will be sent to $\overline{e_1}' = (v'_4, v'_5)$. The most natural choices for the other edges are to send $(v_1, w)$ to $(v'_5, w')$ (since $(v_1, w)$ is an edge in $R$, so it should be sent to the corresponding edge in $R'$) and $(v_4, w)$ to $(v'_4, w')$. These three edges \emph{do} form a cycle, so the final edge will be rejected. Thus, we reach a state where $T$ is no longer possible while $T'$ is, which is precisely what we must rule out.}
Indeed, we have verified computationally that there is no cycle-expanding bijection (as in \S\ref{subRotationTrick}) from a tree with three paths of length 2 joined at a common endpoint to the path of length 6.

\subsubsection{Folding a path}

We introduce a new proof technique that records
partial (probabilistic) information about the randomly generated edge weights. We will define an equivalence relation on weight-orderings (or permutations) $\sigma$, and a bijection $\beta$ on the resulting equivalence classes, so that every equivalence
class~$\rho$ induces $T$ with probability greater than or equal to the probability than $\beta(\rho)$ induces $T'$ (with a strict inequality for at least one~$\rho$). To illustrate this approach with reference to Figure~\ref{fig:PathRotation}: we associate the equivalence class $\{e_1, \overline{e_1}\}$ with the equivalence class $\{e_1', \overline{e_1}'\}$ and remain agnostic to which of the two edges was actually processed first, allowing us to maintain symmetry. More generally, we will introduce several ``folded" concepts here; the term refers to identifying each $e_i$ with its reflection $\bari i$ and $e_i'$ with $\bari i'$ in a manner that suggests folding the path in half.  

\begin{definition}[Folded permutations]\label{def:folded-perm}
Using the notation introduced right after Theorem \ref{thm:PathRotation},  let $\iota$ be the
  involution on $E(G)$ that swaps each edge of the form $(v_i,v_j)$
  with its reflection $(\overline{v_i},\overline{v_j})$ and is
  otherwise the identity on $E(G)$. Likewise, $\iota'$ is defined
  on $E(G')$.
  Consider a permutation of the edges 
  $\sigma \in S(E(G))$, thought of as a map
  $\sigma \colon \{1,\dots,m\} \to E(G)$. We write $E(G) / \iota$ for the set of equivalence classes under $\iota$. Define the corresponding \emph{folded
    permutation} $\sigma/\iota \in S(E(G))/\iota$ to be
  \[
    \rho = \sigma / \iota \colon \{1,\dots,m\} \to E(G) / \iota.
  \]
  \end{definition}
  This is no longer a bijection. The equivalence classes of edges are denoted by $[e]$ and have one or two representatives.
  Indeed,  those with one representative are exactly the $r$ equivalence classes for the edges of the form $(v_i,\overline{v_i})$ which are fixed by the central symmetry of the path. This shows that there are $a = \frac{1}{2}\left(\binom{\ell}{2} - r\right)$ different equivalence classes of size 2 that arise from the complete graph on the vertices of the path, and thus $2^a$ representatives for every folded permutation.

There is a corresponding \emph{folded bijection}  $\beta\colon E(G) / \iota
\to E(G') / \iota$ that takes each $[(u, v)]$ to $[(u', v')]$ except that $[(u, v_i)] \mapsto [(u, \overline{v_i})]$ if $u\in R$ and $v_i\in P$.  Looking ahead, this can be seen in  Figure~\ref{fig:entanglement}.\footnote{For example, the edges numbered 3 in $G'$ form an equivalence class of size 2 and are mapped to edges $\{(v_1, v_3), (v_3, v_5)\}$ in $G$. On the other hand, the edge numbered 6 in $G'$ is in an equivalence class of size 1, and is mapped to the edge in $G$ joining the leftmost vertex of $T$ to $v_2$.}  This means that from a folded permutation $\rho$ on $G$ we obtain a corresponding folded permutation $\rho'=\beta(\rho)$ on $G'$.

%  \begin{definition}[Folded bijections]   
%     % as follows, where, for any vertex $u \in L \cup R$, we denote the corresponding vertex in $L' \cup R'$ as $u'$:
%     % \begin{align*}
%     %     [(u, v)] &\mapsto [(u', v')] & \txt{for } u, v \in L \cup R\\
%     %     [(u, v_i)] &\mapsto [(u', v_i')] & \txt{for } u \in L, \ v_i \in P\\
%     %     [(u, v_i)] &\mapsto [(u', \overline{v_{i}}')] & \txt{for } u \in R, \ v_i \in P\\
%     %     [(v_i, v_j)] &\mapsto [(v_i', v_j')] & \txt{for } v_i, v_j \in P
%     % \end{align*}
% \end{definition}    

If we run Kruskal's algorithm on a folded permutation~$\rho$, the
outcome is no longer a single tree, but a distribution over trees---the first time we see a particular equivalent pair of edges, we choose one of them uniformly at random.  The next time, we take the other. Write $\prob_\rho(T)$
for the probability that $\rho$ induces~$T$, and similarly for
$\prob_{\rho'}(T')$.

With this setup, the key fact to prove
Theorem~\ref{thm:PathRotation} is that, for any folded
permutation $\rho \in S(E(G))/\iota$ and corresponding $\rho' \coloneqq  \beta \circ \rho$, we have a (weak) probability inequality $\prob_\rho(T) \ge \prob_{\rho'}(T')$.
We actually prove a stronger statement, about the probability
distribution of forests at each stage of Kruskal's algorithm, in addition to showing that the inequality is sometimes strict. To state it precisely, we build up some more terminology.

% Given a permutation $\sigma$ and corresponding folded permutation
% $\rho = \sigma / \iota$, define, for $0 \le k \le m$,
% \begin{align*}
%   \sigma_k &= \sigma |_{1,\dots,k} \colon \{1,\dots,k\} \to E(G)\\
%   \rho_k &= \rho |_{1,\dots,k} \colon \{1,\dots,k\} \to E(G)/\iota.
% \end{align*}

\subsubsection{State vector}
Fixing $\rho$ and $k$, consider the probability space where we uniformly sample a random permutation $\sigma$ (the order of edges in a run of Kruskal's algorithm) but condition on $\sigma$ being consistent with $\rho$ on the first $k$ labels $\{1,\dots,k\}$, meaning that $[\sigma(i)]=\rho(i)$ for $i=1,\dots,k$. Let $\mathrm{Kru}_k(\rho)$ be the forest that
Kruskal's algorithm outputs after processing the lowest-weight $k$ edges of~$\sigma$, thought of as a forest-valued random variable. Define $\mathcal{E}_k$ to be the event that $\mathrm{Kru}_k(\rho) \subseteq T$, respectively $\mathcal{E}_k'$.

We prove the probability inequality by induction on $k$. To make the induction go through, we add additional hypotheses tracking the distribution of $\mathrm{Kru}_k(\rho)$ conditioned on $\mathcal{E}_k$, and of $\mathrm{Kru}_k(\rho')$ conditioned on $\mathcal{E}_k'$. These distributions are simultaneously encoded by a state vector $s$
that records the status of the edges $e_1,\dots,e_r$ as the steps of
Kruskal's algorithm are incremented.  For each edge $e_i$ and its
reflection $\overline{e_i}$, it will tell us what we know about their
inclusion in the tree $T$ to that point.
Formally,
$$s\colon \{1,\cdots,r\}  \to \{\Neither, \Both, \Left, \Right\} \cup 
\mathcal P (\{1,\cdots,r\}),$$
where $\mathcal P (\{1,\cdots,r\})$ is the power set of $\{1,\cdots,r\}$.

\begin{algorithm}
\footnotesize
	\caption{\label{algPathRotation}\footnotesize Computes the probabilities of sampling a pair of spanning trees differing by a path rotation.}
	\KwIn{A pair of spanning trees $T = L \cup P \cup R$ and $T' = L' \cup P' \cup R'$ as in Figure \ref{fig:PathRotation}.}
	\KwOut{A pair of probabilities: The respective likelihoods of sampling $T$ and $T'$ under $\MSTZ$.}
	$(p, p') \gets (0, 0)$\;
	\For{$\rho \in \{\textnormal{folded permutations}\}$}
	{
		$(q, q') \gets (1, 1)$\label{linInitQs}\;
		$s \gets (i \mapsto \Neither)$\label{linInitState}\;
		\For{$k \in \{1, 2, \dots, m\}$}
		{\label{linMainFor}
			\uIf{$\rho(k) = [e_{\ell/2}]=\{e_{\ell/2}\}$\label{linInTreeInPathMiddle}}
			{
				%Case 2-a
				$s(\ell/2) \gets \Both$\;
			}
			\uElseIf{$\rho(k) =[e_i]= \{e_i, \overline{ e_i}\}$ where $i < \frac{\ell}{2}$\label{linInTreeInPathGeneric}}
			{
				\uIf{this is the first time we have seen  this $\rho(k)$\label{linFirstTime}}
				{
					%Case 2-b
					$s(i) \gets \{i\}$\;
				}
				\uElse
				{\label{linSecondTime}
					%Case 2-c
					$s(i) \gets \Both $\;
					remove all occurrences of $i$ from sets in $s$\;
				}
			}
			\uElseIf{$\rho(k) = \{(u, v)\} \hbox{\rm ~for~ }  (u,v) \notin T$\label{linUVNotinT}}
			{
				\uIf{the path from $u$ to $v$ in $T$ contains an edge $e$ such that $[e] \notin \{\rho(1), \rho(2), \dots, \rho(k - 1)\}$\label{linDefiniteFail}}
				{
					$(q, q') \gets (0, 0)$\;
				}
				\uElseIf{$u \in V(L)$ and $v \in V(R)$, or $u \in V(R)$ and $v \in V(L)$\label{linCrossWholePath}}
				{
					%Case 4
					\uIf{$s \neq [\Both, \Both, \dots, \Both]$}
					{
						$q' \gets 0$\label{linFailCrossWholePath}\;
					}
				}
				\uElseIf{one vertex from $\{u, v\}$ (without loss of generality $u$) is in $V(L) \cup V(R)$ and the last edge of the path from $u$ to $v$ in $T$ is some $e_i$ or $\overline{e_i}$\label{linSideToPath}}
				{
					%Case 5
					$\tt{side} \gets (\Left \txt{ if } u \in V(L) \txt{, else } \Right)$\;
					\uIf{the last edge was $\overline{e_i}$ and there is some $j\ge i$ with $s(j) \neq \Both$ \label{linNotBothThereSide}}
					{
						$(q, q') \gets (0, 0)$\;
					}
					$M \gets \{s(j) \suchthat j \in \{1, 2, \dots, i\},\ s(j) \txt{ is a set}\}$\;
					$N \gets \{s(j) \suchthat j \in \{1, 2, \dots, i\},\ s(j) \in \{\Left, \Right, \Neither\}\}$\;
					$(q, q') \gets (q/2^{\abs{M}}, q'/2^{\abs{M}})$\label{linUpdateQSide}\;
					\uIf{$N \not\subseteq \{\tt{side}\}$\label{linSideInductiveCollapse}}
					{
						$q' \gets 0$\label{linQPrimeGetsZero}\;
					}
					\For{$j \in \bigcup M$\label{linUpdateStateSide}}
					{
						$s(j) \gets \texttt{side}$\;
					}
				}
			}
			\uElseIf{$\rho(k) = [(v_{i_1}, v_{i_2})]$ for 
   $1 \leq i_1 < i_2 \le \ell$ where 
   $v_{i_1}$ is the lower-indexed vertex of some $e_{I_1}$ or $\overline{e_{I_1}}$ and $v_{i_2}$ is the higher-indexed vertex of some $e_{I_2}$ or $\overline{e_{I_2}}$ \label{linPossibleEntanglement}}
			{
				%Case 6                    
				    $M \gets \{s(j) \suchthat \min(I_1,I_2)\le j\le \max(I_1,I_2)  ,\ s(j) \txt{ is a set}\}$\;
				        $N \gets \{s(j) \suchthat \min(I_1,I_2)\le j\le \max(I_1,I_2) ,\ s(j) \in \{\Left, \Right, \Neither\}\}$\;
                    \uIf{$i_1 \leq \frac{\ell}{2} \leq i_2$ and $s(j) \neq \Both$ for some $j \geq I_1,I_2$\label{linNotBothThereMiddle}}
				{
					$(q, q') \gets (0, 0)$\;
				}
                    \uElse{
				$(q, q') \gets (q/2^{\abs{M}}, q'/2^{\abs{M}})$\;
                    }
				\uIf{$N \neq \emptyset$}
				{
					\uIf{$\Neither \in N$ or $\{\Left, \Right\} \subseteq N$ or this is the second time we have seen $\rho(k)$}
					{
						$(q, q') \gets (0, 0)$\label{linDoubleFailNeither}\;
					}
					\uElse
					{
						$(q, q') \gets (q/2, q'/2)$\;
                        \For{$j \in \bigcup M$\label{linMiddleInductiveCollapse}}
        				{
        					$s(j) \gets$ the unique element of $N$\;
        				}
					}
				}
                \uElse
                {
				    \For{$j \in \bigcup M$\label{linEntanglement}}
				    {
						$s(j) \gets \bigcup M$\;
				    }
                }
			}
		}
		$(p, p') \gets (p + q, p' + q')$\label{linUpdateP}\;
	}
	\KwRet{$(\frac{2^a}{m!} p, \frac{2^a}{m!} p')$} where $a = \frac{1}{2}\left(\binom{\ell}{2} - r\right)$\label{linReturn}\;\vspace{1em}
\end{algorithm}

The meaning of this state vector will be illustrated in several examples, but we first give a formal description via Algorithm~\ref{algPathRotation} and Lemma~\ref{lemPathRotationAlgorithmCorrect}.  
The algorithm describes how to successively update entries and track the probabilities of $\mathcal{E}_k$ and $\mathcal{E}_k'$, and the lemma proves it is correct. Algorithm~\ref{algPathRotation} is written in the standard computer science style called  {\em imperative programming}, with $x \gets v$ denoting an update of variable $x$  to hold value $v$, so that for instance $x\gets x+1$ means to increment by one.  We initialize the probabilities at $(1,1)$ and then track through a series of steps.

\FloatBarrier

\begin{lemma}[State vector and probability updates]\label{lemPathRotationAlgorithmCorrect}
	Fix any folded permutation $\rho$ and index $0 \leq k \leq m$. Consider the probability space where we uniformly sample a random permutation $\sigma$, but condition on $\sigma$ being consistent with $\rho$ on $\{1,\dots,k\}$.  Then, after $\rho(k)$ has been processed in Algorithm~\ref{algPathRotation} (i.e., after $k$ iterations of the inner for-loop on Line~\ref{linMainFor}), the values of the variables $q$ and $q'$ equal the respective probabilities of $\mathcal{E}_k$ and $\mathcal{E}'_k$.  Furthermore, 
conditioned on $\mathcal{E}_k$ and $\mathcal{E}'_k$, respectively, the data in the state vector $s$ describes the distribution of $\mathrm{Kru}_k(\rho) \cap E(T)$ and $\mathrm{Kru}_k(\rho') \cap E(T')$ as follows:
  \begin{itemize}
  \item For $i$ such that $s(i) = \Neither$, neither edges $e_i$ nor $\overline{e_i}$ have been added to $\Kru_k(\rho)$, and neither edges $e'_i$ nor $\overline{e'_i}$ have been added to $\Kru_k(\rho')$.
  \item For $i$ such that $s(i) = \Both$, both edges $e_i$ and $\overline{e_i}$ have been added to $\Kru_k(\rho)$, and both edges $e'_i$ and $\overline{e'_i}$ have been added to $\Kru_k(\rho')$.
  \item For $i$ such that $s(i) = \Left$, edge $e_i$ has been added to $\Kru_k(\rho)$, but not edge $\overline{e_i}$, and likewise edge $e'_i$ has been added to $\Kru_k(\rho')$, but not edge $\overline{e_i'}$.
  
  \item For $i$ such that $s(i) = \Right$, edge $e_i$ has been added to $\Kru_k(\rho)$, but not edge $\overline{e_i}$, while on the other hand edge $\overline{e_i'}$ has been added to $\Kru_k(\rho')$, but not edge $e'_{i}$.
  
  \item For $i$ such that $s(i) = S \subseteq \{1, 2, \dots, r \}$,
    there is a $\frac 12$ probability that all $e_j$ for $j\in S$ have been added
    to $\Kru_k(\rho)$ but no $\bari j$ has.  Otherwise, all $\bari j$ have been added to $\Kru_k(\rho)$ but no $e_j$ has. Likewise, there is a $\frac 12$ probability of all $e_j'$  but no $\bari j'$ (else all $\bari j'$ but no $e'_j$) were added to $\Kru_k(\rho')$.
  \end{itemize}
\end{lemma}

Before proving the lemma, we first illustrate the encoding and updating with an example. To understand the meaning of the entries of $s$, consider an example with $\ell=14$ (so $r=7$), where we might encounter the following state partway through the process:
$s=\left( \Right, \{2\}, \Neither, \{4,5\}, \{4,5\}, \{6\},
  \Both\right)$. (See Appendix~\ref{sec:path-rotation-example} for a visualization.) This indicates that up to that point in the process,
$e_1\in T$ and $\bari 1'\in T'$ rather than their reflections (i.e.,
the right-hand choice has been made in $T'$); exactly one of
$\{e_2,\bari 2\}$ is included and the probabilities are equal and
independent of other indeterminacies to this point; neither $e_3$ nor
$\bari 3$ has been processed yet; $\{e_4,\bari 4\}$ and $\{e_5,\bari
5\}$ are both uncertain, with $1/2$ probability of including both
$e_4$ and $e_5$ and $1/2$ of including their reflections instead and
likewise $e_4',e_5'$; inclusion of $\{e_6,\bari 6\}$, like
$\{e_2,\bari 2\}$, is 50-50 with no
dependencies; and the self-paired edge $e_7$ is included (i.e.,
``both'' $e_7$ and its reflection).
Whenever some $s(j)$ is a subset of $\{1,\cdots,r\}$, $s(j)$ must include $j$, and when its size is greater than one, as with the $\{4,5\}$ entry here, we think of this as an {\em entanglement} of probabilities. 
In  Appendix~\ref{sec:path-rotation-example} we give an extensive discussion showing all of the ways that this state vector could be updated, which is complex enough to illustrate essentially all of the cases in Algorithm~\ref{algPathRotation}. Here, we present a simpler example with $\ell = 5$.

\begin{figure}[bht!]
    \centering
    \includegraphics[width=6in]{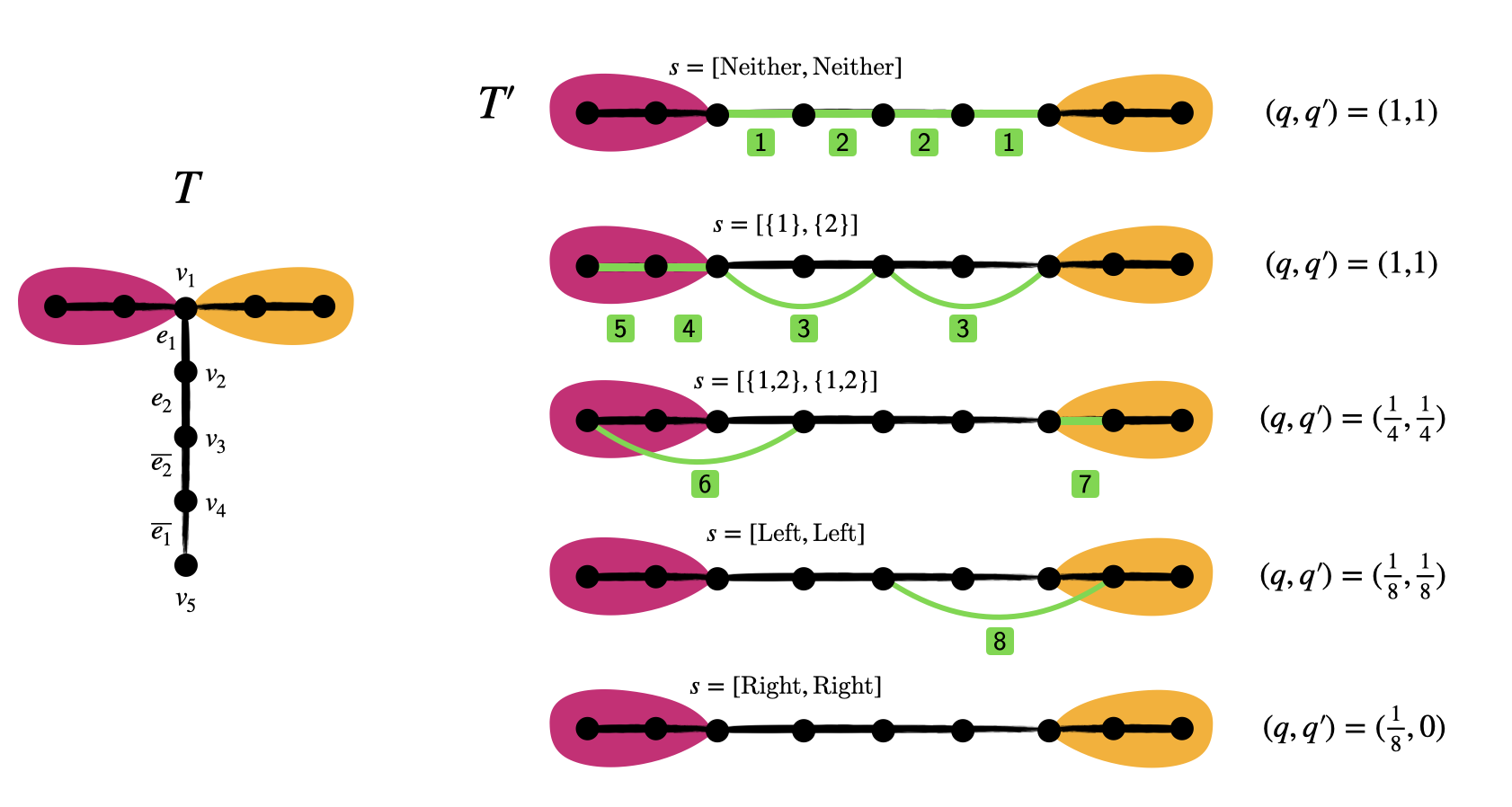}
    \caption{A simple example of processing a folded permutation;  the green numbering on folded edges in $T'$ shows the ordering from $\rho'$ up to $k=8$.
      After these eight edges have been processed, there is a $1/8$ probability that $T$ can be chosen by Kruskal's, while $T'$ has been ruled out.}
    \label{fig:entanglement}
\end{figure}

\begin{example}\label{exaPathRotation}
We work through the example in Figure~\ref{fig:entanglement}. The $k$ values index the weight order encoded by the folded permutations $\rho,\rho'$.  After $k=1,2$, we have definitely added one from each pair and nothing outside the tree, so $q=q'=1$ give the probability that $T,T'$ are still viable at that stage.
After $k=3$, we note two possibilities to continue: on one hand,
perhaps both
of the previously
processed edges are chosen to the left in $T'$ (probability $1/4$), in
which case the new edge must be chosen left as well in order to be
rejected.  The right side is similar, giving total probability
$q'=1/4$.  The situation in $T$ is analogous, so $q=1/4$ as well.  The
$k=4,5$ edges are added without changing $s,q,q'$.  At $k=6$, the only
way for this edge to be rejected is if $\rho'(1)$ was on the left,
which forces $\rho'(2)$ on the left as well and drops the overall
probability by a factor of~$2$ (so $q=q'=1/8$).  Once the $k=7$ edge
is added, the $k=8$
edge is definitely rejected in $G$ because the other three edges on
its broken cycle are already assumed present---that means that $T$ is still possible. However, this edge is
accepted in $G'$, which kills the probability of Kruskal
selecting~$T'$ (so $q=1/8$ but $q'=0$).
\end{example}

Note the asymmetry between the meanings of the bar notations in $T$ versus $T'$. In $T$, the $e_i$ edges are always closer than the $\overline{e_i}$ edges to both the $L$ and $R$ components of the tree, while in $T'$, the $e'_i$ edges are closer to $L$ and the $\overline{e_i}'$ edges are closer to $R$. This induces a crucial asymmetry in the statement of Lemma~\ref{lemPathRotationAlgorithmCorrect}. In the cases where $s(i) \in \{\texttt{Left}, \texttt{Right}\}$, 
we can sometimes end up with one demand on the $G$ side and two demands on the $G'$ side, as in Example~\ref{exaPathRotation}. 
This can lead to updating to
$q' = 0$ while $q \neq 0$ (in Line~\ref{linQPrimeGetsZero}), which gives the strict inequality we need.

\FloatBarrier
\subsubsection{Proof of algorithm correctness, proof of path rotation theorem}

\begin{proof}[Proof of Lemma~\ref{lemPathRotationAlgorithmCorrect}]
  Clearly the stated properties hold at initialization, i.e.\txt{} when $k
  = 0$; the lemma is proved by inductively checking that $q$, $q'$,
  and $s$ are updated correctly in all of the cases of
  Algorithm~\ref{algPathRotation}. Many cases require no explanation.
  We work through several  cases here to illustrate the trickier
  steps. Illustrations of these cases and more are given in
  Figure~\ref{figPathRotationCases} and
  Table~\ref{tabPathRotationExample} in Appendix~\ref{sec:path-rotation-example}.

    First consider Line~\ref{linFirstTime}, processing a pair of edges
    inside the central path for the first time. This does not change
    the probabilities $q$ or $q'$, because those are concerned with
    the likelihood of Kruskal's algorithm adding any edges
    \emph{outside} of $T$, so $(q,q')$ are still accurate.
    Furthermore, it is equally likely that a uniformly random
    permutation $\sigma$ consistent with $\rho$ on
    $\{1,\dots,k\}$
    would have added edge $e_i$ versus $\bari i$ to $T$, and likewise
    for $\beta(\sigma)$ with respect to $e_i',\bari i'$ in $T'$.
    The algorithm sets $s(i) = \{i\}$, and this checks out with the claimed meaning of $s(i)=S$.
    All other entries of $s$ remain unchanged at this stage, and there is no new information about the likelihoods of various edges having been added, and thus the accuracy of $s$ is maintained.

    Next, consider the case on Line~\ref{linSideToPath}, where we process an edge $e \notin T$ and corresponding edge $e' \notin T'$ that both join the central path to one of the sides. For simplicity, we consider the case $i < \frac{\ell}{2}$. To update $q$ and $q'$, the algorithm computes the probability that all of the edges related to $e$ by the cycle relation $R_T$ have been added to $T$, and analogously, the probability that all of the edges related to $e'$ by the cycle relation $R_{T'}$ have been added to $T'$. By the inductive hypothesis, the \Left and \Right  designations in $s$ specify knowledge that, in each tree, we have added a specific one of a given pair of edges. In $T$, this will always be the edge of lower index, so this will always be the edge in the cycle relation since $i \leq \frac{\ell}{2}$. However, for $T'$, we may be on the wrong side; Line~\ref{linSideInductiveCollapse} catches this possibility and accordingly sets $q'$ to zero. Otherwise, the probabilities in $T$ and $T'$ that all edges in the cycle relation with the new edge have been added already is just given by $\frac12$ raised to the power of the number of independent binary events involving those edges existing. By the inductive hypothesis, these can be calculated by taking the number of sets in $s$ involved along the respective paths, which is precisely the size of the variable set $M$. Thus, $q$ and $q'$ are updated correctly in Line~\ref{linUpdateQSide}, maintaining their accuracy. To see that $s$ remains correct as well, observe that the for-loop on Line~\ref{linUpdateStateSide} correctly updates the conditional probability distributions with the following fact: if the new edge was rejected, it must be that all of the edges in the cycle relation (or correlated with them through non-singleton sets in $s$) are on the side of the path from which the new edge came.

    As a third case, we  consider  Line~\ref{linEntanglement}, which is where non-singleton sets in $s$ may originate when the sets in $M$ are merged. Here we have processed a pair of edges that are not in the respective trees $T$ and $T'$, with endpoints in the paths $P$ and $P'$. The logic for why $(q,q')$ are still accurate is similar to the previous case. To see that $s$ is valid, observe that, conditioned on the new edge being rejected, all of the once-indeterminate pairs of edges must have now been resolved, and all to the same side (forming a cycle with the new edge), even though it is equally likely which side it was.

    This kind of case analysis establishes that Algorithm~\ref{algPathRotation} correctly processes each edge $\rho(k)$.
    At the final iteration $k = m$, we conclude that $q = \prob_\rho(T)$ and $q' = \prob_{\rho'}(T')$ for each folded permutation $\rho$. The final lines 
    sum these over the folded permutations and divide by the number of folded permutations $m!/2^a$, where 
    $a=\frac 12 \left( \binom{\ell}{2} -r \right)$ as in the discussion of Definition~\ref{def:folded-perm}.  This amounts to averaging over all folded permutations, so we have correctly computed the probabilities $\prob_{\MSTZ}(T)$ and $\prob_{\MSTZ}(T')$.
\end{proof}

With this, the proof of the path rotation theorem is nearly immediate.
\begin{proof}[Proof of Theorem ~\ref{thm:PathRotation}]
	To see that $\prob_{\MSTZ}(T) >\prob_{\MSTZ}(T')$, we first observe that $\prob_\rho(T) \geq \prob_{\rho'}(T')$ for any folded permutation $\rho$. This is because, whenever $q$ is decreased in Algorithm~\ref{algPathRotation}, $q'$ is decreased by the same factor. Furthermore, there are several cases where $q'$ is set to zero and $q$ is not, in which case we can get a strict inequality. Specifically, consider a folded permutation where we first add all edges in $T \setminus P$, then an edge $(u, v)$ where $u \in L$ and $v \in R$, then the remaining edges in $T$. This sets $q' = 0$ on Line \ref{linFailCrossWholePath}, but we will still have $q = 1$ after all $k$ iterations.  Since the overall calculation is an average over cases, this shows a strict probability inequality overall.
\end{proof}

\FloatBarrier

%%%%%%%%%
%%%%%%%%%

\section{Shifted intervals}\label{secConnectedSupport}

\subsection{Parametrizing shifts}\label{sec:preliminaries_conn_support}

Let $G$ be a graph with $m$ edges labeled $\seq{e}{m}$. For any $\SS=(\seq{s}{m})\in \R^m$, we consider the product measure where the weight on edge $i$ is drawn uniformly from $[s_i, s_i+1]$, denoting the induced distribution on spanning trees by $M_\SS=M_\SS(G) \in \dd(\st(G))$, so that $M_{\sf 0}=\MSTZ$. In this section we seek to understand what measures on spanning trees are achievable by this {\it shifted-interval MST} as the $s_i$ vary.

Consider the map $h\colon\R^m\to \dd(\st(G))$ given by $\SS\mapsto M_\SS$. This map is highly non-injective; for instance translating each interval by the same amount has no effect on the order of the random variables, so no effect on the distribution on trees.  
Likewise, if the interval for one variable is entirely above the interval for a second variable, then the first is always greater than the second, no matter how large the gap.  
In light of this it suffices to restrict attention to those shifts $\SS$ where the gaps between $s_i$ on the number line are no greater than 1 and the sum is fixed.  
We can formalize this constructively.

\begin{lemma}[Closing gaps]\label{lem:closing_gaps} For every $\SS\in \R^m$, there exists $\SS'\in \R^m$ such that $M_\SS=M_{\SS'}$ and $\bigcup_{i=1}^m [s_i',s_i'+1]$ is connected. 
\end{lemma}

\begin{proof}
Given $m$ real numbers $s_1,\dots,s_m$, let $\sigma$ be a permutation that sorts them into non-decreasing order, so that $r_1\le \dots \le r_m$ where $r_{\sigma(i)}=s_i$.
Now consider the first $i$ for which  $r_{i+1}\ge r_i+1$.  Then the weight drawn from the $i+1$st interval must be greater than the weight drawn from the $i$th interval.  Replacing $r_{i+1}$ with $r_i+1$ maintains this order.  Let $t=r_i+1-r_{i+1}<0$ and 
shift $(r_1,\dots r_m)$ by $(0,\dots,0,t,t,\dots,t)$, 
where the negative shift occurs in every position from $i+1$ to $m$.  
It is still the case that these $r_i'$ are non-decreasing, and relative orders for any pair of weights are preserved whether both are below $i$, both are $\ge i$, or one is on each side.
Thus the $s_i'=r'_{\sigma(i)}$ are guaranteed to be in the same order as before.
This update can be iterated from $i=1,\dots,m$, ensuring that there are no gaps greater than $1$ between successive $r_i$ while maintaining the order of the variables drawn from the intervals described by $\SS$.
\end{proof}

\begin{figure}[bht!]
    \centering
\begin{tikzpicture}
    \node at (0,0) {\includegraphics[width=5in]{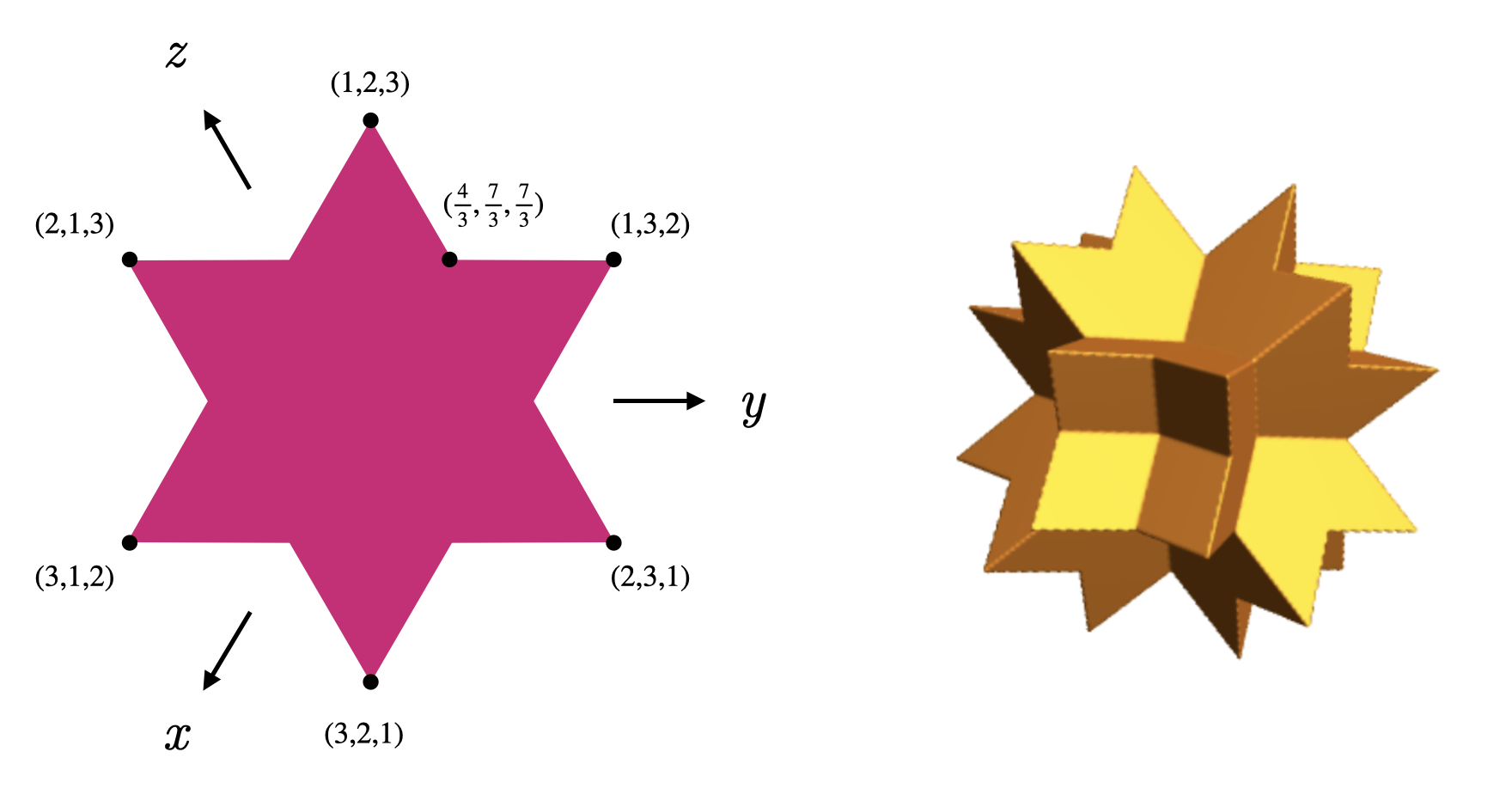}};

\end{tikzpicture}
    
    \caption{The shiftahedra $\Sh(3)$ and $\Sh(4)$.}
    \label{fig:shiftahedra}
\end{figure}

\begin{definition}[Shiftahedron]
For fixed $m$, let 
$$R=\left\{ (r_1,\dots,r_m)\in \R^m : \sum r_i=\textstyle\binom{m}{2} ~\hbox{\rm and}~ r_i\le r_{i+1} \le r_i+1 ~\hbox{\rm for each}~i   \right\}. $$
Then the {\em shiftahedron} for $m$ variables is defined as the orbit
of $R$ under reordering indices:  $$\Sh=\Sh(m)\coloneqq S_m.R=
\left\{  \left(r_{\sigma(1)},\dots,r_{\sigma(m)}\right) : 
(r_1,\dots,r_m)\in R, \sigma\in S_m\right\}.$$
\end{definition}

A few basic properties are easily verified.  Because it is a subset of the closed simplex  $\sum s_i=\binom m2$ defined by non-strict inequalities, $\Sh(m)$ is closed and compact.  
It contains the extreme points 
$S_m.(1,2,\dots,m)$, and is star-shaped  and symmetric with respect to the center point $\SS_0=(\frac{m+1}2,\dots,\frac{m+1}2)$.  
(Here, symmetry means that $\SS\in \Sh \iff 2\SS_0-\SS\in \Sh$.)

\begin{proposition}[Shiftahedron parametrizes all shifts]\label{prop:shiftahedron}
The shiftahedron parametrizes shifted-interval MST:
$h(\Sh(m))=h(\R^m)$.
\end{proposition}

\begin{proof}
The closing gaps lemma (Lemma~\ref{lem:closing_gaps}) implies that $h(\Sh(m))=h(\R^m)$ because it replaces any $\SS$ with an $\SS'$ giving the same distribution on spanning trees such that, after sorting, successive values differ by no more than 1.  Translation suffices to fix the sum of the values, placing them in $\Sh(m)$.
\end{proof}

We note that $h\colon\Sh(m)\to \dd(\st(G))$ is far from surjective, and indeed we will see below that even the much larger locus hit by all product measures does not hit every distribution on spanning trees.
Furthermore, $h$ is far from injective on the boundary of $\Sh(m)$.  The failure of injectivity on the interior can depend on the connection topology of the graph $G$.  In Appendix~\ref{app:theta}, we study this map more closely in some special cases, and thereby confirm that shifts are enough to hit UST on a class of graphs called {\em theta graphs} and on a slightly generalized class we call {\em $\theta$-surgery graphs}.

%%%%%%%%%%%%%%%%%%%%%%%%
%%%%%%%%%%%%%%%%%%%%%%%%
%%%%%%%%%%%%%%%%%%%%%%%%

%%%%
\subsection{Shifts and complete graphs}\label{subShiftsCompleteGraph}

In contrast to the simple case of theta graphs, we show that shifted-interval MST does not suffice to recover the uniform distribution on spanning trees in a complete graph $K_n$ for $n\ge 4$.
We first observe that $\MSTZ\neq \UST$ for $n\ge 4$ and  $M_\SS\neq \UST$ if the intervals $[s_i,s_i+1]$ do not all intersect.

\begin{lemma}[Conditions implying non-uniformity]\label{rmk:non-overlap}
For any graph $G$ and any two distinct, non-separating edges $e_j$ and $e_k$ whose weights are given by random variables $X_j$ and $X_k$, any distribution $D\in\dd(\st(G))$ satisfying $\prob_D(X_j<X_k)=1$ is not uniform:  $D\neq \UST$.
In particular, for $G=K_n$, if $\bigcap_{i=1}^m [s_i,s_i+1]=\emptyset$ then  $M_\SS\neq \UST$.
\end{lemma}
\begin{proof}
The first part is easily seen by selecting any spanning tree $T$ containing $e_k$ but not $e_j$, which must exist if $e_j$ is non-separating.  Since this tree can't be obtained by MST selection using $D$ weights, the distribution does not weight all spanning trees equally.  The same observation carries over to the second part, letting $j$ be the index for the smallest coordinate of $\SS$ and $k$ for the largest.  The fact that not all intervals overlap implies that these intervals do not overlap, so the $k$th weight is guaranteed to be larger than the $j$th.
\end{proof}

Next we observe a monotonicity in probabilities as we shift a single weight.

\begin{lemma}[Monotonicity under interval sliding]\label{lem:monotonicity-slide} 
Given any $\SS\in\Sh(m)$, let $\SS(k,t)$ have all the same coordinates except for a slide in the $k$th coordinate:  $s_k$ is replaced by $s_k+t$.
Then for all $j\neq k$, $\prob_{M_{\SS(k,t)}}(e_j\in T)$ is weakly increasing in $t$, and strictly increasing when the intervals overlap (that is, when $s_j-1<s_k+t<s_j+1$).
\end{lemma}

The weak increase statement holds for any graph $G$ (not just for $G=K_n$). On the other hand, the strict increase does rely on the geometry of the graph---to see this, note that a separating edge always has probability of inclusion equal to $1$.

\begin{proof}
Fix $\SS=(s_1,\dots,s_m)\in \R^m$ and $\ell\in \{1,..,m\}$, and set
$M_t\coloneqq M_{\SS(\ell,t)}$.
Suppose $e$ is an edge with endpoints $v,w$.
In this notation, we have

$$
    \P_{M_t}(e\notin T)  
    = 
    \sum_{\substack{\hat{T}\in ST(K_n)\\ e \notin \hat{T}}} \P_{M_t}(\hat{T})
    =
    \sum_{\substack{P: v\to w\\ |P|\ge 2 }} \P_{M_t}(P\subset T),
$$

\noindent where the last sum is over paths of length $\ge 2$ between the endpoints of $e$, so that each $P$ completes a closed loop with $e$.
Here, $T$ is the tree-valued random variable, and $\hat{T}$ is an instance (a particular tree).  $P\subset T$ means that the path is found within the tree.
For such a path $P$, denote its edges by 
 $e_{i(1)},...,e_{i(k)}$. 

If we write $w,w_1,\dots,w_k$ for a set of draws of weights on the edges $e,e_{i(1)},...,e_{i(k)}$ under $\SS$, 
then we can write $w_i(t)=w_i+\delta_{i\ell}t$ for corresponding weights drawn from $\SS(\ell,t)$. 
 Then $P\subset T$ 
 where $T$ is minimal for the sampled weights 
 if and only if $w\geq \max\{w_{i(1)}(t),...,w_{i(k)}(t)\}\eqqcolon m(t)$.

 So $\P_{M_t}(P\subset T)=\P\left(w> m(t)\right)$. 
If $\ell\notin\{i(1),...,i(k)\}$, then  $\P_{M_t}(P\subset T)$ does not depend on $t$. If $\ell\in\{i(1),...,i(k)\}$, say $\ell=i(1)$, then there are two cases.

If $w<\max\{ w_2,...,w_k\}$, the probability of the event $w> m(t)$ equals zero.
On the other hand, if $w>\max\{ w_2,...,w_k\}$, then the probability of the event $w>m(t)$ equals the probability that 
$w> w_\ell+t$.  This probability is strictly decreasing in $t$ as long as $w$ is in the interior of the interval from which $w_\ell(t)$ is drawn from uniformly at random, i.e. whenever $|s_1-(a_\ell+t)|<1$. 
Integrating over all samples  $(w,w_{i(2)},...,w_{i(k)})$ yields that $\P_{M_t}(P\subset T)$ is decreasing in $t$ whenever $|s_1-(s_\ell+t)|<1$, which finally tells us that $\P(e \in T)$ is increasing as desired.
\end{proof}

\begin{theorem}[Shifts do not suffice] \label{MST_not_UST_Kn}
For $G=K_n$ with $n\ge 4$, there is no $\SS\in\Sh(\binom n2)$ for which  $M_\SS= \UST$.
\end{theorem}

%% U for uniform

\begin{proof}  Given a shift vector $\SS$, 
if $\min s_i < \max s_i$, then applying interval sliding (Lemma~\ref{lem:monotonicity-slide}) shows that an edge chosen from the lowest interval is more likely to be included in a tree chosen from $M_\SS$ than edge chosen from the highest.  Thus $M_\SS\neq \UST$.
On the other hand, if $s_i=s_j$ for all $1\leq i,j \leq m$, then $M_\SS=\MSTZ \neq \UST$, since stars are more likely than paths.
\end{proof}

% \textcolor{magenta}{A: apparently ``equal edge inclusion proability" is now called `` equal edge utilization" and has been moved to the preliminaries. Somehow the observation that UST on $K_n$ (which this Corollary relies on) has this property disappeared completely. Should be added again somewhere.\\
% Similarly, where is the broken-cycle observation for MST?}

\subsection{A limitation of product measures with connected support}\label{sec:limits_of_connected}

To see the limitations of product measures obtained by shifting, we next show that a shift measure has to be extremely complicated to hit given probabilities on trees; in particular, it is hard to hit the uniform distribution.

\begin{proposition}[UST on $K_n$] \label{proConnectedSupportSwap}
    Let $n \geq 4$, and suppose $\{X_e\}_E$ is a product measure on $K_n$, where  each $X_e$ is supported on a connected subset of $\R$.
If there exist any two edges $e$ and $e'$ with a common endpoint for which $X_e$ is distributed identically to $X_{e'}$, then $\{X_e\}_E$ does not induce UST on $K_n$.
\end{proposition}

\begin{proof}
    Since the sets $\supp(X_e)$ for each edge $e$ are all connected intervals in $\R$, there are two possibilities for how they can intersect:
    \begin{enumerate}[label=(\arabic*)]
        \item\label{itmLimitationDisjoint} There is some pair of distinct edges $e_1, e_2 \in E(K_n)$ such that $\supp(X_{e_1})$ and $\supp(X_{e_2})$ are disjoint intervals (possibly overlapping at endpoints), with $w_1 \leq w_2$ for all $w_1 \in \supp(X_{e_1})$ and $w_2 \in \supp(X_{e_2})$.
        \item\label{itmLimitationIntersect} The intersection
        $\bigcap_{e \in E} \supp(X_e)$
        is an interval $[a, b]$ of nonzero length (i.e., $a < b$).
    \end{enumerate}
    
    In case \ref{itmLimitationDisjoint}, observe that it is impossible for the product measure $\{X_e\}_E$ to yield a minimum spanning tree that is a path from one endpoint of ${e_1}$ to the other, passing through $e_2$, because swapping out $e_2$ for $e_1$ would yield a spanning tree of lower weight. (It is indeed \emph{strictly} lower, otherwise the product measure fails to be non-colliding.) Therefore, the product measure does not induce UST, which requires equal weight on all trees.

    In case \ref{itmLimitationIntersect} we will use an edge rotation, observing that the cycle expansion inequality of Proposition~\ref{prop:tri_rotation} holds in a slightly more general context, beyond ordinary MST. The only property we require for the weak inequality to hold is that the cycle-expanding bijection preserve the probability distributions on the edges that are being permuted. In particular, it applies to the bijection that swaps edges $e$ and $e'$ since they have the same distributions by assumption. And the only property we require for the strict inequality to hold is that every permutation has nonzero probability of occurring, which follows from the condition in case~\ref{itmLimitationIntersect}. Thus, we may apply Lemma~\ref{lem:tri-rotation} with $T_1$ and $T_3$ as single vertices and $T_2$ some spanning tree of the rest of $K_n$ (as in the top pair of graphs in Figure~\ref{fig:tri_rotation}), with $\{e_{12}, e_{13}\} = \{e, e'\}$ to conclude that some pair of trees have unequal probabilities of being sampled. (Note that $n \geq 4$ implies there is another edge in $K_n$ between $T_1$ and $T_2$.)
\end{proof}

Consider coloring the edges of a graph with a different color for distinct distributions; in this formulation, the proposition states that a product measure inducing UST would have to be a proper coloring.  
Note that the edge coloring problem for complete graphs requires at least $n-1$ colors if $n$ is even and $n$ colors if $n$ is odd, which means that a pattern of shifts recovering UST has to be highly nontrivial if one exists at all.

%%%%%%%%%
%%%%%%%%%

\section{Arbitrary product measures}\label{secArbitraryProductMeasures}

In this section, we move to the full generality of product measures, exploring the very natural question presented in the introduction: when real-valued random variables are drawn from independent measures, how likely are they to be in each possible order?  That is, what distributions on permutations are induced?
As noted above, the distribution on permutations is a strict generalization of the classic topic of intransitive dice, which just concerns  pairwise comparisons.

%%%%
\subsection{Defining the permutation locus}\label{subArbitraryIntro}

For any integer $m\ge 1$,
let $\cM(\R^m)$ be the set of finite Borel probability measures on $\R^m$. In this section we study the  set $\cM=\cM_m \subsetneq (\cM(\R))^m \subsetneq \cM(R^m)$ consisting of non-colliding product measures, meaning that every $\mu\in\cM$ corresponds to $m$ independent random variables $X_1,\dots,X_m$ on the real line $\R$ with the property that 
$\P(X_i=X_j)=0$ for all $i\neq j$.
The non-colliding property is automatic when the  $X_i$ are continuous random variables, but below we will explore discrete measures with finite support, where the non-colliding property requires the supporting points to be distinct.

Let $\Delta(S_m)$ be the space of probability measures on permutations, which is a simplex with the $m!$ permutations as its extreme points.
Consider the map
$\psi\colon\cM\longrightarrow\Delta(S_m)$ given by 
$$\psi(\mu)(\si):=\mu\left(\left\{x\in\R^m: x_{\si_1}<\cdots<x_{\si_m}\right\}\right).$$
Non-collision implies that any draw from the $X_i$ induces a strict order, so $\psi(\mu)$ is a probability distribution on $S_m$.
The main goal of this section is to understand the image of $\psi$, denoted by $P_m \coloneqq \txt{Im}(\psi)$,  which we call the {\em permutation locus} for product measures.

First, we observe that the image of $\psi$ is a subset of the simplex $\Delta(S_m)$ that contains all $m!$ extreme points.  To see this, note that you can construct a probability measure supported only on the permutation $\sigma$ by having each of the random variables $X_i$ be an atomic measure with support at one point, and arrange the support points in the order needed to get $\sigma$.  

Next, we note that the image is not everything (and so, in particular, is not convex.) In the sections below, we will realize the image of $\phi$ as a semi-algebraic set and will set about describing bounds on its dimension---non-surjectivity follows from a dimension count.  But in addition, it is easy to explicitly describe a point of $\Delta(S_3)$ (say) that is missed by product measures.

\begin{example}[A point missed by product measures]\label{lem:canthit}
No product measure on $\R^3$ induces equal probability $1/3$ for each of the events
$a<b<c$, \quad $b<c<a$, and $c<a<b$.

To see this, name three independent random variables $a,b,$ and $c$ on $\R$. Write $\Phi_a, \Phi_b$ and $\Phi_c$ for the corresponding cumulative distribution functions.

Suppose that $\P(a<b<c)=\P(b<c<a)=\P(c<a<b)=\frac{1}{3}$. Then
$$
0  = \P(b<a<c) =
  \int_{-\infty}^{+\infty} \int_{-\infty}^z\int_{-\infty}^x d\Phi_b(y) d\Phi_a(x) d\Phi_c(z) \\
 = \int_{-\infty}^{+\infty} \int_{-\infty}^z\Phi_b(x) d\Phi_a(x) d\Phi_c(z). 
$$

Since everything is non-negative, we get
$$
 \int_{-\infty}^z\Phi_b(x) d\Phi_a(x) = 0, \qquad\text{for $c$-a.e. $z\in\R$.}
$$
Let $[c_1,c_2]$ be the smallest interval containing the support of $d\Phi_c$. Then
$\Phi_b(x)=0$ for $a$-a.e. $x\in[c_1,c_2]$.
Thus, if $[b_1,b_2]$ is the smallest interval containing the support of $d\Phi_b$, we must have 
$c_2\le b_1$.
By cyclicity, $\P(a<c<b)=\P(c<b<a)=0$ implies that
\[
a_1\le a_2\le c_1 \le c_2 \le b_1 \le b_2 \le a_1,
\]
so $a=b=c$ all have equal one-point support, which contradicts the given description.
\end{example}

Certain equalities and inequalities are easily seen to be satisfied at all points in $P_m$.  These include independence identities like 
$\P(a<b \hbox{\rm ~and~} c<d) = \P(a<b)\cdot \P(c<d)$ and correlation inequalities like $\P(a<b \mid a<c) \ge \P(a<b)$.\footnote{The latter is a simple example of a {\em positive correlation inequality}, which fits in the broader framework of FKG correlation inequalities from statistical physics and percolation theory.}  The equalities cut down the dimension of $P_m$ and the inequalities create its faces.  A fuller study of dimension is found in Section \ref{subsec:Dimension}.

In the remainder of this section, we will construct discrete measures and will use these to express arbitrary product measures and the uniform distribution in particular.  Using this technology, we will bound the dimension of $P_m$ and make a conjecture about its exact value.  

% See Table~\ref{tab:word-lengths-dims} for a summary of the results in
% this section.

% \begin{table}[htb!]
%   \centering
%   \begin{tabular}{llll}
%     \toprule
%     Type & Upper bound & Result & Sharp? \\
%     \midrule\\
%     Universal word for product measures
%          & $m(m-1)m! + 1$
%                  & Corollary~\ref{cor:universal_word}
%                           & Unlikely \\
%     Each word in universal collection
%          & $m(m!)$
%                  & Theorem~\ref{thm:ProductMeasuresDiscrete}
%                           & Unlikely \\
%     Uniform word, all weights equal
%          & $(m!)^{m-1}$
%                  & Theorem~\ref{thm:UniformWordMap}
%                           & Unlikely \\
%     Uniform word, varying weights
%          & $L(m)$, factorial growth
%                  & Proposition~\ref{prop:quad-length}
%                           & Possible \\
%     Dim of $P_m$
%          & $C(m) \sim e(m-1)!$
%                  & Theorem~\ref{thm:dim-Pn-upper}
%                           & Conjectured \\
%     \bottomrule
%   \end{tabular}
%   \caption{Known upper bounds for word lengths and dimensions for constructions presented below. We give references to the precise
%     results, and our guess as to whether the bound is sharp.}
%   \label{tab:word-lengths-dims}
% \end{table}

\FloatBarrier
%%%%%
\subsection{Discrete product measures and word maps}
\label{sec:discrete-prod-measures}

Here, we develop the formalism of {\em weighted words} over a finite alphabet, and we will ultimately prove that these suffice to capture all non-colliding product measures. The terminology of ``words" is meant to suggest parallels to combinatorial group theory, but this presentation stands alone without group-theoretical machinery.  One motivation is to show that
the locus of product measures $P_m$ is semi-algebraic by 
describing it with {\em word maps}---mappings from weights to measures for fixed words.

\begin{definition}[Word maps and weighted words]
Consider a finite set of symbols $\Sigma=\{a_1,\dots,a_m\}$, which we will think of as an alphabet. 
A word of length $r$ is a string of $r$ symbols from the alphabet, $w\in \Sigma^r$.  This word induces a {\em word map}
$F_w:\R_{\ge 0}^r \to \Delta(S_m)$ as follows.  
Let $\alpha\in\R_{\ge 0}^r $ be a weight vector. For a symbol $a_i$,
appearing $s_i$ times in~$w$, suppose 
the weights of these occurrences are $(r_i^1.\dots,r_i^{s_i})$, summing to $r_i$.  For each symbol $a_i$, we will select exactly one of its occurrences in the word $w$, selecting the $k$th occurrence with probability $r_i^k/r_i$.  This makes $F_w(\alpha)$ into a probability distribution on $S_m$, where the probability of $\sigma\in S_m$ is given by the selected symbols coming in the order
$a_{\sigma(1)}a_{\sigma(2)}\dots a_{\sigma(m)}$.  We will call the pair $(w,\alpha)$ a {\em weighted word} over the alphabet $\Sigma$, and let $\cW_{m,r}\subseteq \cM_m$ denote the measures on $S_m$ given by weighted words of length $\le r$.  
\end{definition}

\begin{example}
Suppose the alphabet has only two letters: $\Sigma=\{a,b\}.$
Consider the weighted word given by $w=abab$ and $\al=(2,1,3,5)$, which gives weights $(2,3)$ to copies of $a$ and $(1,5)$ to copies of $b$.  We now describe how to interpret this as a probability distribution on the strings $\{ab,ba\}$.  
The  word $abab$ contains both ordered subsequences:  $ab$ (achieved three ways as
$\underline{\mathbf{a}}\,\underline{\mathbf{b}}\, a\, b$,
$\underline{\mathbf{a}}\,b\,a\,\underline{\mathbf{b}}$, or
$a\,b\,\underline{\mathbf{a}}\,\underline{\mathbf{b}}$) 
and $ba$ (achieved one way as $a\,\underline{\mathbf{b}}\,\underline{\mathbf{a}}\,b$).  The weights tell us that the relative likelihood that the symbols are chosen in each position, so that for instance $b$ is five times as likely to be chosen in the fourth position as in the second.  We calculate that  $ba$ is chosen with probability $\frac 35\cdot \frac 16=\frac 1{10}$, and $ab$ is chosen the rest of the time (probability $9/10$).  We have therefore defined a probability distribution on the permutations of $\{a,b\}$. 
\end{example}

We can think of the orderings $ab$ and $ba$ as {\em draws} from the weighted word $(w,\alpha)$.

A weighted word also has a concrete interpretation as an atomic measure, which for this example is shown in Figure~\ref{fig:atomic}. Fix any $r$ points on the real line, labeled in order as in the word $w$, with probability weights $r_i^k/r_i$ as described above.  This describes a product measure because the random variable $X_i$ is independent from $X_j$; it induces a distribution on permutations via the order of $\{X_1,\dots,X_m\}$.

\begin{figure}[bht!]\centering
\begin{tikzpicture}
    \draw (0,0) -- (8,0);
    \filldraw [alizarin] (2,0) circle (0.1) node [above=5pt] {$\frac 25$};
    \filldraw [alizarin] (5,0) circle (0.1) node [above=5pt] {$\frac 35$};
    \filldraw [teal] (2.8,0) circle (0.1) node [above=5pt] {$\frac 16$};
    \filldraw [teal] (7,0) circle (0.1) node [above=5pt] {$\frac 56$};
\node at (2,-.4) {2};
\node at (2.8,-.4) {2.8};
\node at (5,-.4) {5};
\node at (7,-.4) {7};

\end{tikzpicture}
\caption{The measure $X=\frac 9{10} \delta_{ab}+\frac 1{10}\delta_{ba} \in \Delta(S_2)$ can be written as a product measure in many ways, including the one shown here.  
The measure $X_1$ is supported on $\{2,5\}$ and the measure $X_2$ is supported on $\{2.8,7\}$, with the atoms having probabilities marked above, so that $X_1=\frac 25 \delta_2 +\frac 35 \delta_5$ and $X_2=\frac 16\delta_{2.8}+\frac 56\delta_{7}$.
The probability that $X_1<X_2$ is therefore $9/10$, so the product measure $(X_1,X_2)$ recovers $X$. This corresponds to the word $abab$ weighted by $(2,1,3,5)$.
The positions $2, 2.8, 5, 7$ are arbitrary in this construction; only the alternation between $a$ sites and $b$ sites is prescribed by $w$.}\label{fig:atomic}
\end{figure}

Since we use normalized probabilities $r_i^k/r_i$, the weight vector is scale invariant, and we can use
homogeneous coordinates $\{[r_1^1 : \dots : r_1^{s_i}] \mid r_i^k \ge 0\} \subset \bR\bP^{s_i-1}$ for the weights, allowing us to group by symbol and rewrite the weights $(2,1,3,5)$ in the example above as  $([2,3],[1,5])$.  
We can use this point of view to extend beyond non-negative weights
and work with complexified projective spaces like $\bP\bC^{s_1} \coloneqq
\C\bP^{s_1-1}$. We can likewise do this on the codomain, thinking of
the set $\Delta(S_m)$  of distributions on permutations as sitting in a projective space
$\bP\bC^{m!}=\C\bP^{m!-1}$. Thus, word maps extend to algebraic maps
$$F_w:\bP\bC^{s_1}\times\dots \times\bP\bC^{s_m}\to\bP\bC^{m!}.$$

\begin{theorem}[Bounded-length word maps suffice] \label{thm:ProductMeasuresDiscrete}
For any $m$ there is an $N$ so that for any non-colliding product
measure on $m$ variables, there is a
weighted word $(w,\alpha)$ with the same image in $\Delta(S_m)$, where $w$ has length at most $N$.

We can choose $N=m(m!+1)$, giving  $P_m=\psi(\cW_{m, \ m(m!+1)})$.

\end{theorem}

% The proof relies on Carath\'eodory's theorem, \cite[Theorem 17.1]{rockafellar1970}.

%\footnote{See:\url{https://en.wikipedia.org/wiki/Carath\%C3\%A9odory's\_theorem\_(convex_hull)}}

The proof has two parts, one that represents arbitrary measures with possibly long word maps, and a lemma that reduces the length of the word representation.

\begin{lemma}[Shortening word maps]\label{lemProductMeasuresDiscreteToDiscrete}
Let $(w,\alpha)$ be a word map on $m$ symbols with any number of copies of each symbol.  Then there is a word map 
$(w',\alpha')$ inducing the same distribution on permutations with
$s_i\le m!$, i.e., 
using at most $m!$ copies of each of the $m$ symbols.
\end{lemma}

\begin{proof}
% Given any discrete product measure $(\seq{X}{n})$ inducing some $p \in \Delta(S_n)$, we simply define $f((\seq{X}{n}))$ to be any product measure $(\seq{Y}{n}) \in C_d$ inducing $p$ for the minimal possible $d$.\footnote{If we impose the extra requirement each $Y_i$ is supported on a set of positive integers (which doesn't affect the set of distributions that can be induced), then we can take the lexicographically smallest $(\seq{Y}{n})$, eliminating the need to invoke the axiom of choice.} We thus need only prove that $d = n!$ suffices. Let 
		
We describe a procedure that iteratively modifies the input word map, one index at a time, until it uses no more than $m!$ copies of each symbol.  The main tool is Carath\'eodory's Theorem \cite[Theorem 17.1]{rockafellar1970}, which says in particular that if a point $p$ in $\R^d$ is a convex combination of an arbitrary set of points, then there is a subset of at most $d+1$ points whose convex hull contains $p$.

Fix an index $i$.  Let 
$q_j=r_i^j/r_i$ be the probability that we select the $j$th position for symbol $a_i$.  
Let $p_j(\sigma)$ be the probability that $(w,\alpha)$ induces $\sigma\in S_m$, conditioned on choosing the $j$th position for $a_i$.  
Then we have 
$p(\sigma)=\sum_{j=1}^{r_i} q_j p_j(\sigma)$.

Thinking of $p$ and $p_j$ as vectors in $\R^{m!}$ (where each coordinate is given by a different choice of $\sigma$), we have 
$$p = \sum_{j = 1}^{r_i} q_j p_j,\qquad\text{where } \sum_{j=1}^{r_i} q_j=1\text{ and }0\le q_j\le 1.$$
We see that $p$ is contained in the convex hull of $\{\seq{p}{r_i}\}$.
Since the $p_j$ lie in $(m! - 1)$-dimensional subspace,
Carath\'eodory's theorem implies
that we can write $p$ as a convex combination of only
$m!$ of these points. Since each of the $p_j$ amounted to putting $a_i$ in a particular position, this reduction amounts to deleting all but $m!$ of the positions. We may repeat this process for each index $i$.
\end{proof}

\begin{proof}[Proof of Theorem \ref{thm:ProductMeasuresDiscrete}] 
Let $X = (\seq{X}{m})$ be an arbitrary product measure with distribution $p \in \Delta(S_m)$.
For each positive integer $j$, we will construct a discrete product measure  $\hat{X}^j \coloneqq (\hat{X}^j_1, \hat{X}^j_2, \dots, \hat{X}^j_m)$ so that, for each $i$,  the $\hat{X}^j_i$ approximate $X_i$ as $j\to\infty$.  This is subdivided in three parts: the construction, a crucial estimate, and the proof of convergence.

\medskip

\paragraph{Part I: Construction of $\hat{X}=\hat{X}^j$.}
Write $\Phi_i$ for the cumulative distribution function of the variable $X_i$.
Let the inverse $\Phi_i^{-1} \colon [0,1]\to\R^*=\R\cup\{-\infty\}\cup\{+\infty\}$ be
$\Phi_i^{-1}(u) \coloneqq \inf\left\{x\in\R^*: \Phi_i(x) \ge u\right\}$, so that whenever $0<\Phi_i(x)<1$, we have that $u\le \Phi_i(x)$ iff $\Phi_i^{-1}(u)\le x$.  We then have the standard construction
$X_i = \Phi_i^{-1}(U)$,
where $U$ is uniform on $[0,1]$.

Let $j$ be a positive integer and split the interval $[0,1]$ uniformly into $j$ consecutive half-open intervals:
$$[0,1]=\textstyle \left[0,\frac{1}{j}\right)\cup\left[\frac{1}{j},\frac{2}{j}\right)\cup\cdots\left[\frac{j-1}{j},1\right]$$
Then, sampling from $U$ is equivalent to choosing $k \in \{0,1,\dots,j-1\}$ uniformly at random and then sampling from $U$ conditioned to be in $\left[\frac{k}{j},\frac{k+1}{j}\right)$. In other words, we have the variable decomposition
$X_i = \sum_{k=0}^{j-1} X^j_{i,k}$,
where $X_{i,k}\coloneqq\Phi_i^{-1}\left(U\cdot\ones_{[k/j,(k+1)/j)}\right)$.

These $X^j_{i,k}$ depend on a choice of $j$, but we suppress
that for now for this and related objects.
Then, in view of the variable decomposition above, we can think of the sample space for the product measure $X$ as being the cube
$\Om\coloneqq [0,1]^m$.

For $k=0,1,\dots,j$, set $x_{i,k}\coloneqq\Phi_i^{-1}\left(\frac{k}{j}\right)$
and (for $k < j$) consider the intervals 
$B_{i,k}$ from $x_{i,k}$ to $x_{i,k+1}$, where we include an endpoint
in the interval iff $X_i$ has a point mass there.
(Thus the $B_{i,k}$ for different~$k$ may overlap at their
  endpoints, and may be single points.)
Now we define a
sequence of approximating discrete
product measures $\hat{X}$ on the same sample space $\Om=[0,1]^m$.
For each $i=1,\dots,m$, pick (once) $u_{i,k}$ uniformly randomly
  in the interval $\left[\frac{k}{j},\frac{k+1}{j}\right)$, and set $\hat{X}_i$
to be the discrete product measure equally weighted on the points
$\hat{X}_{i,k}\coloneqq\Phi_i^{-1}(u_{i,k})$.
With probability one, $\hat X_{i,k} \in B_{i,k}$ and the values 
$\hat X_{i,k}$ are distinct for different~$i$, since the product
measure $X$ was assumed non-colliding.

\paragraph{Part II: Estimating the error in the resulting distribution.}
Next, let $\hat{p}=\hat{p}^j\in\Delta(S_m)$ be the distribution induced by $\hat{X}=\hat{X}^j$ and fix a permutation $\si\in S_m$. We want to estimate the difference $\abs{\hat{p}^j(\sigma) - p(\sigma)}$ in terms of the resolution~$j$ of our grid.
To that end, apply the variable decomposition $X_i = \sum_{k=0}^{j-1}
X_{i,k}$ to  each $X_i$, and pick any two terms from different~$X_i$, say $X_{1,k}$ and
$X_{2,\ell}$. These two variables are supported in the intervals $B_{1,
  k}$ and $B_{2, \ell}$, as are the corresponding approximating
variables $\hat{X}_{1,k}$ and $\hat{X}_{2,\ell}$. In particular,
if $B_{1, k}\cap B_{2, \ell}=\emptyset$, then the relative order
of $X_{1,k}$ and $X_{2,\ell}$ is preserved when we pass to
$\hat{X}_{1,k}$ and $\hat{X}_{2,\ell}$. We will give bounds on
the number of possible overlaps.

Now suppose $B_{1,k}\cap B_{2,\ell}\ne\emptyset$. We claim that if
$k'<k$ and $\ell<\ell'$ then
$B_{1,k'}$ and $B_{2,\ell'}$ do not intersect.
Intuitively, $B_{1,k'}$ is to the left of $B_{1,k}$, which
overlaps $B_{2,\ell}$, which is to the left of $B_{2,\ell'}$. More
formally, we have inequalities between the endpoints of these
intervals:
\[
  x_{1,k'+1} \le x_{1,k} \le x_{2,\ell+1} \le x_{2,\ell'}
\]
where the first inequality comes from $k' < k$, the second
inequality from the intersection between $B_{1,k}$ and
$B_{2,\ell}$, and the third from $\ell < \ell'$.
Thus we can only have an intersection between $B_{1,k'}$ and
$B_{2,\ell'}$ if all three inequalities are equalities and
furthermore $X_{1}$ has a point mass at~$x_{1,k}$ and $X_{2}$ has a
point mass at~$x_{2,\ell}$. But this contradicts the
assumption that $X$ is non-colliding.

With $k,\ell\in\{0,\dots,j-1\},$ consider the indices of the non-empty intersections
$\{(k,\ell)\mid B_{1,k} \cap B_{2,\ell} \ne \emptyset\}$. By the claim
above, these intersections are totally ordered by the relation
$(k,\ell) < (k',\ell')$ iff $k < k'$ or $\ell < \ell'$. A maximum
chain of such intersections therefore contains $2j-1$ intersections total, since the sum $k + \ell$ can take integer values between $0$ and $2(j-1)$. In particular, with
probability at least $1-\frac{2j-1}{j^2}$, the variables $\hat X_1$
and~$\hat X_2$ have the same relative ordering as $X_1$ and~$X_2$.
Since this applies for all pairs of variables, by the union bound, for
all $\sigma \in S_m$, we have
$\abs{\hat{p}^j(\sigma) - p(\sigma)} < \binom{m}{2}\cdot\frac{2}{j} <
\frac{m^2}{j}$
and the $\hat p^j$ converge to~$p$.

\paragraph{Part III: Proof of convergence.} 
Consider the sequence of discrete product measures $Y^1, Y^2, Y^3,
\dots$ defined by letting $Y^j$ be the shortening of $\hat X^j$ to use
at most $m!$ of each symbol (using Lemma
\ref{lemProductMeasuresDiscreteToDiscrete}).
Topologically, we may think of each $Y^j$ as lying in a space
$$C\coloneqq F\times \bP_1^{s_1}\times\dots \times\bP_m^{s_m}$$
where $F$
is a finite set encoding the combinatorial information of the relevant
word maps and the probability weights are drawn from the projective
spaces, as before.  Since $F$ is finite and the projective spaces are
compact,  $C$ is compact. Therefore, there is a subsequence $Y^{j_1},
Y^{j_2}, Y^{j_3}, \dots$ converging to a limit $Y^* \in C$ under the
$L_\infty$ norm. That is, $Y^*$ is a discrete product measure where
each $Y^*_i$ is supported on a set of size at most $m!$ agreeing with
each $Y^{j_k}_i$, and furthermore, for any
$\varepsilon > 0$, sufficiently large~$k$, and
value $v$ in the support, $\abs{\P(Y^{*}_i = v) - \P(Y^{j_k}_i = v)} <
\varepsilon$. Since $Y^{j_k}$ induces $p^{j_k}$, it follows that $Y^*$
induces a distribution $p^*$ such that, for all $\sigma \in S_m$,
$\abs{p^*(\sigma) - p^{j_k}(\sigma)} \leq m \varepsilon$.
By the triangle inequality, we have
$$\abs{p^*(\sigma) - p(\sigma)} \leq m \varepsilon + \frac{m^2}{j_k}.$$
It follows that
$p^* = p$, i.e., $Y^*$ induces the same distribution as $X$.
\end{proof}

\begin{corollary}[Universal words]\label{cor:universal_word}
For any word $u$  of length at least $(m!)m(m-1) + 1$ that cyclically repeats all of the letters in the same order, the collection of word maps $(u,\alpha)$ suffice to cover all of $P_m$.
\end{corollary}

We call such a word $u$ a \emph{universal} word. For example, when $m= 3$, the word
$$u=abcabcabcabcabcabcabcabcabcabcabcabca,$$
(which consists of 12 copies of the string $abc$ followed by the letter $a$) is a universal word.

\begin{proof}
From the previous theorem, any product measure can be associated to some word map on a word $w$ with at most $m!$ copies of each symbol, for a total length of $(m!)m$. 
We will find this $w$ within any given universal word $u$ by explaining how to choose the weights.
The first letter of $w$ occurs in position $m$ or earlier of $u$, and each of the remaining letters occur at most $m - 1$ positions down from the previous one. Thus, after cyclically enumerating
$$m + ((m!) m - 1)(m - 1) = (m!) m (m - 1) + 1$$
letters of the universal word $u$, we have realized $w$ as a subsequence. This implies that we can realize $Y^*$ with the appropriate weights over $u$; we simply set the weights of unused letters to be zero.    
\end{proof}

Another consequence of Theorem~\Ref{thm:ProductMeasuresDiscrete} is that
$P_m$ is a semi-algebraic subset of $[0, 1]^{m!}$, meaning that it is
described by a finite set of equalities and inequalities in the
variables $x_\sigma$, as follows.
For a given $w$, the coefficients of the $x_\sigma$ are polynomial expressions in the weights $\alpha$, and letting the weights vary over a universal word parametrizes the entirety of $P_m$. 
Then apply the Tarski-Seidenberg Theorem, which states that the image
of a semi-algebraic set under an algebraic map is also semi-algebraic.

%%%%%
%%%%%
\subsection{Word maps for the uniform distribution}\label{subWordMapExamples}

Because it leverages compactness for convergence, Theorem \ref{thm:ProductMeasuresDiscrete} is non-constructive. In this section we
show that the classical theory of quadrature gives an elegant construction of short words inducing the uniform distribution.  
To motivate this, we start with a more naive construction and observe that it is highly inefficient.

For any permutation $\sigma \in S_m$ and any word $w$ on $m$ symbols, we define $\sigma \circ w$ to be the word where we substitute every letter in $w$ according to $\sigma$. We inductively define words $v$  as follows. Let $v_{1,m}$ be a word containing each of the $m$ letters once in arbitrary order, and for each $k \geq 2$, let
$$v_{k,m} \coloneqq \prod_{\sigma \in S_m} \sigma \circ v_{k - 1,m},$$
where the product symbol refers to the operation of concatenation.

\begin{theorem}\label{thm:UniformWordMap}
    The words $v_{m,m}$, with equal weight on each letter, induce the uniform distribution on $S_m$.
\end{theorem}

\begin{proof}
We prove the following stronger claim by induction: consider a subset $\Sigma_0\subset \Sigma$
containing at most $r$ of the letters from the alphabet.
Restricting $v_{r,m}$ to the letters from $\Sigma_0$, we will show that
the resulting word map (with equal weights) induces the uniform distribution over orderings of $\Sigma_0$.

Fix any $m$. 
The base case, $r = 1$, is immediate. For the inductive step, suppose $r \geq 2$ and the claim holds for $v_{r - 1,m}$. There are two cases to consider. First suppose that all $r$ of the letters in a draw from $v_{r,m}$ happen to be chosen from the same copy of $v_{r - 1,m}$. Conditioning on this random event, each copy of $v_{r - 1,m}$ is equally likely, and hence the corresponding permutation $\sigma \in S_m$ is uniformly random as well. Thus every possible ordering of the $r$ letters is equally likely. Now suppose instead that at most $r - 1$ of the $r$ letters are chosen from any individual copy of $v_{r - 1,m}$. Conditioning on this event, the inductive hypothesis says the letters are ordered uniformly at random within each copy. Thus, the $r$ letters are randomly distributed across the different copies of $v_{r - 1,m}$ and then randomly ordered wherever there is a collision. This process is clearly equivalent to uniformly randomly ordering the $r$ letters. By induction, the claim holds for all $r \geq 1$.
\end{proof}

For $m = 2$ the two possible words $v_{2,2}$ arising from this construction are   $abba$ and $baab$. A word  $v_{3,3}$ for $m = 3$, broken up into the six transformed copies of $v_{2,3}$ indexed by $S_3$, is:
\begin{gather*}
  abcacbbacbcacabcba \ \ acbabccabcbabacbca \ \ bacbcaabcacbcbacab\\
  bcabaccbacababcacb \ \ cabcbaacbabcbcabac \ \ cbacabbcabacacbabc
\end{gather*}
The length 108 of $v_{3,3}$ is significantly longer than the universal word of length 37 given above for $m=3$ (see below Corollary \ref{cor:universal_word}), but it has the benefit of simple weights $(1,1,\dots,1)$. In general, the length coming from this construction is $m(m!)^{m - 1}$, which is far greater than the $m (m!)$ upper bound established by Theorem \ref{thm:ProductMeasuresDiscrete} that would suffice to hit the uniform distribution.  Closing the gap with an explicit construction is the task at hand.

A more efficient construction can be drawn from ideas of 
 \emph{quadrature},
meaning methods of approximating the integral of a
function~$f$ on an interval by evaluating $f$ at a small number of
points. 
\begin{definition}
A {\em quadrature scheme} is an approximation to an integral constructed by a weighted combination (with weights $\alpha_i$) of function values at particular points ($0\le x_i\le 1$):  $\int_0^1 f(x)\ dx \approx
\sum_i \alpha_i f(x_i)$.  A quadrature scheme is said to {\em integrate polynomials of degree $\le d$} if the approximation is an equality for those inputs.
\end{definition}

\begin{theorem}[Quadrature words]\label{thm:quad-words}
Suppose that $\sum_{i=1}^r \alpha_i \delta_{x_i}$ is a degree-$k$
quadrature scheme with $0 \le x_1 < x_2 <\dots < x_r \le 1$ and
$w$ is a uniform word in $a_1,\dots,a_k$.
Then
$$U=w^{x_1-0} {a_{k+1}}^{\alpha_1} w^{x_2-x_1} {a_{k+1}}^{\alpha_2}\dots {a_{k+1}}^{\alpha_r} w^{1-x_r}$$
is a uniform word in $a_1,\dots,a_{k+1}$.
\end{theorem}

(Recall that 
exponents on a letter are used to indicate the weight of that letter in our
word map; we abuse notation to write $w^\alpha$ to mean the word $w$ with
each letter individually scaled.)
We defer the proof to Appendix~\ref{app:quadrature}.

Two examples of quadrature schemes are the approximations
$$\int_0^1 f(x)\, dx \approx \frac{f(0)}{4} + \frac{3 f(2/3)}{4}, \qquad \int_0^1 f(x)\, dx \approx \frac{f(0)}{6} + \frac{2 f(1/2)}{3} + \frac{f(1)}{6}.$$
These are, respectively, \emph{Gauss-Radau quadrature} with weights $(1/4,3/4)$ and nodes $(0,2/3)$ and \emph{Gauss-Lobato quadrature} with weights $(1/6,2/3,1/6)$ and nodes $(0,1/2,1)$.  
The first integrates polynomials of degree up to $2$ and the second integrates polynomials of degree up to $3$. 

%%%
We can now apply the theorem in the case $m=3$.  Since $w = aba$ is a
word in the two variables $a,b$ generating the uniform distribution on those letters,  applying the degree-2 Gauss-Radau quadrature scheme gives
\[ 
  c^{1/4} w^{2/3} c^{3/4} w^{1/3} = c^{1/4} a^{2/3} b^{2/3} a^{2/3}
  c^{3/4} a^{1/3} b^{1/3} a^{1/3} \equiv c a^2 b^2 a^2 c^3 aba
\]
as a uniform word on three letters, which at length 8 is highly efficient.

These constructions extend to higher degree, where Gauss-Radau has $x_1=0$ and integrates polynomials up to degree $2r-2$,  while Gauss-Lobatto has $x_1=0, x_r=1$ and integrates polynomials up to degree $2r-3$, handling the odd and even cases, respectively \cite{recipes}.  
(As a note,  the form of our quadrature theorem means that we will
produce the shortest words when $x_1=0$, $x_r=1$, or both.  This makes
Gauss-Radau and Gauss-Lobato slightly more efficient than the more
popular quadrature scheme known as Gauss-Legendre, whose nodes are
all internal.) As we show in Appendix~\ref{app:quadrature}, this
adds up as follows.

\begin{proposition}[Word lengths from quadrature]\label{prop:quad-length}
Define $L(m)$ recursively by $L(1) = 1$ and
  \begin{align*}
    L(m) &=
      \begin{cases}
        \frac{m+1}{2}\cdot L(m-1)+\frac{m+1}{2}, & m\text{ odd}\\
         \frac{m}{2} \cdot L(m-1)+ \frac {m+2}{2},  & m\text{ even}.
      \end{cases}
  \end{align*}
  Then there is a word map $(w_m,\alpha_m)$, with $w_m$ of length $L(m)$, that generates
  the uniform distribution on $m$ letters.
\end{proposition}

Clearly $L(m)$ grows factorially;  Table~\ref{tab:uniform-words} gives the first few values.

\begin{table}[bht!]
  \centering
  \begin{tabular}{crl}
    \toprule
    $m$& $L(m)$ & $(w_m,\alpha_m)$\\ \midrule
    1 & 1 & $a$\\
    2 & 3 & $bab$ \\
    3 & 8 & $c b^2 a^2 b^2 c^3 bab$ \\
    4 & 19 & $d c b^2 a^2 b^2 c^3 bab d^4 c b^2 a^2 b^2 c^3 bab d$\\
    5 & 60 \\
    6 & 184 \\
    7 & 740 \\
    \bottomrule
  \end{tabular}
  \caption{The first few values of $L(m)$, and corresponding
    uniform words $w_m$.}
  \label{tab:uniform-words}
\end{table}
Note that varying the weights on the uniform word $w_4$ from this construction gives a subset of $P_4$
of dimension at most $3 + 7 + 3 + 2 = 15$, and we will see in the next section that $\dim P_4 = 20$, showing that these are not also universal words.  
However, the length of 19 significantly improves on all constructions discussed previously for uniform words.  In particular, the naive construction $v_{4,4}$ from Theorem~\ref{thm:UniformWordMap} would have required  55,296 letters.

%%%%
%%%%
\subsection{Dimension of \texorpdfstring{$P_m$}{P\_m}}\label{subsec:Dimension}

We may think of each point in the permutation locus $P_m$ as a vector
	$$(x_{(1, 2, 3, 4 \dots, m)}, x_{(2, 1, 3, 4, \dots, m)}, x_{(2, 3, 1, 4, \dots, m)}, \dots, x_{(m, m - 1, m - 2, m - 3, \dots, 1)}) \in [0, 1]^{m!}$$
	where $x_{(i_1, i_2, \dots, i_m)}$ is the probability of realizing the ordering of random variables $X_{i_1} < X_{i_2} < \dots < X_{i_m}$.
In this section, we will sometimes use numerical symbols rather than letters to denote orderings and permutations, and we will switch between horizontal and vertical notation for permutations, as needed for clean exposition.
We have seen that $P_m$ is a semi-algebraic set in $[0, 1]^{m!}$.
 In this final section, we ask the most basic question about $P_m$: what is its dimension at generic points?
 Fuller definitions and proofs for this section appear in Appendix~\ref{app:dimension}; e.g., see Definition~\ref{def:generic} for the notion of ``generic''.
	
For instance, $P_3$ sits in $\Delta(S_3)$, a simplex on 6 extreme points, so the most naive upper bound on its dimension is 5. This turns out to be tight: while $P_3$ does not contain every point in this simplex, it is still full-dimensional. However, that is no longer true for $P_4$. This is due to additional constraints that hold of all distributions in $P_4$, such as the {\em independence constraints} mentioned in Section~\ref{subArbitraryIntro}. For instance, the event $X_1 < X_2$ must be independent from the event  $X_3 < X_4$:
$$\P([X_1 < X_2] \txt{ and } [X_3 < X_4]) = \P(X_1 < X_2) \cdot \P(X_3 < X_4).$$
Translating to permutation notation, the left-hand side has six terms, while
the probabilities on the right expand to twelve terms each:
$$x_\abcd 1234 + x_\abcd 1324 +x_\abcd 1342
+x_\abcd 3124 + x_\abcd 3142 +x_\abcd 3412$$
$$=\left(x_\abcd 1234 + x_\abcd 1324 + x_\abcd 1342 + x_\abcd 3124
+x_\abcd 3142 + x_\abcd 3412 + x_\abcd 1243 + x_\abcd 1423
+x_\abcd 1432 + x_\abcd 4123 + x_\abcd 4132 + x_\abcd 4312\right)$$
$$\cdot \left(x_\abcd 1234 + x_\abcd 1324 + x_\abcd 1342 + x_\abcd 3124
+x_\abcd 3142 + x_\abcd 3412 + x_\abcd 2134 + x_\abcd 2314
+x_\abcd 2341 + x_\abcd 3214 + x_\abcd 3241 + x_\abcd 3421\right).$$
	% \begin{align*}
	% 	&x_{(1, 2, 3, 4)} + x_{(1, 3, 2, 4)} + x_{(1, 3, 4, 2)} + x_{(3, 1, 2, 4)} + x_{(3, 1, 4, 2)} + x_{(3, 4, 1, 2)}\\
	% 	&= (x_{(1, 2, 3, 4)} + x_{(1, 3, 2, 4)} + x_{(1, 3, 4, 2)} + x_{(3, 1, 2, 4)} + x_{(3, 1, 4, 2)} + x_{(3, 4, 1, 2)}\\
	% 	&\hspace{.5em} + x_{(1, 2, 4, 3)} + x_{(1, 4, 2, 3)} + x_{(1, 4, 3, 2)} + x_{(4, 1, 2, 3)} + x_{(4, 1, 3, 2)} + x_{(4, 3, 1, 2)})\\
	% 	&\hspace{.6em} \cdot (x_{(1, 2, 3, 4)} + x_{(1, 3, 2, 4)} + x_{(1, 3, 4, 2)} + x_{(3, 1, 2, 4)} + x_{(3, 1, 4, 2)} + x_{(3, 4, 1, 2)}\\
	% 	&\hspace{.5em} + x_{(2, 1, 3, 4)} + x_{(2, 3, 1, 4)} + x_{(2, 3, 4, 1)} + x_{(3, 2, 1, 4)} + x_{(3, 2, 4, 1)} + x_{(3, 4, 2, 1)})
	% \end{align*}
Since all vectors in $P_4$ satisfy this polynomial equation, we know that that $\dim(P_4)\le \dim(\Delta(S_4)) - 1 = 22$. In fact, this is one of three similar equality constraints (partitioning $\{1, 2, 3, 4\}$ into two sets of size 2), and the true dimension works out to be 20.
	
The main result of this section (Theorem~\ref{thm:dim-Pn-upper}) is a general upper bound on $\dim(P_m)$, which we conjecture to be tight (Conjecture~\ref{conj:dim-Pn}). Our basic approach is to construct a large enough number of equality constraints that all points in $P_m$ satisfy and show that
their gradients are linearly independent at generic points. Since the
constraints  are all polynomials in the
$x_{\sigma}$ variables, it suffices to identify a
\emph{single} point at which the gradients of each constraint are
linearly independent; at generic points, the number of independent
constraints can only be larger. (See 
Lemma~\ref{lemGenericPoints}.) We
will thus choose to evaluate derivatives at the uniform distribution over permutations,
where each $x_{\sigma} = \frac{1}{m!}$.
	
For the construction, we will not use independence constraints that directly generalize the example above; instead, we employ constraints like the following.
Let $A$ be the event $X_1<X_2$ and $B$ be $X_3<X_4$.  Then one identity can be nicely expressed by referring to the matrix
\begin{equation}\label{eq:eo-examp}
  \left( \begin{array}{cc} \P(A\wedge B) & \P(A\wedge \lnot B) \\
    \P(\lnot A \wedge B) & \P(\lnot A \wedge \lnot B) \end{array}\right).
\end{equation}
We see that the determinant of this matrix is zero because independence lets us express each conjunction as a product, so that both diagonal products are equal to $\P(A)\cdot\P(\lnot A)\cdot\P(B)\cdot\P(\lnot B)$.  Note that the main diagonal contains terms with an even number of $A,B$ events and the anti-diagonal contains terms with an odd number of $A,B$ events.  

This determinant identity generalizes via multilinear algebra to any number of events. If sets of events $\mathcal{E}^+_i$ for $i \in [k]$ are independent, then the product of the probabilities of all ways where an even number of the events happen equals the product of the probabilities of all ways where an odd number of the events happen. Equality still holds if you condition on another global event. In the most general setting,  consider a  collection of independent events $\mathcal{E}_1, \mathcal{E}_2, \dots, \mathcal{E}_k$, each partitioned into disjoint sets $\mathcal{E}_i = \mathcal{E}^+_i \sqcup \mathcal{E}^-_i$, not necessarily complementary. (In the previous example, there was no conditioning; $\mathcal{E}_1$ and $\mathcal{E}_2$ were both the entire universe, with $\mathcal{E}^-_i$ equal to the complement of $\mathcal{E}^+_i$.) By expanding conjunctions into products and rearranging terms as before, one can prove the following even/odd constraint holds for any such collection of independent events; see Lemma~\ref{lemIndependenceConstraintsGeneral} for the formal proof.\footnote{Russell Lyons has communicated to us an elegant proof of the converse: if the product-of-even probabilities equals the product-of-odd probabilities for all subsets of~$I$, then the events $\mathcal{E}_i$ are independent. We will not apply this converse statement here.} 
\begin{equation}\label{EO}\tag{EO}
\prod_{\substack{s\colon [k] \to \{+, -\}\\\textnormal{such that }  
|s^{-1}(+)|
\textnormal{ is even}}} \P\left[\bigwedge_{i \in [k]} \mathcal{E}_i^{s(i)}\right] =
\prod_{\substack{s\colon [k] \to \{+, -\}\\\textnormal{such that } |s^{-1}(+)| \textnormal{ is odd}}} \P\left[\bigwedge_{i \in [k]} \mathcal{E}_i^{s(i)}\right].
\end{equation}

%For a given even/odd constraint, we can take the ratio of left-hand side to right-hand side (so that the constraint reads LHS/RHS$=1$).
For $m \geq 5$, we will take the gradient of Equation~(\ref{EO}) with respect to events involving more than just pairs of variables. For instance, in one of our constraints, we will use the following collection of events:
\begin{align*}
    \E_1^+ &:= [X_1 < X_2 < X_3] \vee [X_3 < X_2 < X_1] & \E_2^+ &:= [X_4 < X_5]\\
    \E_1^- &:= [X_2 < X_1 < X_3] \vee [X_3 < X_1 < X_2] & \E_2^- &:= [X_5 < X_4]
\end{align*}
We will end up associating the corresponding \ref{EO} constraint with the permutation $(123)(45)$, just as the \ref{EO} constraint from \eqref{eq:eo-examp} is associated with the permutation $(12)(34)$. For any $m$, we will prove that we have a linearly independent vector in the cotangent space of $P_m$ at the uniform distribution for every permutation on $\{1, \dots m\}$ that has at least two nontrivial cycles. Counting the number of such permutations will therefore give us a lower bound on the dimension of the cotangent space, and thus an upper bound on the dimension of the tangent space, which is the dimension of $P_m$.

We now need to introduce just enough notation to compactly write down these cotangent vectors (i.e., gradients of these constraints at the uniform distribution).
We will give the key ideas here, saving full notation and detail for Appendix~\ref{app:dimension}.

We will use $\Sigma\subset \{1,\dots,m\}$ for a set of symbols, and will write $O(\Sigma)$ for orderings on the symbols, such as $\ordabc 132$ as an ordering on $\{1,2,3\}$; we let $O(m)$ denote orderings of the full set $\{1,\dots,m\}$.  Orderings are in natural bijection with permutations, but we maintain distinct notation for readability;  we will treat $\R^{O(\Sigma)}$ as a vector space of formal combinations of orderings.\footnote{In fact, in many of the places where $S_m$ is used in this paper, permutations are really appearing as orderings; however, we only introduce this notation here because it helps with disambiguation.} We think of the cotangent space of $P_m$ as lying in $\rr^{O(m)}$.

We define an \emph{ordering map} $F\colon S(\Sigma) \to \R^{O(\Sigma)}$ 
carrying permutations to combinations of orderings.  
A formal definition is found in Definition~\ref{def:order-map}, but we give examples here that suggest the structure. 
We first define $F$ on permutations which are pure cycles. For instance,
$F( (12) ) = \ordab 12 - \ordab 21$ and 
$F( (123) ) =\ordabc 123 + \ordabc 321 - \ordabc 213 - \ordabc 312$.
Products of disjoint cycles are combined by riffling the orderings together with a \emph{shuffle operator} $\shuffle$, so that, for example,
$$F((12)(34))=F((12))\shuffle F((34)) = \ordab 12 \shuffle \ordab 34 - \ordab 12 \shuffle \ordab 43 - \ordab 21 \shuffle \ordab 34 + \ordab 21 \shuffle \ordab 43,$$
where we can expand
$$\ordab 12 \shuffle \ordab 34 = 
\ordabcd 1234 + \ordabcd 1324 +
\ordabcd 1342 + \ordabcd 3124 + \ordabcd 3142
+\ordabcd 3412,$$
and so on. For permutations $\pi\in S_m$, the vector $F(\pi)$ always has an equal number of terms with a positive sign and terms with a negative sign, because the same is true for images of cycles $F(\sigma_i)$.  
The general definition of the map $F$ makes use of the formalism of Lie brackets and shuffle products, and uses a lexicographic normal form on permutations so that the map is well defined.  The expressions $F(\pi)$ suggest the corresponding sets of events using indices from the $\sigma_i$.

We pull the following key statement from two works of Reutenauer and collaborators \cite[Section 6.5.1]{Reutenauer93:FreeLie} \cite[Remark 6]{MR89:LyndonShuffles}, who are working in the more general setting of free associative algebras.  
Our setting has a simplifying feature:  we work with vectors that are homogeneous of degree 1 in each variable. This allows us to make the needed statement very succinct.

\begin{lemma}[Lie shuffle basis]\label{lem:shuf-basis}
The  set $\{F(\pi) : \pi\in S_m \}$ is a basis for $\R^{O(m)}$.
\end{lemma}

We call this basis the \emph{Lie shuffle basis}.
With this notation, we are ready to compute gradients in 
$\R^{O(m)}$.

\begin{restatable}[Gradient expression]{lemma}{gradLem}\label{lem:gradient-body}
Let $\pi$ be a permutation composed of disjoint cycles $c_1, c_2, \dots, c_\ell$, and let $u_i = F(c_i)$. Then there are independent events $\mathcal{E}_i = \mathcal{E}_i^+ \cup \mathcal{E}_i^-$ for each $1 \leq i \leq \ell$ where $c_i$ is nontrivial (has length at least 2) such that the gradient of Equation~(\ref{EO}) at the uniform distribution is a scalar multiple of
$u_1 \shuffle u_2 \shuffle \dots \shuffle u_k$.
\end{restatable}

Together, these facts give us enough linearly independent cotangent vectors to get a dimension bound.

\begin{theorem}[Dimension upper bound]\label{thm:dim-Pn-upper}
Write $C(m)=\sum_{k=2}^m \frac{m!}{k\cdot (m-k)!}$ for the number of {\em pure cycles} in $S_m$, i.e., permutations with exactly one nontrivial cycle and all other points fixed.
Then for all $m$, the dimension of $P_m$ is at most $C(m)$.
\end{theorem}

\begin{proof}
Consider the set
$$B := \{F(\pi) \suchthat \pi \txt{ contains at least two nontrivial cycles}\}.$$
By Lemma~\ref{lem:gradient-body}, each element of $B$ is a scalar multiple of a gradient of Equation~(\ref{EO}) at the uniform distribution for some collection of at least 2 disjoint events. By Lemma~\ref{lemIndependenceConstraintsGeneral}, these gradients are all cotangent vectors. Since the elements of $B$ are scalar multiples of a subset of the Lie shuffle basis, these gradient vectors are all linearly independent by Lemma~\ref{lem:shuf-basis}.
Thus, so far we have found a collection of independent cotangent vectors, one for each permutation with at least two nontrivial cycles. There is one more constraint: In $P_m\subset \Delta(S_m)$, all elements are linear combinations of permutations whose coefficients sum to one.  
Thus, in the tangent space, vectors can be regarded as linear combinations of permutations with coefficients summing to zero. In other words, we have the all-ones vector as an additional vector in the cotangent space, which is independent of the others. We define this vector to be $F$ of the identity permutation.
Hence, the codimension of the tangent space of $P_m$ at the uniform
distribution is at least the number of permutations that have \emph{at least two}
nontrivial cycles, plus one for the identity permutation (which has \emph{zero} nontrivial cycles). Since this holds at the uniform distribution,  Lemma \ref{lemGenericPoints} lets us conclude that this holds at all points in $P_m$. Therefore, the dimension of $P_m$ is at most the number of permutations with \emph{exactly one} nontrivial cycle.
\end{proof}

This sequence $C(m)$  appears in OEIS as entry \oeis{A006231} \cite{EIS}.  
We note that $C(m)\sim e(m-1)!$, while $\Delta(S_m)$ itself has dimension $m!-1$, so asymptotically this dimension tends to a fraction $e/m$ of the ambient dimension.

\begin{conj}\label{conj:dim-Pn}
For all $m$, the dimension of $P_m$ is exactly $C(m)$.
\end{conj}

We now describe our computational approach to verifying that this upper bound is tight up to $m = 7$.

\begin{definition}[Draw matrix]\label{def:drawmatrix}
A word $w$ on $m$ symbols that uses each symbol at least once has some 
ordering of the $m$ symbols
as substrings; pairing the word with weights $\alpha$  in a word map $(w,\alpha)$ induces a distribution on $S_m$ in the usual way, where the permutations are realized as draws from the word $w$ using weights from $\alpha$.  
The {\em draw matrix} $M_w$ for a word $w$  length $r$  on $m$ symbols is the $r \times m!$ matrix with columns indexed by $S_m$, where entry $(i, \sigma)$ of $M$ is the number of ways to draw $\sigma$ as a substring of $w$ that use the $i\tth$ letter of $w$.
\end{definition}

For example, the draw matrix $M_w$ for the word $w = abcabcba$ is:
	$$\begin{blockarray}{ccccccc}
		& abc & acb & bac & bca & cab & cba\\
		\begin{block}{c[cccccc]}
			\underline{\mathbf{a}}bcabcba & 3 & 3 & 0 & 0 & 0 & 0\\
			a\underline{\mathbf{b}}cabcba & 2 & 0 & 1 & 3 & 0 & 0\\
			ab\underline{\mathbf{c}}abcba & 1 & 2 & 0 & 2 & 2 & 2\\
			abc\underline{\mathbf{a}}bcba & 1 & 1 & 1 & 1 & 2 & 0\\
			abca\underline{\mathbf{b}}cba & 2 & 1 & 0 & 1 & 1 & 1\\
			abcab\underline{\mathbf{c}}ba & 3 & 2 & 1 & 2 & 0 & 1\\
			abcabc\underline{\mathbf{b}}a & 0 & 3 & 0 & 0 & 1 & 2\\
			abcabcb\underline{\mathbf{a}} & 0 & 0 & 0 & 3 & 0 & 3\\
		\end{block}
	\end{blockarray}$$
The permutation $abc$ can be realized in four ways as a draw from $abcabcba$, three using the first $a$ and one using the second $a$; this is reflected in the first column of $M_w$, in the first and fourth rows.

\begin{lemma}[Matrix rank as a lower bound]\label{lemDimensionLowerBound}
For any word $w$ on $m$ symbols, the draw matrix gives a lower bound for the dimension of $P_m$:
		$$\dim(P_m) \geq \rank(M_w) - 1.$$
	\end{lemma}
		
\begin{proof}
		Define $\widetilde{P_m} \coloneqq \{cx \suchthat c \in \R_{\geq 0},\ x \in P_m\}$. Clearly, the dimension of $\widetilde{P_m}$ is at most one more than the dimension of $P_m$; it thus suffices to show that $\dim(\widetilde{P_m}) \geq \rank(M)$. For this we lower-bound the dimension of the subset $S \subseteq \widetilde{P_m}$ defined as the set of non-normalized probability distributions (measures) over permutations of the $m$ letters obtained by word maps $(w, \alpha)$ for \emph{arbitrary} $\alpha \in \R_{\geq 0}^r$ (not necessarily summing to one). Observe that $S$ is the image of a map $f$ from $\R_{\geq 0}^r$ to $\R_{\geq 0}^{O(m)}$. We claim that the matrix $M$ is the Jacobian of $f$ evaluated at $\alpha = (1, 1, \dots, 1)$. For any permutation $\sigma$, the map $f$ defines
		\[
        x_\sigma = \sum_{\substack{i_1, i_2, \dots, i_m \in [r] \\ w_{i_1}w_{i_2} \dots w_{i_m} = \sigma}}\alpha_{i_1} \alpha_{i_2} \dots \alpha_{i_m}.
        \]
		Taking the derivative of $x_\sigma$ with respect to $\alpha_i$, we see that the number of nonzero terms is precisely the number of subsequences of $w$ yielding $\sigma$ that use $w_i$, and furthermore, evaluating at $\alpha = (1, 1, \dots, 1)$, these terms are all 1. Thus, $M$ is the Jacobian of $f$, so its rank is the dimension of $S$ at $\alpha = (1, 1, \dots, 1)$. By Lemma~\ref{lemGenericPoints} this gives us the desired global lower bound on $\dim(\widetilde{P_m})$, and thus on $\dim(P_m)$ as well.
	\end{proof}

\begin{proposition}[Dimensions for small $m$]\label{prop:dim-Pn}
$\dim(P_m)=C(m)$ holds for $m \le 7$, giving dimensions
		$\dim P_m = 0,1,5,20,84,409,2365$ for $m=1,2,3,4,5,6,7$.
\end{proposition}

For the example $w=abcabcba$ above, the reader can verify that the rank of the draw matrix is 6, implying the tight bound of $\dim(P_3) \geq 5$. Our approach for the rest of the verifications in Proposition~\ref{prop:dim-Pn} is simply to try sufficiently long words (roughly 10\% longer than the anticipated rank of $P_m$) and compute ranks. Even for very long words, it is possible to very efficiently compute each entry of $M$ using dynamic programming. However, the matrix $M$ has thousands of rows and columns starting at $m = 7$; this is the largest size we can handle with current computational resources.  Replication code can be found in \cite{GitHub}.

%%%%%%%%%
%%%%%%%%%

\section{Final remarks}\label{secConclusion}

We have explored some of the quantitative properties of
ordinary MST, its successive extensions to shifted intervals, and then the full generality of product measures.

For ordinary MST, we have presented procedures for calculating the probability of any labeled tree in an arbitrary ambient graph $G$,  and we have developed rotation moves that let us distinguish $\MSTZ$ from UST on random graphs and that let us show that paths and stars have extremal $\MSTZ$ probability in complete graphs.   For shifted-interval MST, we have defined a combinatorially natural object called a shiftahedron as a parameter space for studying the induced distribution on permutations, and from there the distribution on trees.   We included examples of distributions that are and are not realizable by shifts.
Finally, for fully general product measures, we have made significant progress towards describing what distributions on permutations they induce, which enriches the well-known topic of {\em intransitive dice}.  We established that every product measure has a corresponding word map of bounded length (associated to a discrete product measure with finite support) that induces the same probability weight on all permutations.  This lets us tie the study of word maps to the theory of integration, which we can leverage to find particular word maps with nice properties.  Finally, we have motivated and stated a conjecture that the dimension of the locus in $\Delta(S_m)$ hit by product measures is exactly the number of pure cycles in $S_m$.  This is verified for $1\le m\le 7$, and the upper bound is established for all $m$.  

The results here give several ways of quantifying how ordinary MST is different from UST---for instance, the degree distribution skews to higher degrees because high-degree nodes create short broken cycles (cf. Theorem~\ref{thm:rd-global}).
But more than that, we build theory for the broad generalization to product measures.  While the full space of distributions on $S_m$ has dimension $m!-1$, our results show that product measures fill out a locus whose dimension tends to $e\cdot (m-1)!$; indeed, just the part that is hit by shifted intervals has dimension at most $m-1$ and is already flexible enough for many applications (as in \S\ref{subsec:ReCom-motivation}).

%%%%%%%%%
%%%%%%%%%

\bibliographystyle{plain}
\bibliography{bibliography}

\clearpage

\appendix

%%%
\section{Probability and runtime}
\label{sec:example-computations}

We can make a few additional observations about computing the probability of spanning trees using internal and external formulas. For complete graphs we can write
$\#\partial(H)$ more simply, noting that if two connected components have $a$ and $b$ vertices, respectively, then there are $ab$ missing edges between them.  
Since any missing edge in $H$ is between some two of its components, we get 
\[
\#\partial(H) = e_2(\text{comp-size}(H))
\]
where $\text{comp-size}(H)$ is the multi-set of sizes of the
connected components of~$H$ and, for a multi-set $S$ of integers, $e_2(S)$ is
the second elementary symmetric function of~$S$:
\[
e_2(x_1,\dots,x_k) = \sum_{1 \le i < j \le k} x_i x_j.
\]
Furthermore, all forests $F \subset K_n$ that are isomorphic to each other as graphs have the same probability by symmetry, so we only need to compute one number for each shape of forest.

\begin{figure}[htb!]
	\centering
	\begin{tikzpicture}[x=3cm,y=2cm,vertex/.style={circle,fill}]
		\node[rectangle,rounded corners,draw] (star4) at (0,0) {
			\tikz[x=3mm,y=3mm,inner sep=1pt] {
				\node[vertex] (a) at (0,0) {};
				\node[vertex] (b) at (-1,0) {};
				\node[vertex] (c) at (0,-1) {};
				\node[vertex] (d) at (1,0) {};
				\node[vertex] (e) at (0,1) {};
				\draw (b)--(a)--(c); \draw (e)--(a)--(d);
			}
		};
		\node[rectangle,rounded corners,draw] (fork) at (1,0) {
			\tikz[x=3mm,y=3mm,inner sep=1pt] {
				\node[vertex] (a) at (0.5,0.87) {};
				\node[vertex] (b) at (0.5,-0.87) {};
				\node[vertex] (c) at (1,0) {};
				\node[vertex] (d) at (2,0) {};
				\node[vertex] (e) at (3,0) {};
				\draw (a)--(c)--(d)--(e);
				\draw (b)--(c);
			}
		};
		\node[rectangle,rounded corners,draw] (path5) at (2,0) {
			\tikz[x=3mm,y=3mm,inner sep=1pt] {
				\node[vertex] (a) at (0,0) {};
				\node[vertex] (b) at (1,0) {};
				\node[vertex] (c) at (2,0) {};
				\node[vertex] (d) at (3,0) {};
				\node[vertex] (e) at (4,0) {};
				\draw (a)--(b)--(c)--(d)--(e);
			}
		};
		
		\node[rectangle,rounded corners,draw] (star3) at (0,-1) {
			\tikz[x=3mm,y=3mm,inner sep=1pt] {
				\node[vertex] (a) at (0,0) {};
				\node[vertex] (b) at (-1,0) {};
				\node[vertex] (c) at (0,-1) {};
				\node[vertex] (d) at (1,0) {};
				\node[vertex] at (1,-1) {};
				\draw (b)--(a)--(c); \draw (a)--(d);
			}
		};
		\node[rectangle,rounded corners,draw] (path4) at (1,-1) {
			\tikz[x=3mm,y=3mm,inner sep=1pt] {
				\node[vertex] (a) at (0,1) {};
				\node[vertex] (b) at (1,1) {};
				\node[vertex] (c) at (2,1) {};
				\node[vertex] (d) at (3,1) {};
				\node[vertex] at (1.5,0) {};
				\draw (a)--(b)--(c)--(d);
			}
		};
		\node[rectangle,rounded corners,draw] (path32) at (2,-1) {
			\tikz[x=3mm,y=3mm,inner sep=1pt] {
				\node[vertex] (a) at (0,1) {};
				\node[vertex] (b) at (1,1) {};
				\node[vertex] (c) at (2,1) {};
				\node[vertex] (d) at (0.5,0) {};
				\node[vertex] (e) at (1.5,0) {};
				\draw (a)--(b)--(c); \draw (d)--(e);
			}
		};
		
		\node[rectangle,rounded corners,draw] (path3) at (0.5,-2) {
			\tikz[x=3mm,y=3mm,inner sep=1pt] {
				\node[vertex] (a) at (0,1) {};
				\node[vertex] (b) at (1,1) {};
				\node[vertex] (c) at (2,1) {};
				\node[vertex] at (0.5,0) {};
				\node[vertex] at (1.5,0) {};
				\draw (a)--(b)--(c);
			}
		};
		
		\node[rectangle,rounded corners,draw] (path22) at (1.5,-2) {
			\tikz[x=3mm,y=3mm,inner sep=1pt] {
				\node[vertex] (a) at (0,1) {};
				\node[vertex] (b) at (1,1) {};
				\node[vertex] (c) at (2,1) {};
				\node[vertex] (d) at (3,1) {};
				\node[vertex] at (1.5,0) {};
				\draw (a)--(b); \draw (c)--(d);
			}
		};
		
		\node[rectangle,rounded corners,draw] (path2) at (1,-3) {
			\tikz[x=3mm,y=3mm,inner sep=1pt] {
				\node[vertex] (a) at (0.5,1) {};
				\node[vertex] (b) at (1.5,1) {};
				\node[vertex] at (0,0) {};
				\node[vertex] at (1,0) {};
				\node[vertex] at (2,0) {};
				\draw (a)--(b);
			}
		};
		
		\node[rectangle,rounded corners,draw] (triv) at (1,-4) {
			\tikz[x=3mm,y=3mm,inner sep=1pt] {
				\node[vertex] at (0,0) {};
				\node[vertex] at (1,0) {};
				\node[vertex] at (2,0) {};
				\node[vertex] at (3,0) {};
				\node[vertex] at (4,0) {};
			}
		};
		\draw (path2) -- (triv);
		\draw (path3) -- node[below left=-2pt,cdlabel] {2} (path2);
		\draw (path22) -- node[below right=-2pt,cdlabel]{2} (path2);
		\draw (star3) -- node[below left=-2pt,cdlabel]{3} (path3);
		\draw (path4) -- node[above left=-2pt,cdlabel]{2} (path3);
		\draw (path4) -- (path22);
		\draw (path32) -- (path3);
		\draw (path32) -- node[below right=-2pt,cdlabel]{2} (path22);
		\draw (star4) -- node[left,cdlabel]{4} (star3);
		\draw (fork) -- (star3);
		\draw (fork) -- node[left,cdlabel]{2} (path4);
		\draw (fork) -- (path32);
		\draw (path5) -- node[pos=0.3,above left=-2pt,cdlabel]{2} (path4);
		\draw (path5) -- node[right,cdlabel]{2} (path32);
		
		\node[right] at (triv.east) {
			$\begin{smallmatrix*}[l]
				\scriptstyle\prob = 1\\[1pt]
				%        \scriptstyle\partial=10\\[1pt]
				\scriptstyle\wt{\prob}=\frac{1}{10}
			\end{smallmatrix*}$};
		\node[right] at (path2.east) {
			$\begin{smallmatrix*}[l]
				\scriptstyle\prob = \frac{1}{10}\\[1pt]
				%        \scriptstyle\partial=9\\[1pt]
				\scriptstyle\wt{\prob}=\frac{1}{90}
			\end{smallmatrix*}$};
		\node[right] at (path3.east) {
			$\begin{smallmatrix*}[l]
				\scriptstyle\prob = \frac{1}{45}\\[1pt]
				%        \scriptstyle\partial=7\\[1pt]
				\scriptstyle\wt{\prob}=\frac{1}{315}
			\end{smallmatrix*}$};
		\node[right] at (path22.east) {
			$\begin{smallmatrix*}[l]
				\scriptstyle\prob = \frac{1}{45}\\[1pt]
				%        \scriptstyle\partial=8\\[1pt]
				\scriptstyle\wt{\prob}=\frac{1}{360}
			\end{smallmatrix*}$};
		\node[right] at (star3.east) {
			$\begin{smallmatrix*}[l]
				\scriptstyle\prob = \frac{1}{105}\\[1pt]
				%        \scriptstyle\partial=4\\[1pt]
				\scriptstyle\wt{\prob}=\frac{1}{420}
			\end{smallmatrix*}$};
		\node[right] at (path4.east) {
			$\begin{smallmatrix*}[l]
				\scriptstyle\prob = \frac{23}{2520}\\[1pt]
				%        \scriptstyle\partial=4\\[1pt]
				\scriptstyle\wt{\prob}=\frac{23}{10080}
			\end{smallmatrix*}$};
		\node[right] at (path32.east) {
			$\begin{smallmatrix*}[l]
				\scriptstyle\prob = \frac{11}{1260}\\[1pt]
				%        \scriptstyle\partial=6\\[1pt]
				\scriptstyle\wt{\prob}=\frac{11}{7560}
			\end{smallmatrix*}$};
		\node[right] at (star4.east) {
			$\scriptstyle\prob = \frac{1}{105}$};
		\node[right] at (fork.east) {
			$\scriptstyle\prob = \frac{127}{15120}$};
		\node[right] at (path5.east) {
			$\scriptstyle\prob = \frac{113}{15120}$};
	\end{tikzpicture}
	\caption{Probabilities for all forest shapes in~$K_5$. Edges show
		inclusions $F' \subset F$ obtained by deleting one edge, with
		multiplicity according to the number of ways to delete the edge.}
	\label{fig:prob-K5-forests}
\end{figure}

We can improve the running time of Kruskal's algorithm by combining some ingredients already assembled. Specifically, for
a graph $G$ and a forest $F \subset G$, consider the probability
$\bP(F)$ that $F$ appears at some point in Kruskal's algorithm on~$G$.
We have an inductive formula for these probabilities.
For $F\subset G$ a forest that is not a
tree, define
\[
\wt{\bP}(F) \coloneqq \frac{\bP(F)}{\#\partial(F)}.
\]

\begin{proposition}\label{prop:forest-induction-gen}
	For a graph $G$ and forest $F \subset G$,
	\[
	\bP(F) = \sum_{e \in \mathrm{Edges}(F)}
	\wt{\bP}(F \setminus e).
	\]
\end{proposition}

This follows from the internal formula by breaking up
the sum over $S_{n-1}$ according to the last-added edge,
$e_{\pi(n-1)}$. 
This suggests an efficient method for computing
$\bP(T)$ for a tree $T$: compute $\wt{\bP}(F)$ for all $F \subsetneq T$ in order from
the fewest number of edges to the most, using
Proposition~\ref{prop:forest-induction-gen}. This requires time $O(2^n)$,
as opposed to time $O(n!)$ from directly applying the general internal formula.

If $G$ is the complete graph $K_n$, then we can further simplify by
noting that every labeling of a forest has the same probability. We
can then apply Proposition~\ref{prop:forest-induction-gen} on unlabeled
forests as in  Figure~\ref{fig:prob-K5-forests}, at the cost of checking for graph isomorphisms.

\FloatBarrier
%%%%

\section{Extended example for path rotation algorithm}
\label{sec:path-rotation-example}

\begin{figure}[bht!]
	\centering
	\includegraphics[width=6in]{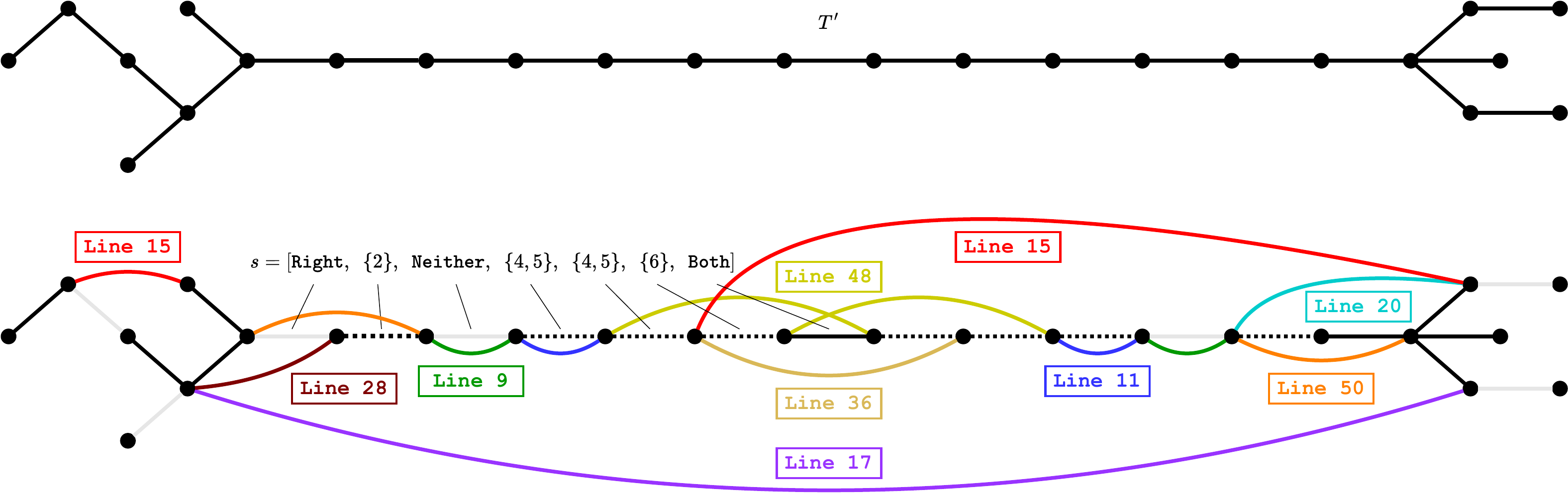}
	\caption{Above, a tree $T'$ containing a path $P'$ of length $\ell=14$, giving $r=7$ and six sets of paired edges.  The state vector $s$ has seven entries.  
		Below, a possible state $s$ midway through the inner for-loop of Algorithm~\ref{algPathRotation}. Solid black edges have  been confirmed to be added to the tree, dashed black edges have been added with probability $\frac12$, and gray edges have been confirmed to be excluded. The colored, curved edges are possible edges that could be processed on the next iteration. The effects of processing these edges are shown in Table~\ref{tabPathRotationExample}.}
	\label{figPathRotationCases}
\end{figure}

{\tiny
	\begin{table}[bht!]
		\centering
		\begin{tabular}{l | l | c | c}
			After processing\dots & New state $s$, with updates colored & $q$ & $q'$\\
			\hline
			(initially) & $[\texttt{Right}, \{2\}, \texttt{Neither}, \{4,5\}, \{4,5\}, \{6\}, \texttt{Both}]$ & $1/4$ & $1/4$\\
			{\color{myGreen}the pair of edges triggering Line~\ref{linFirstTime}} & $[\texttt{Right}, \{2\}, {\color{myGreen}\{3\}}, \{4,5\}, \{4,5\}, \{6\}, \texttt{Both}]$ & $1/4$ & $1/4$\\
			{\color{myBlue}the pair of edges triggering Line~\ref{linSecondTime}} & $[\texttt{Right}, \{2\}, \texttt{Neither}, {\color{myBlue}\texttt{Both}}, {\color{myBlue}\{5\}}, \{6\}, \texttt{Both}]$ & $1/4$ & $1/4$\\
			{\color{myRed}{\color{myRed}either} edge {\color{myRed}triggering} Line~{\color{myRed}\ref{linDefiniteFail}}} & (does not matter) & {\color{myRed}0} & {\color{myRed}0}\\
			{\color{myPurple}the edge triggering Line~\ref{linCrossWholePath}} & (no change) & $1/4$ & {\color{myPurple}0}\\
			{\color{myTurquoise}the edge triggering Line~\ref{linSideToPath}} & $[\texttt{Right}, {\color{myTurquoise}\texttt{Right}}, \texttt{Neither}, \{4,5\}, \{4,5\}, \{6\}, \texttt{Both}]$ & {\color{myTurquoise}$1/8$} & {\color{myTurquoise}$1/8$}\\
			{\color{myMaroon}the edge triggering Line~\ref{linSideInductiveCollapse}} & (no change) & $1/4$ & {\color{myMaroon}0}\\
			{\color{myBrown}the edge triggering Line~\ref{linNotBothThereMiddle}} & (does not matter) & {\color{myBrown}0} & {\color{myBrown}0}\\
			{\color{myGold}the pair of edges triggering Line~\ref{linEntanglement}} & $[\texttt{Right}, \{2\}, \texttt{Neither}, {\color{myGold}\{4,5,6\}, \{4,5,6\}, \{4,5,6\}}, \texttt{Both}]$ & {\color{myGold}$1/16$} & {\color{myGold}$1/16$}\\
			{\color{myOrange}the pair of edges triggering Line~\ref{linMiddleInductiveCollapse}} & $[\texttt{Right}, {\color{myOrange}\texttt{Right}}, \texttt{Neither}, \{4,5\}, \{4,5\}, \{6\}, \texttt{Both}]$ & {\color{myOrange}$1/16$} & {\color{myOrange}$1/16$}\\
			{\color{myRed} the pair of Line \ref{linFirstTime} edges, followed by}&&\\
			\hfill{\color{myRed}the Line \ref{linDefiniteFail} edge on the top-right,} & (does not matter) & {\color{myRed}$0$} & {\color{myRed}$0$}\\
			\hfill{\color{myRed}which would trigger Line \ref{linNotBothThereSide}}&&
		\end{tabular}
		\caption{States and conditional probabilities after processing each of the colored, curved edges for the example of the graph depicted in Figure~\ref{figPathRotationCases}. Note that, for the pairs of edges that trigger the conditions in Lines~\ref{linFirstTime}, \ref{linEntanglement}, and \ref{linMiddleInductiveCollapse}, this would be the first time we have seen the given pair of edges; for the pair that triggers the condition in Line~\ref{linSecondTime}, it would be the second time.}
		\label{tabPathRotationExample}
	\end{table}
}

\FloatBarrier
%%%%%
\section{Hitting \texorpdfstring{$\MST=\UST$}{MST=UST} with shifted intervals}\label{app:theta}

\subsection{Shifts and theta graphs}\label{subShiftsThetaGraph}

In the introduction, it was noted that when $G$ is the square with a diagonal, then there exists $\SS$ such that $M_\SS=\UST$.  

In this section we prove that shifts suffice to recover the uniform distribution on a larger class of graphs called theta graphs.

\begin{figure}[htb!]\centering
	\def\svgwidth{0.85\columnwidth}
	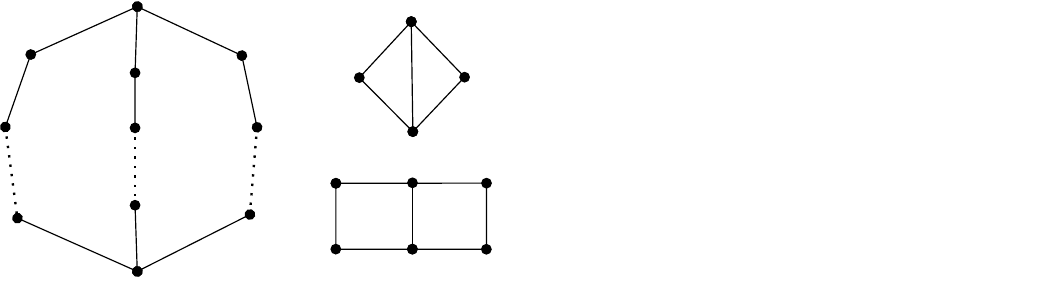
	\caption{Left: The general theta graph $\theta(r,s,t)$ and two specific examples, $\theta(2,1,2)$ (the square with a diagonal) and $\theta(3,1,3)$. Right: Examples of trees of type $R$, $S$, and $T$  for $\theta(3,2,2)$.}\label{fig:thetagraphs_and_treetypes}
\end{figure}

\begin{definition} A {\em theta graph} $\theta(r,s,t)$
	is formed by connecting two base vertices with disjoint paths of length $r$, $s$, and $t$, respectively, so that there are $m=r+s+t$ edges overall. We call these the $R$-path, the $S$-path, and the $T$-path, and refer to their edges as $R$-edges, $S$-edges, and $T$-edges; see left-hand side of Figure~\ref{fig:thetagraphs_and_treetypes}.

	We can  define shifted interval MST on theta graphs by using the same shifts on the edges of each type:
	$$\Sh(r,s,t)=\left\{
	(\alpha,\dots,\alpha,\beta,\dots,\beta,\gamma,\dots,\gamma): (\alpha,\beta,\gamma)\in \Sh(3)  \right\}.$$
\end{definition}

We note that every spanning tree of $\theta(r,s,t)$ must be missing exactly two edges:  one edge each from two of the three paths in $G$.  A tree of $R$ type contains all of the $R$ path but is missing one $S$-edge and one $T$-edge; likewise for trees of $S$ and $T$ type; see right-hand side of Figure~\ref{fig:thetagraphs_and_treetypes}.  Given $\SS\in\Sh(r,s,t)$, let  $p_R+p_S+p_T=1$ be the probabilities that $M_\SS$ is of type $R,S,T$, respectively. 
Any two trees of the same type occur with equal probability. Hence, every individual $R$-tree occurs with probability $\frac{p_R}{st}$; similarly for $S$-trees and $T$-trees.  
Thus the distribution is uniform if and only if $rP_R=sP_S=tP_T$, which occurs at exactly one $(P_R,P_S,P_T)$, given by 
$P_R=u_R\coloneqq st/(st+rt+rs)$, with similar expressions for $P_S$ and $P_T$.

%NEW COMBINED FIGURE

\begin{theorem}[UST on theta graphs]\label{prop:theta}
	On theta graphs, the uniform distribution is achievable by shifted intervals:
	for any $(r,s,t)$, there exists $\SS\in \Sh(r,s,t)$ such that $M_\SS=\UST$ on $\theta(r,s,t)$.
\end{theorem}

\begin{proof}
	Consider $\Delta(S_3)$, the triangle (2-simplex) parameterizing $P_R+P_S+P_T=1$.  There is a map $\Sh(3)\to \Delta(S_3)$ given by the construction above, where the weight interval for $R$ edges is $[\alpha,\alpha+1]$ and likewise $S$ edges and $T$ edges have weight drawn from the $\beta,\gamma$ intervals.  As we move around the boundary of $\Sh(3)$ (shown in Figure~\ref{fig:shiftahedra}), we  traverse the boundary of $\Delta(S_3)$ one time.  So the map is continuous and is degree 1 on the boundary (as a map $S^1\to S^1$), which implies that it is surjective.
\end{proof}

In fact, we can say more.  Consider the external formula from Theorem~\ref{thm:rd-global}.  Under $\MSTZ$, the probability of obtaining a tree of $R$-type is $m_R\coloneqq \frac{1}{r+s}\cdot\frac{1}{r+s+t}+ \frac{1}{r+t}\cdot\frac{1}{r+s+t}$, and likewise for $S$ and $T$.  
Under the wlog assumption $r\ge s\ge t$, we have 
$$u_R-m_R=\frac{rst(r^2-st)}{(r+s)(r+t)(rs+rt+st)},$$ which is strictly positive unless $r=s=t$.  

\begin{corollary}\label{cor:theta} 
	$\MSTZ=\UST$ if and only if $(r,s,t)=(r,r,r)$.
\end{corollary}

With a bit more work one can show that the map $(\alpha,\beta,\gamma)\mapsto (P_R,P_S,P_T)$ is non-singular on the interior of the region defined by 
$0=\alpha \le \beta\le \gamma\le 1$, which tells us that there is in fact a unique way to hit $(u_R,u_S,u_T)$ with this class of shifts.  

Finally, we note that the work in this section extends readily to {\it generalized theta graphs} $\theta(k;r_1,\dots,r_k)$, which can have any number $k\ge 3$ of paths connecting a pair of base vertices; only the notation gets worse. In particular, the following hold: for a generalized theta graph $\theta(r_1,\dots,r_k)$, $\UST=\MSTZ$ if and only if  $r_1=\dots=r_k$.

%%%%%
\subsection{Snowmen and \texorpdfstring{$\theta$}{θ}-surgery graphs}

For this section, we will call a theta graph $\theta(r,s,t)$ with $(r,s,t)\neq (r,r,r)$ a {\it snowman}; any graph that does not contain a snowman as a subgraph is called {\it snowman-free}.

%\begin{remark} A graph $G$ is called a {\it snowman} if it consists of two cycles of unequal length that share at least one edge. Figure \ref{fig:snowman} shows a snowman that consists of cycles of length $6$ and $8$ that share one edge (or equivalently: cycles of length $12$ and $8$ that share $5$ edges). Equivalently, in the language of Section~\ref{subShiftsThetaGraph}), a snowman is a theta graph $\theta(r,s,t)$ for which $(r,s,t)\neq (r,r,r)$. 

%\begin{figure}[htp]
%   \centering
% \includegraphics[width=7cm]{snowman_combo}
%  \caption{The snowman graph $\theta(7,1,5)$.}
%   \label{fig:snowman}
%\end{figure}

\begin{figure}[htb!] \centering
	\def\svgwidth{0.90\columnwidth}
	%% Creator: Inkscape 1.0.2 (e86c8708, 2021-01-15), www.inkscape.org
%% Accompanies image file 'snowman_and_surgery.pdf' (pdf, eps, ps)
\begingroup%
  \makeatletter%
  \providecommand\color[2][]{%
    \errmessage{(Inkscape) Color is used for the text in Inkscape, but the package 'color.sty' is not loaded}%
    \renewcommand\color[2][]{}%
  }%
  \providecommand\transparent[1]{%
    \errmessage{(Inkscape) Transparency is used (non-zero) for the text in Inkscape, but the package 'transparent.sty' is not loaded}%
    \renewcommand\transparent[1]{}%
  }%
  \providecommand\rotatebox[2]{#2}%
  \newcommand*\fsize{\dimexpr\f@size pt\relax}%
  \newcommand*\lineheight[1]{\fontsize{\fsize}{#1\fsize}\selectfont}%
  \ifx\svgwidth\undefined%
    \setlength{\unitlength}{514.27708616bp}%
    \ifx\svgscale\undefined%
      \relax%
    \else%
      \setlength{\unitlength}{\unitlength * \real{\svgscale}}%
    \fi%
  \else%
    \setlength{\unitlength}{\svgwidth}%
  \fi%
  \global\let\svgwidth\undefined%
  \global\let\svgscale\undefined%
  \makeatother%
  \begin{picture}(1,0.3150798)%
    \lineheight{1}%
    \setlength\tabcolsep{0pt}%
    \put(0,0){\includegraphics[width=\unitlength,page=1]{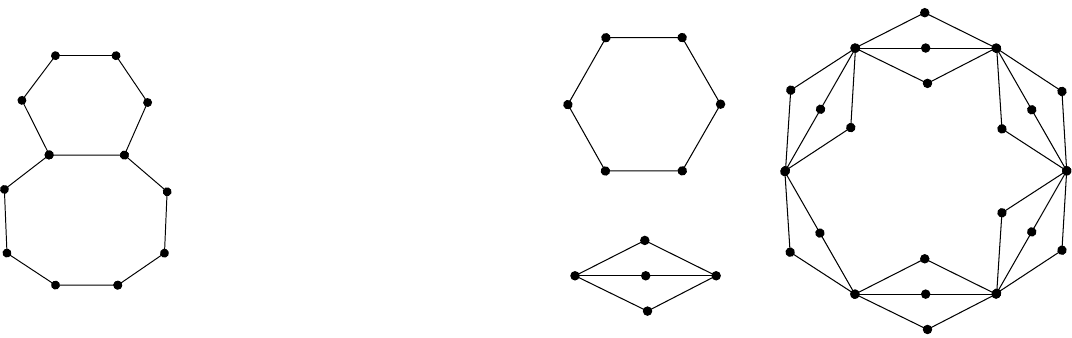}}%
    \put(0.5754351,0.21099427){\color[rgb]{0,0,0}\makebox(0,0)[lt]{\lineheight{1.25}\smash{\begin{tabular}[t]{l}$G_0$\end{tabular}}}}%
    \put(0.84862452,0.15079994){\color[rgb]{0,0,0}\makebox(0,0)[lt]{\lineheight{1.25}\smash{\begin{tabular}[t]{l}$G$\end{tabular}}}}%
    \put(0.84862452,0.15079994){\color[rgb]{0,0,0}\makebox(0,0)[lt]{\lineheight{1.25}\smash{\begin{tabular}[t]{l}$G$\end{tabular}}}}%
    \put(0.636787,0.08597538){\color[rgb]{0,0,0}\makebox(0,0)[lt]{\lineheight{1.25}\smash{\begin{tabular}[t]{l}$\theta$\end{tabular}}}}%
    \put(0,0){\includegraphics[width=\unitlength,page=2]{snowman_and_surgery.pdf}}%
  \end{picture}%
\endgroup%

	\caption{Left: The snowman graph $\theta(7,1,5)$. Right: the $\theta$-surgery graph $G$ built from the cycle $G_0$ of length $6$ and the $\theta$-graph $\theta=\theta(2,2,2)$ (that is, $r=2$, $k=3$).}\label{fig:snowman_and_surgery}
\end{figure}

%Corollary \ref{cor:theta} implies that, on a snowman $G$, MST cannot equal UST. They are guaranteed to be equal distributions on snowman-free graphs.
%\end{remark}

\begin{proposition}[Snowman-free graphs]\label{prop:snowmanfree}
	For a snowman-free graph $G$, $\MSTZ=\UST$.\end{proposition}

\begin{figure}[htb!] \centering
	\def\svgwidth{0.75\columnwidth}
	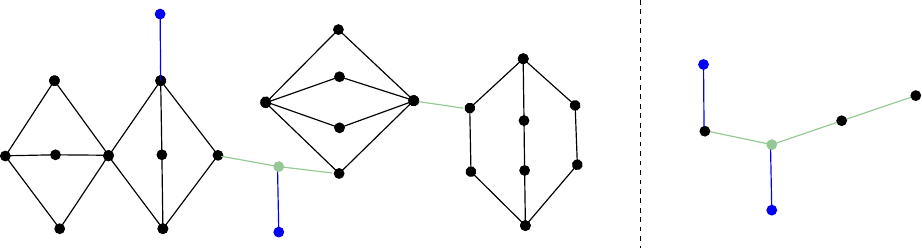
	\caption{Condition {\it no new loops arise}. The left-hand side shows the snowman-free graph $G$ consisting of four theta graphs (black), connected by paths (green), and two additional leaves (blue). The right-hand side shows the tree obtained from $G$ by collapsing the theta graphs into points. 
	}\label{fig:snowmanfree}
\end{figure}

\begin{proof}
	First, observe that every snowman-free graph $G$ is either a tree or a finite collection of generalized theta graphs $\theta(r,r,r,...,r)$  (for various $r$) that are connected by paths of edges so that no new loops arise; possibly with additional leaves. The condition {\it no new loops arise}, more formally, means that if we collapsed each theta graph in a snowman-free graph, the resulting graph would be a tree; see Figure \ref{fig:snowmanfree}.  
	For a snowman-free graph $G$, every spanning tree must contain all edges of the connecting paths between the theta graphs as well as all leaves. So the fact that $\MSTZ=\UST$ on a theta graph $\theta(r,r,r)$ (Corollary \ref{cor:theta}) implies that $\MSTZ=\UST$ on $G$. \end{proof} 

However, the converse of Proposition \ref{prop:snowmanfree} is false. To see this, consider the following construction:
Let $G_0$ be a graph for which $\MSTZ=\UST$ and let $r,k\in N$. Define $G$ to be the graph obtained from $G_0$ by replacing each edge in $G_0$ by $k$ paths of $r$ edges. In other words, we replace every edge of $G_0$ by a $\theta(r,...,r)$, with the two poles of the theta graph placed at the endpoints of the replaced edge. We call such a graph $G$ a {\it $\theta$-surgery graph}. 

\begin{proposition}\label{prop:thetasurgery}
	For a $\theta$-surgery graph $G$, $\MSTZ=\UST$.
\end{proposition}

Figure \ref{fig:snowman_and_surgery} depicts an example of a $\theta$-surgery graph. Notice that this graph fails to be snowman-free (for instance, it contains a $\theta(2,10,10)$ subgraph) but satisfies $\MSTZ=\UST$ by Proposition~\ref{prop:thetasurgery}.
In an exactly similar way, we can define a {\em shift surgery graph} $G$ in which the base graph $G_0$ and the surgery graph $G_1$ have the property that shifts can recover UST.  This property then passes to the the compound graph $G$.

%%%%%%%%%%%%%%
%%%%%%%%%%%%%%

\section{Details on quadrature constructions}
\label{app:quadrature}

To prove Theorem~\ref{thm:quad-words}, we first observe that a
quadrature scheme of degree~$k$ gives a way to make one measure
discrete out of several copies of the uniform measure.

\begin{proposition}\label{prop:quad-uniform}
	If $Y = \sum_{i=1}^r \alpha_i \delta_{x_i}$ is a degree-$k$ quadrature
	scheme and $\mathrm{Leb}$ is the uniform measure on $[0,1]$, then the
	measures
	\[
	X_i \coloneqq
	\begin{cases}
		\mathrm{Leb} & 1 \le i \le k\\
		Y & i = k+1
	\end{cases}
	\]
	is a product measure giving the uniform distribution on $S_{k+1}$.
\end{proposition}

\begin{proof}
	For
	$0 \le i \le k$ and $0 \le x \le 1$, let $P_j(x)$ be the probability
	that, in $k$ draws from~$\mathrm{Leb}$, exactly $j$ of them are less
	than~$x$. This makes $P_j(x)$ a polynomial in $x$ of degree~$k$. The
	average value of $P_j$ is $1/(k+1)$, because it is equal to the probability
	that, in $k+1$ draws from~$\mathrm{Leb}$, the last draw is in rank $j$. Thus,
	by the quadrature property, the probability that $X_{k+1}$ appears
	as the $j$th letter in a draw from the $(X_i)_{i=1}^{k+1}$ is
	\[
	\sum_i \alpha_i P_j(x_i) = \int_0^1 P_j(x)\,dx = \frac{1}{k+1},
	\]
	as desired.
	By symmetry, the remaining letters occur in each possible order
	equally likely.
\end{proof}

\begin{remark}
	Since the polynomials $P_j(x)$ span the space of degree~$k$
	polynomials, the converse of Proposition~\ref{prop:quad-uniform} is
	also true (after suitable normalization). 
\end{remark}

To go further, we formulate a criterion to
replace pieces of measures without disturbing the overall distribution
on~$S_m$.

\begin{proposition}\label{prop:measure-replace}
	Suppose we are given a non-colliding product measure
	$X = (X_1,\dots, X_m)$ and an open interval
	$I \subset \mathbb{R}$. Suppose that $X_1,\dots,X_k$ have positive mass on
	$I$ and the remainder $X_{k+1}, \dots, X_m$ do not. For $j =
	1,\dots,k$, decompose $X_j$ as
	\[
	X_j = Y_j + R_j
	\]
	where $Y_j(I) = X_j(I) > 0$, so that $R_j$ is the remainder.
	Let $Y_j'$ for $j=1,\dots,k$ be any other measures with support
	contained in $I$ so that
	\begin{itemize}
		\item $Y_j(I) = Y_j'(I)$;
		\item the $(Y_j')_{j=1}^k$ are non-colliding on~$I$; and
		\item $Y = (Y_j)_{j=1}^k$ and $Y' = (Y_j')_{j=1}^k$ give the same distribution on $S_k$.
	\end{itemize}
	Then the product measure $X' = (X_1',\dots,X_m')$
	defined by
	\[
	X_j' \coloneqq
	\begin{cases}
		Y_j' + R_j & j \le k\\
		X_j & k < j \le m
	\end{cases}
	\]
	gives the same distribution on~$S_m$ as $(X_j)_{j=1}^m$.
\end{proposition}
\begin{proof}
	Consider a draw from $X$ or $X'$. Some subset $T
	\subset \{1,\dots,k\}$ of
	the $X_j/X_j'$ variables comes from $Y_j/Y_j'$ rather than $R_j$.
	Since $Y_j(I) = Y_j'(I)$, the probability of each subset $T$
	occurring is equal. If we condition on a particular $T$ occurring,
	since $Y$ and $Y'$ induce the same measure on $S_k$, they also
	induce the same measure on orderings of $T$. Putting these together
	we get the result.
\end{proof}

\begin{proof}[Proof of Theorem~\ref{thm:quad-words}]
	Start with $k+1$ copies of Lebesgue measure $\mathrm{Leb}$. By
	Proposition~\ref{prop:quad-uniform}, we can replace the last copy
	with the discrete measure $\sum_{i=1}^r \alpha_i \delta_{x_i}$. Set
	also $x_0 = 0$ and $x_{r+1} = 1$. For each non-empty interval among
	the $I_i = (x_i,x_{i+1})$, the word $w^{x_{i+1}-x_{i}}$ gives a
	discrete measure with the same
	total mass in each variable as $\mathrm{Leb}(I_i)$;
	rescale the domain of $w^{x_{i+1}-x_i}$ so that it lies entirely within $I_i$.
	Since $w$ was assumed to be uniform on $k$ letters, we thus form
	measures $Y'_i$ as in the hypotheses of
	Proposition~\ref{prop:measure-replace}. If we do this for each $0
	\le i \le r$, we get a discrete measure corresponding to the word~$U$
	in the statement.
\end{proof}

%%%
\begin{proof}[Proof of Proposition \ref{prop:quad-length}]
	Applying this inductive construction, we use the quadrature schemes discussed above.
	\begin{itemize}
		\item \emph{Gauss-Radau} quadrature of degree $2r-2$  has $r$ nodes and weights, and $r$ interval widths,  giving a $(2r-2,1)$-uniform word; and
		\item \emph{Gauss-Lobatto} quadrature of degree $2r-3$ has $r$ nodes and weights and $r-1$ interval widths, giving a $(2r-3,1)$-uniform word.
	\end{itemize}
	This lets us get an appropriate scheme for each parity.  
	Defining $L(m)$ recursively by $L(1) = 1$ and
	\begin{align*}
		L(m) &=
		\begin{cases}
			\frac{m-1}{2}\cdot L(m-1)+\frac{m-1}{2}, & m\text{ odd}\\
			\frac{m-2}{2} \cdot L(m-1)+ \frac m2,  & m\text{ even},
		\end{cases}
	\end{align*}
	we have built uniform words of length $L(m)$ on $m$ symbols for every $m$.
\end{proof}

%%%%%%%%%%%%%%
%%%%%%%%%%%%%%%
\section{Details for dimension bounds}
\label{app:dimension}

We begin with an explicit proof of the even/odd identities discussed in  \S\ref{subsec:Dimension}, which are constraints on $P_m$.

\begin{lemma}[General independence constraints]\label{lemIndependenceConstraintsGeneral}
	Fix an integer $k \geq 2$ and consider a collection of independent events events $\E_1, \E_2, \dots, \E_k$, each partitioned into disjoint sets $\E_i = \E_i^+ \cup \E_i^-$.
	Then Equation (\ref{EO}) holds:
	\begin{equation}
		\prod_{\substack{s\colon [k] \to \{+, -\}\\\textnormal{such that }  
				|s^{-1}(+)|
				\textnormal{ is even}}} \P\left[\bigwedge_{i \in [k]} \E_i^{s(i)}\right] =
		\prod_{\substack{s\colon [k] \to \{+, -\}\\\textnormal{such that } |s^{-1}(+)| \textnormal{ is odd}}} \P\left[\bigwedge_{i \in [k]} \E_i^{s(i)}\right]. \tag{\ref{EO}}
	\end{equation}
\end{lemma}

\begin{proof}
	By independence, we may expand each conjunction as a product of probabilities across $i$. Switching the order of the two products, the left-hand side of (\ref{EO}) then becomes
	\begin{equation*}
		\prod_{i \in [k]} \prod_{\substack{s\colon [k] \to \{+, -\}\\\textnormal{such that }  
				|s^{-1}(+)|
				\textnormal{ is even}}} \P\left[\E_i^{s(i)}\right]
	\end{equation*}
	Next, for each fixed choice of $i$ we may decompose the second product into two pieces, based on whether $s(i) = +$, obtaining
	\begin{align*}
		\prod_{i \in [k]} \left(\prod_{\substack{s\colon [k] \setminus \{i\} \to \{+, -\}\\\textnormal{such that }  
				|s^{-1}(+)|
				\textnormal{ is odd}}} \P\left[\E_i^{+}\right]\right)\left(\prod_{\substack{s\colon [k] \setminus \{i\} \to \{+, -\}\\\textnormal{such that }  
				|s^{-1}(+)|
				\textnormal{ is even}}} \P\left[\E_i^{-}\right]\right).
	\end{align*}
	Observe that the probabilities do not depend on the choices of $s$; all that matters is the numbers of terms. And since $k \geq 2$, the two products over $s$ each have the same number of terms, namely, $2^{k - 1}$. Hence, we may rewrite the expression as
	\begin{align*}
		\prod_{i \in [k]} \left(\P\left[\E_i^{+}\right]\right)^{(2^{k - 1})}\left(\P\left[\E_i^{-}\right]\right)^{(2^{k - 1})}.
	\end{align*}
	We may analogously rewrite the right-hand side and obtain the exact same expression. This concludes the proof.
\end{proof}

For the gradient computation itself, 
it is helpful to use the notation of shuffle products first introduced by Eilenberg-Mac Lane in the 1950s. For two disjoint orderings $u_1$ and $u_2$ of  distinct elements from $\{1,\dots,m\}$, the \emph{shuffle product} $u_1 \shuffle u_2$ is the formal sum of all ways of interleaving the numbers, as in the example given previously:
$$\ordab 12 \shuffle \ordab 34 = 
\ordabcd 1234 + \ordabcd 1324 + \ordabcd 3124 + \ordabcd 3142
+\ordabcd 3412.$$

The \emph{Lie bracket} $[u_1, u_2]$ is the concatenation of $u_1$ with $u_2$, minus the concatenation of $u_2$ with $u_1$. For example,
$$\bigl[\ordab 12, \ordab 34\bigr] = \ordabcd 1234 - \ordabcd 3421.$$
We extend the shuffle product and commutators linearly,
distributing over addition.
We also adopt left-associated notation for nested
brackets by $[u_1, u_2, u_3] \coloneqq  [[u_1,
u_2], u_3]$, and so on, with the convention that $[u] = u$.

\begin{definition}[Ordering map and Lie shuffle basis]\label{def:order-map}
	Define a linear map $F\colon S_m \to \R^{O(m)}$ as follows. Put each permutation into a lexicographic normal form where each of its cycles $c_i$ is ordered with its lowest index first, and then the cycles are ordered by their lowest indices. We denote the elements of each cycle using superscript notation, $c_i = (c_i^1, c_i^2, \dots, c_i^{r_i})$, where $r_i$ is the length of the $i\tth$ cycle.
	For an individual cycle in lexicographic normal form, we define
	$F\left( c_i \right) \coloneqq [c_i^1, c_i^2, \dots, c_i^{r_i}]$.  Then for
	a general permutation, $F$ is the shuffle product of the brackets
	for the constituent cycles.  Then $\{F(\pi): \pi\in S_m\}$ is called the {\em Lie shuffle basis}.
\end{definition}

Throughout the rest of this section, we maintain the convention of writing cycles with the lowest index first, because it is important that a fixed element be left-most so that it is deepest in the nested brackets, which is used in the bracket arithmetic. We now prove Lemma~\ref{lem:gradient-body}, restated below.

\gradLem*

\begin{proof}
	For each nontrivial cycle $c_i$ with length $r_i$, we define events $\E_1^\pm$ as follows. Consider the
	expansion of $F(c_i)$; each of the $2^{r_i-1}$ terms (which are
	orderings of the indices appearing in the cycle) appears at most
	once, with coefficient $+1$ or $-1$. Let $\E_i^+$ (resp.\ $\E_i^-$)
	be the event that those indices occur in an ordering with
	coefficient~$+1$ (resp.\ $-1$), so that at the uniform distribution
	$\P(\E_i^+) = \P(\E_i^-) \eqqcolon p_i$ (with
	$p_i = 2^{r_i-2}/(r_i)!$), and set $\E_i = \E_i^+ \vee \E_i^-$ be
	the union of these events.
	
	We now look at the gradient of
	$Q \coloneqq \textit{LHS}-\textit{RHS}$ of \eqref{EO} with respect
	to these events, which by
	Lemma~\ref{lemIndependenceConstraintsGeneral} is a cotangent vector
	to $P_n$. For an arbitrary ordering $u \in O(m)$, the variable $x_u$
	appears on at most one side in~$Q$. It appears at all only if it
	satisfies all $\E_i$, and which side it appears on is determined by
	the parity of the number of $\E_i^+$ it satisfies; this is the same
	as the sign of $x_u$ in $F(\pi)$. The coefficient of $x_u$ in the
	\emph{gradient} of $Q$ is determined by the other probabilities it
	is multiplied by on the same side of \eqref{EO}. Since each such
	event $\bigvee \E_i^{s(i)}$ has the same probability $\prod p_i$,
	all these coefficients are the same, giving the desired scalar
	multiple.
\end{proof}

Finally, we state and prove a fact used in our arguments about the
dimension of $P_m$, namely that it suffices to find a single point
giving the right dimension bound.

\begin{definition}[Generic point]\label{def:generic}
	A \emph{generic point} of a simplex $\Delta^n$ is a point where the
	coordinates are algebraically independent. More generally, in an
	(irreducible)
	semi-algebraic set~$S$ defined over~$\bQ$, a generic point is a
	point that satisfies no algebraic relations with rational
	coefficients other than those satisfied by every element of~$S$.
\end{definition}

Recall that an irreducible semi-algebraic set like $P_m$ is in general
singular, but always has the structure of a stratified space as a
union of strata that are smooth manifolds (with boundary) of possibly varying
dimension. The \emph{dimension} of the semi-algebraic set is the
maximal dimension of a stratum, and all generic points
are necessarily in this stratum.

\begin{lemma}[Dimension at generic points]\label{lemGenericPoints}
	Let $M(x_1, x_2, \dots, x_m)$ be a matrix-valued function where
	each entry is polynomial in the $m$ variables. If there
	exists a point $p$ such that $M(p)$ has rank $r$, then at all
	generic points $x$, $M(x)$ has rank at least $r$.
	
	More generally,
	the same is true for an algebraic function $M$ on an irreducible
	semi-algebraic set~$S$.
\end{lemma}

\begin{proof}
	Since $M(p)$ has rank $r$, we know that $M(x)$ has an $r \times r$
	sub-matrix $S(x)$ that has full rank at $p$; i.e., $\det(S(p))
	\neq 0$. Note that $\det(S(x))$ is a polynomial function of $r^2$
	variables which is nonzero at $p$. Therefore, we cannot have
	$\det(S(x)) = 0$ at generic points, for otherwise $\det(S(x))$
	would be the zero polynomial. This means that, at generic points
	$x$, $S(x)$ is invertible, implying that $M(x)$ has rank at least
	$r$.
	
	The statement about a general semi-algebraic set~$S$ follows
	similarly; we give an argument that works in our case for
	completeness. Working in coordinates, suppose we have
	$S \subset \mathbb{R}^m$, and that $M$ is a polynomial function on
	$\mathbb{R}^m$. Then a generic point~$x$ in~$S$ is not usually
	generic in~$\mathbb{R}^m$; but by definition any equation like
	those above that is zero at~$x$ is also zero at every point~$p$.
\end{proof}

We will apply this lemma to help us with both lower and upper bounds for $\dim(P_m)$.  
For the lower bound, we consider maps {\em into} $P_m$ and for the upper bound we consider maps {\em from} $P_m$.

For the lower bound, we let $M$ be the Jacobian of the word map from a
product of simplices to $P_m$. The
rank of $M$ at any point gives a lower bound on the rank of $M$ at a
generic point, which is a lower bound on $\dim P_m$ (with equality at the
last step for
long enough word maps).

For the upper bound, we let the matrix $M$ be gradients of
independence constraints, as a function on 
$\R^{O(m)}\supset
P_m$. The rank of $M$ at the uniform distribution gives a
lower bound on the rank of $M$ at any point in $P_m$, as desired.
(Note that the rank of $M$ at a generic point in $\R^{O(m)}$ might
be larger still, but that is not relevant to $P_m$ itself.)

%%%%

%%% Local Variables:
%%% mode: latex
%%% TeX-master: "main"
%%% End:

\end{document}